\documentclass[11pt]{amsart}

\usepackage{amsfonts,amssymb}
\usepackage{amsmath}
\usepackage{xcolor}
\usepackage{pstcol,pst-node}
\usepackage{graphics}
\usepackage{epic}
\usepackage{curves}
\usepackage[matrix,arrow,frame,curve]{xy}
\CompileMatrices

\def\appendix#1{
\addtocounter{section}{1} \setcounter{equation}{0}
\renewcommand{\thesection}{\Alph{section}}
\section*{Appendix \thesection\protect\indent\quad
#1}
}
\renewcommand{\theequation}{\thesection.\arabic{equation}}

\catcode`\@=11
\def\marginnote#1{}

\newcount\hour
\newcount\minute
\newtoks\amorpm
\hour=\time\divide\hour by60 \minute=\time{\multiply\hour by60
\global\advance\minute by-\hour}
\edef\standardtime{{\ifnum\hour<12 \global\amorpm={am}%
        \else\global\amorpm={pm}\advance\hour by-12 \fi
        \ifnum\hour=0 \hour=12 \fi
        \number\hour:\ifnum\minute<10 0\fi\number\minute\the\amorpm}}
\edef\militarytime{\number\hour:\ifnum\minute<100\fi\number\minute}

\newcommand{\tcb}{\textcolor{blue}}

\def\draftlabel#1{{\@bsphack\if@filesw {\let\thepage\relax
      \xdef\@gtempa{\write\@auxout{\string
          \newlabel{#1}{{\@currentlabel}{\thepage}}}}}\@gtempa \if@nobreak
    \ifvmode\nobreak\fi\fi\fi\@esphack} \gdef\@eqnlabel{#1}}
    \def\@eqnlabel{}
\def\@vacuum{}
\def\draftmarginnote#1{\marginpar{\raggedright\scriptsise\tt#1}}

\def\draft{
  \oddsidemargin -.5truein
  \def\@oddfoot{\footnotesise \sl preliminary draft \hfil
    \rm\thepage\hfil\sl\today\quad\militarytime}
  \let\@evenfoot\@oddfoot \overfullrule 3pt
    \let\label=\draftlabel
    \let\marginnote=\draftmarginnote
  \def\@eqnnum{(\theequation)\rlap{\kern\marginparsep\tt\@eqnlabel}%
    \global\let\@eqnlabel\@vacuum}

  }

\newcommand{\beq}{\begin{equation}}
\newcommand{\eeq}{\end{equation}}
\newcommand{\tr}{\,{\rm Tr}\,}

\def\be{\begin{equation}}
\def\ee{\end{equation}}
\def\bea{\begin{eqnarray}}
\def\eea{\end{eqnarray}}
\def\<{\langle}
\def\>{\rangle}

\def\nn{\nonumber}

\def\Tr{{\rm Tr}}
\def\one#1{#1^{\raise5pt\hbox{$\scriptstyle\!\!\!\!1$}}\,{}}
\def\two#1{#1^{\raise5pt\hbox{$\scriptstyle\!\!\!\!2$}}\,{}}
\def\onetwo#1{#1^{\raise5pt\hbox{$\scriptstyle\!\!\!\!\!{12}$}}\,{}}

\newtheorem{theorem}{Theorem}[section]

\newtheorem{prop}[theorem]{Proposition}
\theoremstyle{definition}
\newtheorem{df}[theorem]{Definition}

\newtheorem{remark}[theorem]{Remark}

\theoremstyle{remark}

\allowdisplaybreaks

\begin{document}
\title[{\it Painlev\'e manifolds, cusped character varieties, cluster algebras}]
{Painlev\'e monodromy manifolds, decorated character varieties and cluster algebras.}
\author{Leonid Chekhov$^\ast$, Marta Mazzocco$^\dagger$, Vladimir Rubtsov{$^\star$}}\thanks{$^\ast$Steklov Mathematical Institute, Moscow, Russia. Email: chekhov@mi.ras.ru.$^\dagger$Department of Mathematical Sciences, Loughborough University, UK. Email: m.mazzocco@lboro.ac.uk, Phone: +44 (0)1509 223187, 
Fax: +44 (0)1509 223969. $^\star$Mathematics Department, University of Angers, France. Email: volodya@tonton.univ-angers.fr.}

\maketitle

\begin{abstract}
In this paper we introduce the concept of decorated character variety for the Riemann surfaces arising in the theory of the Painlev\'e differential equations. Since all Painlev\'e differential equations (apart from the sixth one) exhibit Stokes phenomenon, we show that it is natural to consider Riemann spheres with holes and bordered cusps on such holes. The decorated character variety is considered here as complexification of the bordered cusped Teichm\"uller space introduced in arXiv:1509.07044. We show that the decorated character variety of a Riemann sphere with s holes and $n\geq 1$ bordered cusps is a Poisson manifold of dimension $3 s+ 2 n-6$ and we explicitly compute the Poisson brackets which are naturally of cluster type. We also show how to obtain the confluence procedure of the Painlev\'e differential equations in geometric terms.
\end{abstract}

\section{Introduction}

The Painlev\'e differential equations describe monodromy preserving deformations of auxiliary linear systems of two first order ODEs on the punctured Riemann sphere. The monodromy and Stokes matrices of this linear system are encoded in the so-called {\it monodromy manifolds.}\/  For example in the case of the sixth Painlev\'e equation, that describes the isomonodromic deformations of a flat $SL_2$--connection on $\mathbb P^1$ with four regular singularities,  the monodromy manifold is realised as the Fricke-Klein affine cubic surface \cite{J}. Many mathematicians have obtained similar formulae for all other Painlev\'e equations, and recently Saito and van der Put  \cite{SvdP}  have provided a unified approach in which
they showed  the existence of ten families of affine
cubic surfaces that can be realised as monodromy manifolds for the Painlev\'e differential equations. One of the aims of our paper is to give a geometric interpretation of this classification and to prove that the confluence procedure of the Painlev\'e differential equations corresponds to a new type of surgery on Riemann surfaces called {\it chewing-gum}\/ \cite{ChM1}.

It is well known that the sixth Painlev\'e monodromy manifold is the $SL_2(\mathbb C)$ character variety of a 4 holed Riemann sphere. The real slice of this character variety is the decorated Teichm\"uller space of a 4 holed Riemann sphere, and can be combinatorially described by a fat-graph and shear coordinates. By complexifying the shear coordinates, flat coordinates for the character variety of a 4 holed Riemann sphere were found in \cite{ChM}.  For the other Painlev\'e equations, the interpretation of their monodromy manifolds as ``character varieties" of a Riemann sphere with boundary is still an 
extremely difficult problem due to the fact that the linear problems associated to the other Painlev\'e equations exhibit Stokes phenomenon. This implies that some of the boundaries have {\it bordered cusps}  on them \cite{ChM1}. Being on the boundary, these bordered cusps escape the usual notion of character variety leading to the necessity of introducing a decoration. 

In this paper we present a decoration which truly encodes the geometry of each cusped boundary. On the real slice of our decorated character variety, this decoration corresponds to choosing some horocycles to associate a $\lambda$-length to each bordered cusp.\footnote{We use the term bordered cusp meaning a vertex 
of an ideal triangle
in the Poincar\'e metric in order to distinguish it from standard cusps (without borders)
associated to punctures on a Riemann surface.}
This geometric description allows us to introduce flat coordinates in the corresponding {\it bordered cusped Teichm\"uller space} (see \cite{ChM1} for the definition of this notion) and, by complexification, on the decorated character variety.

This leads us to define explicitly a set of coordinates on the decorated character variety of the Riemann spheres with bordered cusps  which arise in the theory of the Painlev\'e differential equations and to compute the Poisson brackets in these coordinates. Such Poisson brackets coincide with the cluster algebra Poisson structure as predicted in \cite{ChM1}. 

We note that another approach to this problem was developed in \cite{PB} where the definition of wild character variety was proposed following a construction by Gaiotto, Moore  and Neitzke \cite{GMN} which consisted in introducing spurious punctures at the points of intersection between the Stokes lines and some fixed circles around each irregular singularity. This description does not seem compatible with the confluence procedure of the Painlev\'e equations nor with the case of isomonodromic deformations in which all singularities are non-simple poles (such as teh one associated to $PIII^{D_8}$), which is one of our motivations to propose a new approach.

We show that, if we exclude $PVI$, we have nine possible Riemann surfaces with bordered cusps, for which we define the decorated character variety. We
 show that in each case there is a specific Poisson sub-algebra 
that is the coordinate ring of an affine variety - the monodromy manifold of the given Painlev\'e differential equation. These monodromy manifolds can all be described by affine cubic surfaces in $\mathbb C^3$ defined by the zero locus of the corresponding polynomials in
 $\mathbb C[x_1,x_2,x_3]$ given in Table 1, where  $\omega_1,\dots,\omega_4$ are some constants related to the parameters appearing in the corresponding Painlev\'e equation as described in Section \ref{se:unitary}. 
 
 \begin{table}[h]
\begin{center} 
\begin{tabular}{|c||c|c|} \hline 
P-eqs & Polynomials \\ \hline 
$PVI$ & $x_1 x_2 x_3 +x_1^2+x_2^2+x_3 ^2+\omega_1 x_1+\omega_2 x_2+\omega_3 x_3+\omega_4$ \\ \hline 
$PV$ & $x_1 x_2 x_3 +x_1^2+x_2^2+\omega_1 x_1+\omega_2 x_2+\omega_3 x_3+1+\omega_3^2-\frac{\omega_3(\omega_2+\omega_1\omega_3)(\omega_1+\omega_2 \omega_3)}{(\omega_3^2-1)^2}$ \\ \hline
$PV_{deg}$ & $x_1 x_2 x_3 +x_1^2+x_2^2+\omega_1 x_1+\omega_2 x_2+\omega_1-1$\\ \hline 
$PIV$ & $x_1 x_2 x_3 +x_1^2+\omega_1 x_1+\omega_2 (x_2+ x_3)+\omega_2(1+\omega_1-\omega_2)$\\ \hline 
${PIII^{D_6}}$ & $x_1 x_2 x_3 +x_1^2+x_2^2+\omega_1 x_1+\omega_2 x_2+\omega_1-1$\\ \hline 
$PIII^{D_7}$ & $x_1 x_2 x_3 +x_1^2+x_2^2+\omega_1 x_1-x_2$\\ \hline 
$PIII^{D_8}$ & $x_1 x_2 x_3 +x_1^2+x_2^2-x_2$\\ \hline 
$PII^{JM}$ & $x_1 x_2 x_3 - x_1+ \omega_2 x_2- x_3-\omega_2+1$\\ \hline 
$PII^{FN}$& $ x_1 x_2 x_3 + x_1^2 +\omega_1  x_1-x_2-1$\\ \hline 
$PI$& $x_1 x_2 x_3-x_1-x_2+1$ \\ \hline
\end{tabular}
\vspace{0.2cm}
\end{center}
\caption{}
\label{tab:sing}
\end{table}

Note that in Table 1, we distinguish ten different monodromy manifolds, the $PIII^{D_6}$, $PIII^{D_7}$ and $PIII^{D_8}$ correspond to the three different cases of the third Painlev\'e equation according to Sakai's classification \cite{sakai}, and  the two monodromy manifolds  $PII^{FN}$ and $PII^{JM}$ associated to the same second Painlev\'e equation correspond to the two different isomonodromy problems found by Flaschka--Newell  \cite{FN} and Jimbo--Miwa \cite{MJ1} respectively.

\begin{figure}[h]
{\psset{unit=0.5}
\begin{pspicture}(-2,-2)(2,3.5)
\newcommand{\PATTERN}{%
{\psset{unit=1}
\rput{135}(-3,-1){\psellipse[linecolor=blue, linewidth=1pt](0,0)(.75,.5)}
\rput{45}(3,-1){\psellipse[linecolor=blue, linewidth=1pt](0,0)(.75,.5)}
\psbezier[linecolor=blue](-2.5,-1.5)(-2,-1)(2,-1)(2.5,-1.5)
\psbezier[linecolor=blue](-3.5,-.5)(-3,0)(-2,0.5)(-2,1.5)
\psbezier[linecolor=blue](3.5,-.5)(3,0)(2,0.5)(2,1.5)
\rput(0,1.5){\psellipse[linecolor=blue, linewidth=1pt](0,0)(2,0.5)}
\rput{45}(0,1.5){\psframe[linecolor=white, fillstyle=solid, fillcolor=white](-1.2,-0.5)(1.2,.5)}
}
}
\rput(0,0){\PATTERN}
\psbezier[linecolor=blue](-1.2,1.1)(-0.1,0.85)(-.5,3)(0.2,3)
\psbezier[linecolor=blue](0.15,1.05)(-0.25,1.05)(-.2,3.1)(0.5,3.1)
\psbezier[linecolor=blue](1.05,1.87)(0.25,2.07)(1,3.1)(0.5,3.1)
\psbezier[linecolor=blue](-0.1,1.95)(0.3,1.95)(.75,3)(0.25,3)
%
%
\psbezier[linecolor=red](0,2.8)(-0.7,.8)(-.5,-1.1)(0,-1.1)
\psbezier[linecolor=red, linestyle=dashed](0.45,1.05)(0.3,0)(.5,-1.1)(0,-1.1)
\psbezier[linecolor=red](0.45,1.05)(0.5,1.8)(.7,2.8)(0.4,3)
%
\end{pspicture}
\begin{pspicture}(-6,-2)(2,3.5)
\newcommand{\PATTERN}{%
{\psset{unit=1}
\rput{135}(-3,-1){\psellipse[linecolor=blue, linewidth=1pt](0,0)(.75,.5)}
\rput{45}(3,-1){\psellipse[linecolor=blue, linewidth=1pt](0,0)(.75,.5)}
\psbezier[linecolor=blue](-2.5,-1.5)(-2,-1)(2,-1)(2.5,-1.5)
\psbezier[linecolor=blue](-3.5,-.5)(-3,0)(-2,0.5)(-2,1.5)
\psbezier[linecolor=blue](3.5,-.5)(3,0)(2,0.5)(2,1.5)
\rput(0,1.5){\psellipse[linecolor=blue, linewidth=1pt](0,0)(2,0.5)}
\rput{45}(0,1.5){\psframe[linecolor=white, fillstyle=solid, fillcolor=white](-1.2,-0.5)(1.2,.5)}
}
}
\rput(0,0){\PATTERN}
\psbezier[linecolor=blue](-1.2,1.1)(-0.1,0.85)(-0.2,2.5)(-0.2,3)
\psbezier[linecolor=blue](0.15,1.05)(-0.25,1.05)(-0.2,2.5)(-0.2,3)
\psbezier[linecolor=blue](1.05,1.87)(0.25,2.07)(0.5,3)(0.5,3.4)
\psbezier[linecolor=blue](-0.1,1.95)(0.3,1.95)(.5,3)(0.5,3.4)
%
%
\psbezier[linecolor=red](-0.2,2)(-0.2,1.4)(-.5,-1.1)(0,-1.1)
\psbezier[linecolor=red, linestyle=dashed](0.45,1.05)(0.3,0)(.5,-1.1)(0,-1.1)
\psbezier[linecolor=red](0.45,1.05)(0.5,1.8)(.4,1.8)(0.4,2.6)
\rput(0.5,3.4){\pscircle[linecolor=green, linestyle=dashed,linewidth=1.5pt](0,0){0.4}}
\rput(-0.2,2.9){\pscircle[linecolor=green, linestyle=dashed,linewidth=1.5pt](0,0){0.4}}
\end{pspicture}
}
\caption{\small The process of confluence of two holes on the Riemann sphere with four holes: as a result
we obtain a Riemann sphere with one less hole, but with two new cusps on the boundary of this
hole. The geodesic line which was initially closed becomes infinite, therefore two horocycles (the dashed circles) must be introduced in order to measure its length.}
\label{fi:chewinggum}
\end{figure}
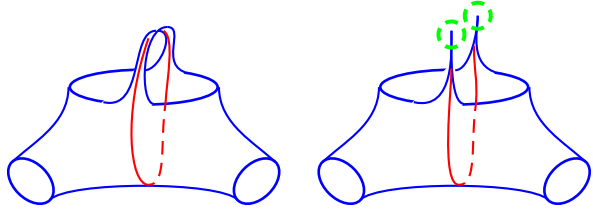

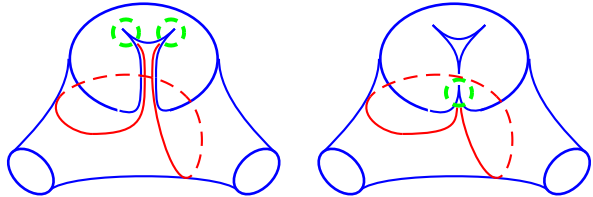
\begin{figure}[h]
{\psset{unit=0.5}
\begin{pspicture}(-2,-1.5)(2,3.5)
\newcommand{\PATTERN}{%
{\psset{unit=1}
\rput{135}(-3,-1){\psellipse[linecolor=blue, linewidth=1pt](0,0)(.75,.5)}
\rput{45}(3,-1){\psellipse[linecolor=blue, linewidth=1pt](0,0)(.75,.5)}
\psbezier[linecolor=blue](-2.5,-1.5)(-2,-1)(2,-1)(2.5,-1.5)
\psbezier[linecolor=blue](-3.5,-.5)(-3,0)(-2,1)(-2,2)
\psbezier[linecolor=blue](3.5,-.5)(3,0)(2,1)(2,2)
\rput(0,2){\psellipse[linecolor=blue, linewidth=1pt](0,0)(2,1.5)}
\rput(0,0.5){\psframe[linecolor=white, fillstyle=solid, fillcolor=white](-.7,-0.3)(.7,.3)}
}
}
\rput(0,0){\PATTERN}
\psbezier[linecolor=blue](-0.77,.62)(-0.23,0.47)(-.2,0.7)(-0.2,1.5)
\psbezier[linecolor=blue](0.7,.67)(0.2,0.52)(.2,0.7)(0.2,1.5)
\psbezier[linecolor=blue](-0.7,2.8)(-0.2,2.3)(-.2,2.1)(-0.2,1.5)
\psbezier[linecolor=blue](0.7,2.8)(0.2,2.3)(.2,2.1)(0.2,1.5)
\psbezier[linecolor=blue](-0.7,2.8)(-0.2,2.3)(.2,2.3)(0.7,2.8)
\rput(-0.6,2.7){\pscircle[linecolor=green, linestyle=dashed,linewidth=1.5pt](0,0){0.4}}
\rput(0.6,2.7){\pscircle[linecolor=green, linestyle=dashed,linewidth=1.5pt](0,0){0.4}}
%
%
\psbezier[linecolor=red](-0.3,2.4)(-0.1,2.2)(-.1,2)(-0.1,1.5)
\psbezier[linecolor=red](0.3,2.4)(0.1,2.2)(.1,2)(0.1,1.5)
\psbezier[linecolor=red](-1.5,0)(-0.1,0)(-.1,0.5)(-0.1,1.5)
\psbezier[linecolor=red](-1.5,0)(-2,0)(-2.75,0.5)(-2.35,1)
\psbezier[linecolor=red, linestyle=dashed](1,1)(0,2)(-1.95,1.5)(-2.35,1)
\psbezier[linecolor=red, linestyle=dashed](1,1)(1.7,0.3)(1.4,-1.17)(1,-1.17)
\psbezier[linecolor=red](0.1,1.5)(0.1,.5)(.6,-1.17)(1,-1.17)
%
\end{pspicture}
\begin{pspicture}(-6,-1.5)(2,3.5)
\newcommand{\PATTERN}{%
{\psset{unit=1}
\rput{135}(-3,-1){\psellipse[linecolor=blue, linewidth=1pt](0,0)(.75,.5)}
\rput{45}(3,-1){\psellipse[linecolor=blue, linewidth=1pt](0,0)(.75,.5)}
\psbezier[linecolor=blue](-2.5,-1.5)(-2,-1)(2,-1)(2.5,-1.5)
\psbezier[linecolor=blue](-3.5,-.5)(-3,0)(-2,1)(-2,2)
\psbezier[linecolor=blue](3.5,-.5)(3,0)(2,1)(2,2)
\rput(0,2){\psellipse[linecolor=blue, linewidth=1pt](0,0)(2,1.5)}
\rput(0,0.5){\psframe[linecolor=white, fillstyle=solid, fillcolor=white](-.7,-0.3)(.7,.3)}
}
}
\rput(0,0){\PATTERN}
\psbezier[linecolor=blue](-0.77,.62)(-0.23,0.47)(0,0.7)(0,1.3)
\psbezier[linecolor=blue](0.7,.67)(0.2,0.52)(0,0.7)(0,1.3)
\rput(0,0.1){\psbezier[linecolor=blue](-0.7,2.8)(-0.2,2.3)(0,2.1)(0,1.5)
\psbezier[linecolor=blue](0.7,2.8)(0.2,2.3)(0,2.1)(0,1.5)
\psbezier[linecolor=blue](-0.7,2.8)(-0.2,2.3)(.2,2.3)(0.7,2.8)}
\rput(0,1.1){\pscircle[linecolor=green, linestyle=dashed,linewidth=1.5pt](0,0){0.4}}
%
%
\psbezier[linecolor=red](-1.5,0)(-0.1,0)(-0.08,0.3)(-0.05,.7)
\psbezier[linecolor=red](-1.5,0)(-2,0)(-2.75,0.5)(-2.35,1)
\psbezier[linecolor=red, linestyle=dashed](1,1)(0,2)(-1.95,1.5)(-2.35,1)
\psbezier[linecolor=red, linestyle=dashed](1,1)(1.7,0.3)(1.4,-1.17)(1,-1.17)
\psbezier[linecolor=red](0.05,.7)(0.09,.1)(.6,-1.17)(1,-1.17)
\end{pspicture}
}
\caption{\small The process of breaking up a Riemann surface with boundary cusps: by grabbing together two cusps and pulling we 
tear apart an ideal triangle.}
\label{fi:chewinggum3}
\end{figure}

Our methodology consists in reproducing the famous confluence scheme for the Painlev\'e equations:
$$
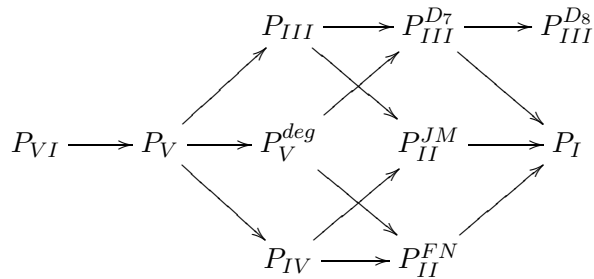
\xymatrix
 {&&P_{III}\ar[dr]\ar[r]&P_{III}^{D_7}\ar[dr]\ar[r]&P_{III}^{D_8}\\
 P_{VI}\ar[r]&P_{V}\ar[r]\ar[ur]\ar[dr]&P_{V}^{deg}\ar[dr]\ar[ur]& P_{II}^{JM}\ar[r]&P_I\\
 &&P_{IV}\ar[ur]\ar[r] & P_{II}^{FN}\ar[ur]\\
}
$$
in terms of the following two {\it chewing-gum operations}\/  on the underlying Riemann sphere:
\begin{itemize}
\item \tcb{\it Hole--hooking:} hook two holes together and stretch to infinity by keeping the area between them finite (see Fig.~\ref{fi:chewinggum}).
\item \tcb{\it Cusps removal:} pull two cusps on the same hole away by tearing off an ideal triangle (see Fig.~\ref{fi:chewinggum3}).
\end{itemize}

\begin{figure}[h]
{\psset{unit=0.4}
\begin{pspicture}(-2,-8)(2,9.5)
\newcommand{\ARROW}[2]{%
{\psset{unit=#1}
\pcline[linewidth=4pt,linecolor=white](-1,0)(0.5,0)
\pcline[linewidth=2pt,linecolor=#2](-1,0)(1,0)
\pcline[linewidth=2pt,linecolor=#2](0.4,0.6)(1,0)
\pcline[linewidth=2pt,linecolor=#2](0.4,-0.6)(1,0)
\pscircle*[linecolor=#2,fillcolor=#2](-1,0){0.2}
\pscircle*[linecolor=#2,fillcolor=#2](1,0){0.2}
\pscircle*[linecolor=#2,fillcolor=#2](0.4,0.6){0.2}
\pscircle*[linecolor=#2,fillcolor=#2](0.4,-0.6){0.2}
}
}
\newcommand{\LONGARROW}[2]{%
{\psset{unit=#1}
\pcline[linewidth=4pt,linecolor=white](-2.4,0)(2,0)
\pcline[linewidth=2pt,linecolor=#2, linestyle=dashed](-2.4,0)(2.4,0)
\pcline[linewidth=2pt,linecolor=#2](1.8,0.6)(2.4,0)
\pcline[linewidth=2pt,linecolor=#2](1.8,-0.6)(2.4,0)
\pscircle*[linecolor=#2,fillcolor=#2](-2.4,0){0.2}
\pscircle*[linecolor=#2,fillcolor=#2](2.4,0){0.2}
\pscircle*[linecolor=#2,fillcolor=#2](1.8,0.6){0.2}
\pscircle*[linecolor=#2,fillcolor=#2](1.8,-0.6){0.2}
}
}
\newcommand{\PATTERN}{%
{\psset{unit=1}
\rput{135}(-1.5,0){\psellipse[linewidth=1pt](0,0)(.5,.25)}
\rput{45}(1.5,0){\psellipse[linewidth=1pt](0,0)(.5,.25)}
\psbezier[linewidth=1pt](-1.15,-.35)(-0.8,0)(0.8,0)(1.15,-.35)
\psbezier[linewidth=1pt](-1.85,.35)(-1.5,.7)(-1.5,1)(-1.5,1.5)
\psbezier[linewidth=1pt](1.85,.35)(1.5,.7)(1.5,1)(1.5,1.5)
}
}
\newcommand{\ELLUP}{%
\rput(0,0){\psellipse[linewidth=1pt](0,0)(1.5,.5)}
\rput(0,0){\psframe[linecolor=white, fillstyle=solid, fillcolor=white](-1.6,-0.6)(1.6,0)}
}
\newcommand{\ELLIPSE}{%
\rput(0,0){\psellipse[linestyle=dashed, linewidth=1pt](0,0)(1.5,.5)}
\rput(0,0){\psframe[linecolor=white, fillstyle=solid, fillcolor=white](-1.6,-0.6)(1.6,0)}
\pcline[linewidth=1pt](-1.5,0)(-1.5,1.5)
\pcline[linewidth=1pt](1.5,0)(1.5,1.5)
}
\newcommand{\PONE}{%
\psbezier[linewidth=1pt](-1.5,0)(-1.5,0.5)(-0.75,0.5)(-0.75,1.5)
\psbezier[linewidth=1pt](1.5,0)(1.5,0.5)(-0.75,0.5)(-0.75,1.5)
}
\newcommand{\PTWO}{%
\psbezier[linewidth=1pt](-1.5,0)(-1.5,0.5)(-0.75,0.5)(-0.75,1.5)
\psbezier[linewidth=1pt](.25,1.5)(.25,0.5)(-0.75,0.5)(-0.75,1.5)
\psbezier[linewidth=1pt](1.5,0)(1.5,0.5)(0.25,.5)(0.25,1.5)
}
\newcommand{\PINVTWO}{%
\psbezier[linewidth=1pt](-1.5,0)(-1.5,-0.5)(-0.25,-0.5)(-0.25,1)
\psbezier[linewidth=1pt](-.25,1)(-.25,-0.5)(0.75,-0.5)(0.75,1)
\psbezier[linewidth=1pt](1.5,0)(1.5,-0.5)(0.75,-.5)(0.75,1)
}
\newcommand{\ANG}{%
\psarc[linewidth=1pt](1.5,0.866){0.866}{150}{270}
}
\newcommand{\ANGFIVE}{%
\psarc[linewidth=1pt](1.5,0.93){1.03}{162}{270}
}
\rput(-1,0){
\rput{180}(-12.2,3){\PATTERN}
\rput(-11.9,0){\PATTERN}
}
\rput(-6,3){\pscircle[linecolor=magenta](-0.75,0){0.3}
\pscircle[linecolor=magenta](0.25,0){0.3}}
\rput{180}(-6,1.5){\ELLUP}
\rput(-6,1.5){\PTWO}
\rput(-5.86,0){\PATTERN}
\rput(0,9){\pscircle[linecolor=magenta](-0.75,0){0.3}
\pscircle[linecolor=magenta](0.25,0){0.3}}
\rput{180}(0,4.5){\pscircle[linecolor=blue](-0.75,0){0.3}
\pscircle[linecolor=blue](0.25,0){0.3}}
\rput{180}(0,7.5){\ELLUP}
\rput(0,6){\ELLIPSE}
\rput(0,7.5){{\PTWO}}
\rput{180}(0,6){\PTWO}
\rput(0,3){\pscircle[linecolor=magenta](-0.75,0){0.3}}
\rput{180}(0,1.5){\ELLUP}
\rput(0,1.5){\PONE}
\rput(0.14,0){\PATTERN}
\rput(0,-2.9){\pscircle[linecolor=magenta](-0.75,0){0.3}
\pscircle[linecolor=magenta](0.25,0){0.3}}
\rput(0.5,-3.4){\pscircle[linecolor=green](-0.75,0){0.3}
\pscircle[linecolor=green](0.25,0){0.3}}
\rput{180}(0,-6){\ELLUP}
\rput(0,-6){\psellipse[linestyle=dashed, linewidth=1pt](0,0)(1.5,.5)}
\rput(0,-6){
\pcline[linewidth=1pt](-1.5,0)(-1.5,1.5)
\pcline[linewidth=1pt](1.5,0)(1.5,1.5)
}
\rput(0,-4.5){\PTWO}
\rput(0,-4.5){\PINVTWO}
\rput(6,9){\pscircle[linecolor=magenta](-0.75,0){0.3}}
\rput{180}(6,4.5){\pscircle[linecolor=blue](-0.75,0){0.3}
\pscircle[linecolor=blue](0.25,0){0.3}}
\rput{180}(6,7.5){\ELLUP}
\rput(6,6){\ELLIPSE}
\rput(6,7.5){{\PONE}}
\rput{180}(6,6){\PTWO}
\rput(6,1){\pscircle[linecolor=magenta](1.5,0){0.3}}
\rput{60}(6,1){\pscircle[linecolor=magenta](1.5,0){0.3}}
\rput{120}(6,1){\pscircle[linecolor=blue](1.5,0){0.3}}
\rput{180}(6,1){\pscircle[linecolor=blue](1.5,0){0.3}}
\rput{240}(6,1){\pscircle[linecolor=green](1.5,0){0.3}}
\rput{300}(6,1){\pscircle[linecolor=green](1.5,0){0.3}}
\rput(6,1.0){\ANG}
\rput{60}(6,1.0){\ANG}
\rput{120}(6,1.0){\ANG}
\rput{180}(6,1.0){\ANG}
\rput{240}(6,1.0){\ANG}
\rput{300}(6,1.0){\ANG}
\rput(6,-2.9){\pscircle[linecolor=magenta](-0.75,0){0.3}}
\rput(6.5,-3.35){\pscircle[linecolor=green](-0.75,0){0.3}
\pscircle[linecolor=green](0.25,0){0.3}}
\rput{180}(6,-6){\ELLUP}
\rput(6,-6){\psellipse[linestyle=dashed, linewidth=1pt](0,0)(1.5,.5)}
\rput(6,-6){
\pcline[linewidth=1pt](-1.5,0)(-1.5,1.5)
\pcline[linewidth=1pt](1.5,0)(1.5,1.5)
}
\rput(6,-4.5){\PONE}
\rput(6,-4.5){\PINVTWO}
\rput(12,9){\pscircle[linecolor=magenta](-0.75,0){0.3}}
\rput{180}(12,4.5){\pscircle[linecolor=blue](-0.75,0){0.3}}
\rput{180}(12,7.5){\ELLUP}
\rput(12,6){\ELLIPSE}
\rput(12,7.5){{\PONE}}
\rput{180}(12,6){\PONE}
\rput{-3}(12,1){\pscircle[linecolor=magenta](1.5,0){0.3}}
\rput{69}(12,1){\pscircle[linecolor=blue](1.5,0){0.3}}
\rput{141}(12,1){\pscircle[linecolor=blue](1.5,0){0.3}}
\rput{213}(12,1){\pscircle[linecolor=green](1.5,0){0.3}}
\rput{285}(12,1){\pscircle[linecolor=green](1.5,0){0.3}}
\rput(12,1.0){\ANGFIVE}
\rput{72}(12,1.0){\ANGFIVE}
\rput{144}(12,1.0){\ANGFIVE}
\rput{216}(12,1.0){\ANGFIVE}
\rput{288}(12,1.0){\ANGFIVE}
\rput(-9.5,1.5){\LONGARROW{0.5}{red}}
\rput(-3,1.5){\ARROW{0.5}{green}}
\rput{45}(-3,4.5){\LONGARROW{0.5}{red}}
\rput{-45}(-3,-1.5){\LONGARROW{0.5}{red}}
\rput{45}(3,4.5){\LONGARROW{0.5}{red}}
\rput{-45}(3,4.5){\LONGARROW{0.5}{red}}
\rput{45}(3,-1.5){\LONGARROW{0.5}{red}}
\rput{-45}(3,-1.5){\LONGARROW{0.5}{red}}
\rput(3,7){\ARROW{0.5}{green}}
\rput(3,-5){\ARROW{0.5}{green}}
\rput(9,7){\ARROW{0.5}{green}}
\rput{-45}(9,4.5){\LONGARROW{0.5}{red}}
\rput(9,1){\ARROW{0.5}{green}}
\rput{45}(9,-1.5){\LONGARROW{0.5}{red}}
\end{pspicture}
}
\caption{\small The table of confluences of Riemann surfaces from the Painlev\'e perspective. The long dashed arrows correspond to chewing-gum moves, the short solid ones to cusp removal.}
\label{fi:confluences}
\end{figure}
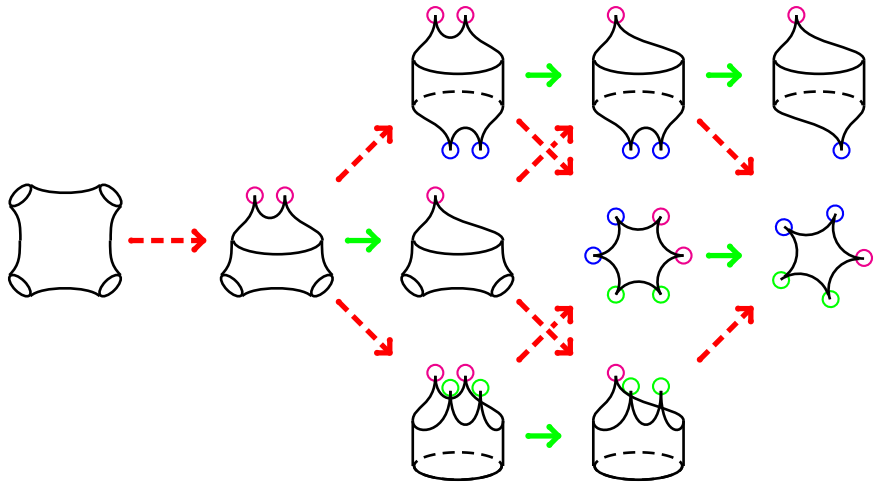

As shown by the first two authors in \cite{ChM1}, these two operations correspond to certain asymptotics in the shear coordinates and perimeters. We will deal with such asymptotics in Section \ref{se:ch-P-eqs}. The confluence process on the underlying Riemann spheres with cusped boundaries is described in Fig.~\ref{fi:confluences}.

In our work, cluster algebras appear naturally when describing the bordered cusped Teichm\"uller space of each Riemann sphere with bordered cusps. Indeed, as shown in \cite{ChM1}, when bordered cusps  arise, it is possible to introduce a {\it generalized lamination} on the Riemann surface consisting only of geodesics which start and terminate at the cusps. The geodesic length functions (well defined by fixing horocycles at each cusp) in this lamination are the coordinates in the bordered cusped Teichm\"uller space, while the decoration itself is given by the choice of horocycles. In the Poisson structure given by the Goldman bracket, these coordinates satisfy the cluster algebra Poisson bracket. This is due to the fact that the geodesics in the lamination do not intersect in the interior of the Riemann sphere, but come together asymptotically in the bordered cusps. 

We also study the corresponding cluster mutations and show that in the case of a Riemann sphere with four holes they correspond to the procedure of analytic continuation for solutions to the sixth Painlev\'e equation, thus showing that  this procedure of analytic continuation satisfies the Laurent phenomenon. For the other Painlev\'e equations, the cluster algebra mutations correspond to the action of the Mapping Class Group on the cusped lamination.

Since our decorated character variety is  the complexification of the bordered cusped Teichm\"uller space, by complexifying 
the coordinates of the latter given by the generalized laminations we obtain coordinates on the decorated character varieties.  We show that in the case of the Painlev\'e differential equations, the decorated character variety is a Poisson manifold of dimension $3 s+ 2 n-6$, where $s$ is the number of holes and $n\geq 1$ is the number of cusps. 
We show that in each case the decorated character variety admits a special Poisson sub--manifold defined by the set of functions which Poisson commute with the frozen cluster variables. This sub--manifold is defined as a 
cubic surface $\mathcal M_\phi:= {{\rm Spec}}(\mathbb C[x_1,x_2,x_3]\slash\langle\phi=0\rangle)$, where $\phi$ is one of the polynomials in Table 1, with the natural Poisson bracket defined by:
\begin{equation}\label{eq:nambu}
\{x_1,x_2\}=\frac{\partial\phi}{\partial x_3},\quad \{x_2,x_3\}=\frac{\partial\phi}{\partial x_1},\quad 
\{x_3,x_1\}=\frac{\partial\phi}{\partial x_2}.
\end{equation}
Interestingly, when confluencing the decorated character varieties, the short solid arrows in Fig. \ref{fi:confluences} must be {\it reversed,}\/ while the long dashed arrows remain pointing in the same direction. In this way we obtain a graph explaining the inclusions of the Poisson algebras on each character variety:
$$
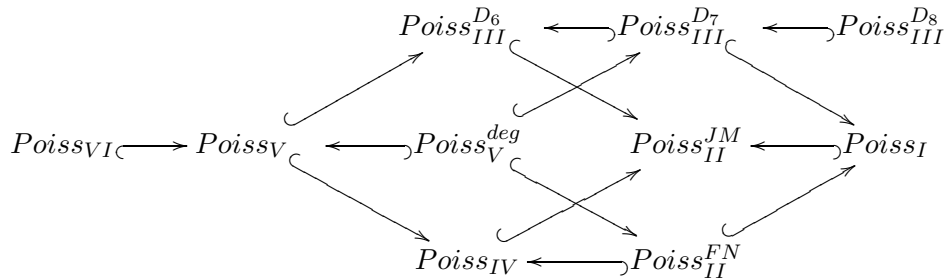
\xymatrix
 {&&Poiss_{III}^{D_6}\quad\ar@{^{(}->}[rd]&Poiss_{III}^{D_7} \ar@{^{(}->}[l]\quad\ar@{^{(}->}[rd]&Poiss_{III}^{D_8} \ar@{^{(}->}[l]\\
 Poiss_{VI}\ar@{_{(}->}[r]&Poiss_{V}\ar@{^{(}->}[ru] \quad\ar@{^{(}->}[rd]&Poiss_{V}^{deg}\ar@{^{(}->}[ru]\ar@{^{(}->}[l]\ar@{^{(}->}[rd]&Poiss_{II}^{JM}&Poiss_I\ar@{^{(}->}[l]\ \\
 &&Poiss_{IV}\ar@{^{(}->}[ru]&Poiss_{II}^{FN}\ar@{^{(}->}[l]\ar@{^{(}->}[ru]\\
}
$$
Therefore the most general Poisson algebra is the one associated to a Riemann sphere with one hole and $6$ bordered cusps (the one associated to $PII^{JM}$), and all the other Poisson algebras can be obtained as Poisson sub-algebras of this one.\footnote{We are grateful to the referee for asking to clarify this point.} 

We note that in \cite{MM} the monodromy manifolds arising in the case of the Painlev\'e differential equations were quantized to obtain the spherical sub-algebras of certain confluent versions of the Cherednik algebra of type $\check C_1 C_1$. The role of cluster algebras in the Cherednik algebra
setting will be investigated further in subsequent publications.

This paper is organized as follows: in Section \ref{se:unitary}, we recall the link between the parameters  $\omega_1,\dots,\omega_4$ and the Painlev\'e parameters $\alpha,\beta,\gamma$ and $\delta$ in each Painlev\'e equation and discuss the natural Poisson bracket (\ref{eq:nambu}) on each cubic.  In Section \ref{se:background}, we remind some important notions on the combinatorial description on the bordered cusped Teichm\"uller space. In Section \ref{se:character}, we introduce the notion of decorated character variety.
In Section \ref{se:ch-P-eqs}, we present the flat coordinates for each cubic and describe the laminations and the corresponding cluster algebra structure. In Section 
\ref{se:P-cluster}, we explain the generalized cluster algebra structure appearing in the case of $PVI$, $PV$, $PIII^{D_6}$ and $PIV$. 
In the Appendices we discuss the geometric meaning of the {\it Katz invariant}  and the  interpretation of each Painlev\'e monodromy manifold as versal deformation of a certain singularity.

\vskip 2mm \noindent{\bf Acknowledgements.} The authors are grateful to
A. Alexeev, P. Clarkson, M. Kontsevich, O.Lisovyy, V.Pestun, P. Severa, Y. Soibelman and T. Sutherland  for helpful discussions. We are thankful to B.V. Dang for his help with SINGULAR package and to the two referees for their insightful questions which have resulted in a much enhanced paper.  
The work of L.O.Ch. was partially supported by the center of excellence grant ``Centre for Quantum Geometry of Moduli Spaces'' from the Danish National Research Foundation (DNRF95) and by the Russian Foundation for Basic Research (Grant Nos. 14-01-00860-a and 13-01-12405-ofi-m2).
This research was supported by the EPSRC Research Grant $EP/J007234/1$, by the Hausdorff Institute, by ANR ``DIADEMS",  by RFBR-12-01-00525-a, by RFBR-15-01-05990,  MPIM (Bonn) and SISSA (Trieste).

\section{The Painlev\'e monodromy manifolds and their Poisson structure}\label{se:unitary}

According to \cite{SvdP}, the monodromy manifolds $\mathcal M^{(d)}$ have all the form
\be\label{eq:mon-mf}
x_1 x_2 x_3 + \epsilon_1^{(d)} x_1^2+ \epsilon_2^{(d)} x_2^2+ \epsilon_3^{(d)} x_3^2 + \omega_1^{(d)} x_1  + \omega_2^{(d)} x_2 + \omega_3^{(d)} x_3+
\omega_4^{(d)}=0,
\ee
where $d$ is an index running on the list of the Painlev\'e cubics $PVI, PV,PV_{deg},PIV$, $PIII^{D_6},PIII^{D_7}$, $PIII^{D_8},PII^{JM},PII^{FN},PI$  and the parameters $ \epsilon^{(d)}_{i},\, \omega^{(d)}_{i}$, $i=1,2,3$ are given by:
 \begin{eqnarray}
&&
 \epsilon^{(d)}_{1} =\left\{\begin{array}{ll}
 1&\hbox{ for } d= PVI, PV,PIII^{D_6},PV_{deg}, PIII^{D_7}, PIII^{D_8},PIV, PII^{FN} ,\\
0&\hbox{ for } d= PII^{JM} , PI ,\\
 \end{array}\right.\nn\\
 &&
 \epsilon^{(d)}_{2} =\left\{\begin{array}{ll}
 1&\hbox{ for } d= PVI, PV,PIII^{D_6},PV_{deg}, PIII^{D_7}, PIII^{D_8}\\
0&\hbox{ for } d= PIV, PII^{FN} , PII^{JM} , PI ,
 \end{array}\right.\nn\\
 &&
  \epsilon^{(d)}_{3} =\left\{\begin{array}{ll}
 1&\hbox{ for } d= PVI,\\
0&\hbox{ for } d= PV,PIII^{D_6},PV_{deg}, PIII^{D_7}, PIII^{D_8},PIV, PII^{FN} , PII^{JM} , PI .
 \end{array}\right.\nn
 \eea
 while 
 \bea\label{eq:omega}
 &&
 \omega^{(d)}_{1} =
-G_1^{(d)}G_\infty^{(d)}-\epsilon^{(d)}_1 G_2^{(d)} G_3^{(d)} ,\quad \omega^{(d)}_{2} =
-G_2^{(d)}G_\infty^{(d)}-\epsilon^{(d)}_2 G_1^{(d)} G_3^{(d)},\nn\\
&&
 \omega^{(d)}_{3} =
-G_3^{(d)}G_\infty^{(d)}-\epsilon^{(d)}_3 G_1^{(d)} G_2^{(d)},\\
&&
 \omega^{(d)}_{4} =\epsilon^{(d)}_2\epsilon^{(d)}_3\left(G_1^{(d)}\right)^2+\epsilon^{(d)}_1\epsilon^{(d)}_3\left(G_2^{(d)}\right)^2
 +\epsilon^{(d)}_1\epsilon^{(d)}_2\left(G_3^{(d)}\right)^2+\left(G_\infty^{(d)}\right)^2+\nn\\
 &&\qquad\quad+G_1^{(d)} G_2^{(d)}
 G_3^{(d)} G_\infty^{(d)}-4\epsilon^{(d)}_1\epsilon^{(d)}_2\epsilon^{(d)}_3,
\nn\end{eqnarray}
 where $G_1^{(d)}, G_2^{(d)}, G_3^{(d)}, G_\infty^{(d)}$ are some constants related to the parameters appearing in the Painlev\'e equations as follows:
 \begin{eqnarray}\label{eq:G-formula}
&&
 G_1^{(d)}=\left\{\begin{array}{lc}
 2\cos\pi\theta_0& d= PVI, PV,PIII^{D_6},PV_{deg}, PIV, PII^{FN} \\ 
    1&d= PIII^{D_8}, PII^{JM} , PI \\
     \infty& d= PIII^{D_7},\\
       0& d= PIII^{D_8},\\
 \end{array}\right.\nn\\
 &&
 G_2^{(d)}=\left\{\begin{array}{lc}
 2\cos\pi\theta_1& d= PVI, PV, \\ 
 2\cos\pi\theta_\infty& d=PIII^{D_6},PV_{deg},PIV, \\
  e^{i\pi\theta_0} &d= PII^{JM} \\
      1&d= PIII^{D_8}, PII^{FN} \, PI \\
          \infty& d= PIII^{D_7}, PIII^{D_8},\\
 \end{array}\right.
\nn\\
&&
 G_3^{(d)}=\left\{\begin{array}{lc}
 2\cos\pi\theta_t& d= PVI,\\
  2\cos\pi\theta_\infty& d= PV,PIV \\
      1&d= PII^{JM} ,\\
        0&d=PIII^{D_6},PV_{deg},  PIII^{D_7}, PIII^{D_8}, PII^{FN} ,  PI \\
 \end{array}\right.\\
 &&
 G_\infty^{(d)}=\left\{\begin{array}{lc}
 2\cos\pi\theta_\infty& d= PVI,PIV \\ 
        1&d= PV, PV_{deg},D_8, PII^{JM} , PII^{FN} , PI \\
  e^{i\pi \theta_0}&d=PIII^{D_6}\\
       0& d= PIII^{D_7}, PIII^{D_8}.\\
  \end{array}\right.
\nn\end{eqnarray}
 where the parameters $\theta_0,\theta_1,\theta_t,\theta_\infty$ are related to the Painlev\'e equations parameters in the usual way \cite{MJ1}. Note that for $ PIII^{D_7}$ the parameters $G_1$ and $G_2$ tend to infinity - we take this limit in such a way that  $\omega_1=-G_1G_\infty$ and $\omega_2=-G_2 G_\infty$ are not zero, while $\omega_4=0$. Similarly for $ PIII^{D_8}$.

 \begin{remark}\label{rem:PII}
Observe that in the article  \cite{SvdP} the cubic corresponding to the Flaschka--Newell isomonodromic problem \cite{FN} is  in the form $x_1 x_2 x_3+x_1 -x_2 +x_3 + 2 \cos{\pi\theta_0}=0$. This can be obtained from our cubic $ x_1 x_2 x_3 +x_1^2 + \omega_1 x_1-x_2 +1$ by the following diffeomorphism (away from $x_2 x_3=0$):
$$x_1\to - s x_1,\qquad x_2\to  \frac{1}{s} x_2,\quad 
x_3\to\frac{s^2 x_1^2-\frac{1+x_1 x_2}{s} x_3}{ x_1 x_2},
$$
where $s= 2 \cos{\pi\theta_0}$. The reason to choose the cubic in the form $PII^{FN} $ will be clear in Section  \ref{se:ch-P-eqs}.
 \end{remark}

\begin{remark}
Note that the  $ PIII^{D_7}$, $ PIII^{D_8}$ and $ PI $ cubics have different signs in \cite{SvdP}, which can both be obtained by a trivial rescaling of the variables $x_1,x_2,x_3$. 
 \end{remark}

\subsection{Natural Poisson bracket on the monodromy manifold}\label{suse:natPF}

We would like to address here some natural facts that arise when comparing the various descriptions of 
 family of affine cubics surfaces with 3 lines at infinity  (\ref{eq:mon-mf}).

First of all,  the projective completion of the family of cubics \ref{eq:mon-mf} with $\epsilon^{(d)}_i \neq 0$ for all $ i = 1,2,3$ has singular points
only in the finite part of the surface and  if any of  $\epsilon^{(d)}_i, i = 1,2,3$ vanish, then $\mathcal M^{(d)}$ is singular at infinity with singular points in 
homogeneous coordinates $X_i =1$ and $X_j = 0, j\neq i$ (\cite{Obl}). Here $x_i=\frac{X_i}{X_0}$.

One can consider this family of cubics as a variety ${\mathcal S} =\{(\bar x,\bar\omega)\in \mathbb C^3 \times\Omega): S(\bar x,\bar \omega)=0\}$ where $\bar x=(x_1, x_2, x_3),\quad \bar\omega = (\omega_1,\omega_2, \omega_3, \omega_4)$ and the ``$\bar x-$forgetful" projection $\pi :{\mathcal S}\to \Omega$ such that $\pi(\bar x, \bar \omega) = \bar\omega.$ This projection defines a family of affine cubics with generically non--singular fibres $\pi^{-1}( \bar\omega)$ (we will discuss the nature of these singularities in Appendix B).

The cubic surface $S_{\bar \omega}$ has a volume form $\vartheta_{\bar \omega}$ given by the 
Poincar\'e residue formulae:
\be\label{eq:volform}
\vartheta_{\bar \omega}=\frac{dx_1\wedge dx_2}{(\partial \phi)/(\partial x_3)}=
\frac{dx_2\wedge dx_3}{(\partial \phi)/(\partial x_1)}=\frac{dx_3\wedge dx_1}{(\partial \phi)/(\partial x_2)},
\ee
where $\phi$ is given in Table \ref{tab:sing}.

The volume form is a holomorphic 2-form on the non-singular part of $S_{\bar \omega}$ and it
has singularities along the singular locus. This form defines  the Poisson brackets on the surface
in the usual way as:
\be\label{eq:poiss1}
\{x_1, x_2\}_ {\bar \omega}=\frac{\partial \phi}{\partial x_3},\quad \{x_2, x_3\}_ {\bar \omega}=\frac{\partial \phi}{\partial x_1},\quad \{x_3, x_1\}_ {\bar \omega}=\frac{\partial \phi}{\partial x_2}.
\ee
By computing the derivative of $\phi$ we obtain the volume form (\ref{eq:volform}) as
\be\label{eq:volform1}
\vartheta_{\bar \omega}=\frac{dx_i\wedge dx_j}{(\partial \phi)/(\partial x_k)}=
\frac{dx_i\wedge dx_j}{(x_i x_j +2\epsilon_{i}^d x_k +\omega_{i}^d)},
\ee
where $(i,j,k)=(1,2,3)$ up to cyclic permutation.

In a special case of $PVI$ , i.e. the $\widetilde D_4$ cubic with parameters $\omega_i=0$ for $i=1,2,3$ and $\omega_4=-4$, there is an isomorphism $\pi : \mathbb C^*\times \mathbb C^*/\eta \to \phi:$ \cite{CantLor} 
\be\label{iso}
\pi (u,v) \to (x_1,x_2,x_3)=(-u-1/u,-v-1/v,-uv-1/uv),
\ee 
where $\eta$ is the involution of $\mathbb C^*\times \mathbb C^*$ given by
$u\to 1/u,\quad v\to 1/v.$ The log-canonical 2-form  $\bar\vartheta = \frac{du\wedge dv}{uv}$ defines a symplectic structure on $\mathbb C^*\times \mathbb C^*$ which is invariant with respect the involution $\eta$ and therefore defines a symplectic structure on the non-singular part of the cubic surface $S_{\bar\omega}$ for  $\omega_i=0$ for $i=1,2,3$ and $\omega_4=-4$. 

The relation between the log-canonical 2-form $\bar\vartheta = \frac{du\wedge dv}{uv}$ and the Poisson brackets on the surface $S_{\bar \omega}$ can be extended to all values of the parameters $\bar\omega$ and for all the 
Painlev\'e cubics as we shall show in this paper. In fact the flat coordinates that we will introduce in Section \ref{se:ch-P-eqs} are such that their exponentials satisfy the log-canonical Poisson bracket. 

 \begin{remark}
The cubic $\mathcal M^{(PVI)}$ appears in many different contexts outside of the Painlev\'e theory. For example, it was studied in 
Oblomkov' s work (see \cite{Obl}) in relation to Cherednik algebras and M. Gross, P. Hacking and S.Keel (see Example  5.12 of \cite{GrossHK}) claim that the family
 \ref{eq:mon-mf} can be ``uniformized" by some analogues of theta-functions related to toric mirror data on log-Calabi-Yau surfaces. 
More precisely, the projectivisation $Y$ of  \ref{eq:mon-mf}  with the cubic divisor $\Delta: X_1X_2X_3=0$ is an example of so called {\it Looijenga pair} and $Y\setminus \Delta$ is a log-symplectic Calabi-Yau variety with the holomorphic 2-form \ref{eq:volform}. We shall quantize this log-Calabi-Yau variety in the subsequent paper \cite{CMR}.
  \end{remark}

\section{Combinatorial description of the bordered cusped Teichm\"uller space}\label{se:background}

Let us start with the standard case, i.e. when no cusps are present. According to Fock~\cite{Fock1}~\cite{Fock2}, 
the fat graph associated to a Riemann
surface  $\Sigma_{g,s}$ of genus $g$ and with $s$ holes is a connected three--valent
graph drawn without self-intersections on $\Sigma_{g,s}$
with a prescribed cyclic ordering
of labelled edges entering each vertex; it must be a maximal graph in
the sense that its complement on the Riemann surface is a set of
disjoint polygons (faces), each polygon containing exactly one hole
(and becoming simply connected after gluing this hole).
In the case of a Riemann sphere $\Sigma_{0,4}$ with $4$ holes, the fat--graph is represented in Fig.~\ref{fgPVI}.

The geodesic length functions, which are traces of hyperbolic elements in the Fuchsian group $\Delta_{g,s}$ such that 
$$
\Sigma_{g,s}\sim\mathbb H\slash \Delta_{g,s}
$$
are obtained by decomposing each hyperbolic matrix $\gamma\in \Delta_{g,s}$ into a
product of the so--called {\it right, left and edge matrices:}
\begin{equation}\nn
R:=\left(\begin{array}{cc}1&1\\-1&0\\
\end{array}\right), \qquad
L:=\left(\begin{array}{cc}0&1\\-1&1\\
\end{array}\right),\qquad
X_{s_i}:=\left(\begin{array}{cc}0&-\exp\left({\frac{s_i}{2}}\right)\\
\exp\left(-{\frac{s_i}{2}}\right)&0\end{array}\right),
\label{eq:generators}
\end{equation}
where $s_i$ is the shear coordinate associated to the $i$-th edge in the fat graph.

In \cite{ChM1} the notion of fat-graph was extended to allow cusps. Here we present this definition adapted to the special cases dealt in the current paper:

\begin{df}\label{def-graph-cusp}
We call {\it cusped fat graph} (a graph with the prescribed cyclic ordering of edges
entering each vertex) ${\mathcal G}_{g,s,n}$ a {\em spine of the Riemann surface} $\Sigma_{g,s,n}$
with $g$ handles, $s $ holes and
$n>0$ decorated bordered cusps if
\begin{itemize}
\item[(a)] this graph can be embedded  without self-intersections in $\Sigma_{g,s,n}$;
\item[(b)] all vertices of ${\mathcal G}_{g,s,n}$ are three-valent except exactly $n$
one-valent vertices (endpoints of the open edges), which are placed at the corresponding
bordered cusps;
\item[(c)] upon cutting along all \emph{nonopen} edges of ${\mathcal G}_{g,s,n}$ the Riemann surface
$\Sigma_{g,s,n}$ splits into $s$ polygons each containing exactly one hole and being
simply connected upon contracting  this hole.
\end{itemize}
\end{df}

\begin{df}\label{def:geom-laminations}
We call geometric 
\emph{cusped geodesic lamination} (CGL) on a bordered cusped Riemann surface a set of nondirected curves up to a homotopic equivalence such that
\begin{itemize}
\item[(a)] these curves are either closed curves ($\gamma$) or \emph{arcs} ($\mathfrak a$) that start and terminate at bordered cusps 
(which can be the same cusp);
\item[(b)] these curves have no (self)intersections inside the Riemann surface (but can be incident to the same bordered cusp);
\item[(c)] these curves are not empty loops or empty loops starting and terminating at the same cusp.
\end{itemize} 
 \end{df}

In the case of arcs, the geodesic length functions are now replaced by the signed
geodesic lengths of the parts of arcs contained between two horocycles decorating the corresponding bordered
cusps; the sign is negative when these horocycles intersect. 

Combinatorially speaking this corresponds to calculating the lengths of such arcs by associating to each arc a matrix in $SL_2(\mathbb R)$ 
in the same way as before, i.e. by taking products of left, right and edge matrices,
but then by taking the trace of such product of matrices {\it multiplied by the cusp matrix:}
$K=\left(
\begin{array}{cc}
0  &  0   \\
-1 &   0  
\end{array}
\right) $ 
{\it at the right hand side of the whole expression.}\/ For example the arc $b$ in Fig.~\ref{fi:PV} has length $l_b$ such that
$$
\exp\left(\frac{l_b}{2}\right) = \Tr\left(X(k_1) R X(s_3)R X(s_2)R X(p_2) RX(s_2)L X(s_3)L X(k_1)K\right).
$$
Note that in all fat graphs in this paper we distinguish the shear coordinates $s_1,s_2,s_3$ which correspond to the edges in the central $T$ shaped part of the graph and  the shear coordinates $k_1,\dots,k_6$ correspond to the cusps which arise when breaking holes.

In \cite{ChM1} it is proved that for every cusped fat-graph with the additional property that the polygons containing holes with no cusps are  monogons, there exists a complete cusped geodesic lamination which consists only of arcs and simple loops around the un-cusped holes. Loosely speaking, this means that all lengths of any closed geodesic or of any arc in the Riemann surface is a Laurent polynomial of the lengths of the elements in the lamination.  

The Poisson brackets between lengths of arcs and closed geodesics can be computed by  using the Weil--Petersson bracket, which in shear coordinates becomes 
\cite{ChF1,ChF2}
\be
\label{WP-PB}
\bigl\{f({\mathbf Z}),g({\mathbf Z})\bigr\}=\sum_{{\hbox{\small 3-valent} \atop \hbox{\small vertices $\alpha=1$} }}^{4g+2s+n-4}
\,\sum_{i=1}^{3\,{\rm mod}\, 3}
\left(\frac{\partial f}{\partial Z_{\alpha_i}} \frac{\partial g}{\partial Z_{\alpha_{i+1}}}
- \frac{\partial g}{\partial Z_{\alpha_i}} \frac{\partial f}{\partial Z_{\alpha_{i+1}}}\right),
\ee
where $Z_\alpha $ are the shear coordinates on each edge and 
 the sum ranges all three-valent vertices of a graph and
$\alpha_i$ are the labels of the cyclically (clockwise)
ordered ($\alpha_4\equiv \alpha_1 $) edges incident to the vertex
with the label $\alpha$. 

This bracket gives rise to the {\em Goldman
bracket} on the space of geodesic length functions \cite{Gold} and in \cite{ChM1} it is proved that on the
lengths of the elements of a complete cusped geodesic lamination which consists only of arcs and simple loops this Poisson bracket gives rise to the cluster algebra Poisson structure. 

In order to describe this Poisson structure more explicitly, notice that for cusped fat-graph with the additional property that the polygons containing holes with no cusps are  monogons,  every hole with no associated bordered cusps is contained inside a closed loop, which is an edge
starting and terminating at the same three-valent vertex. Vice versa, every such closed loop corresponds 
to a hole with no associated bordered cusps. Therefore every open edge  corresponding to a bordered cusp 
``protrudes'' towards the interior of some face of the graph, and we have exactly one hole contained inside this face. 

As a consequence of these facts, we can fix an orientation of the fat graph and of each open edge which allows us to determine a natural partition of the set of bordered cusps into nonintersecting (maybe empty)
subsets $\delta_k$, $k=1,\dots,s$ of cusps incident to the corresponding holes, and to set a
 cyclic ordering in every such subset. This means that all arcs in the lamination are uniquely determined by $4$ indices, telling us in which cusp they originate and terminate and in what order they enter or exit the cusp. For example, if we orient the fat graph in Fig.~\ref{fi:PV} 
  counterclockwise so that the arc $d$ originates in cusp $2$ and is the first arc in that cusp, then it terminates in cusp $1$ and it is the eight arc in that cusp (we count arcs starting from the side of open edge that goes into the fat-graph), we can denote $d=g_{2_1,1_8}$. Analogously $b=g_{1_3,1_4}$ and  so on. The formula for the Poisson brackets is then completely combinatorial:  
\be\label{eq:comb-p}
\{g_{s_i,t_j},g_{p_r,q_l}\}=g_{s_i,t_j} g_{p_r,q_l}\frac{\epsilon_{i-r}\delta_{s,p}+\epsilon_{j-r}\delta_{t,p}+\epsilon_{i-l}\delta_{s,q}+\epsilon_{j-l}\delta_{t,q}}{4},
\ee
where $\epsilon_k:={\rm sign}(k)$.

In \cite{ChM1} it is proved that the abstract bracket defined by  (\ref{eq:comb-p}) is indeed a Poisson bracket.

\section{Decorated character variety}\label{se:character}

The classical character varieties are moduli spaces of monodromy data of regular or singular connections, which can be considered like  representation spaces of the fundamental group of a Riemann surface. Goldman proved that they are endowed with a holomorphic symplectic structure \cite{Gold}. 

It is well-known that so-called Stokes data should be added to the classical monodromy in the case of non-fuchsian irregular singularities. That is why we want to generalize the previous representation space description
to define an appropriate generalisation of the classical (or ``tame") character variety. 
Various descriptions of generalized character varieties as spaces of representations of a ``wild fundamental groupoid" \cite{PR}, ``Stokes groupoid"  \cite{Crelle2015} or as ``fissions" varieties of Stokes representations associated with a complex reductive linear algebraic group \cite{PB}. We will discuss these results at the end of this section.

In this paper we propose a different notion of decorated character variety which is based on the combinatorial description of Teichm\"uller space explained in the previous section. Our construction is based on the fact that topologically speaking a Riemann surface  $\Sigma_{g,s,n}$  with $s$ holes, with $n$ bordered cusps is equivalent to a Riemann surface $\tilde\Sigma_{g,s,n}$ of genus $g$, with $s$ holes and $n$ marked points $m_1,\dots,m_n$ on the boundaries. 

We introduce {\it the fundamental groupoid of arcs}\/ $\mathfrak U$ as the set of all directed paths $\gamma_{ij}:[0,1]\to\tilde \Sigma_{g,s,n}$ such that $\gamma_{ij}(0)=m_i$ and $\gamma_{ij}(1)=m_j$ modulo homotopy. The groupoid structure is dictated by the usual path--composition rules. 

{For each $m_j$, $j=1,\dots,n$, the isotopy group }
$$
{
\Pi_j=\{\gamma_{jj}| \gamma_{jj}:[0,1]\to\tilde \Sigma_{g,s,n},\, \gamma_{jj}(0)=m_j,\, 
\gamma_{jj}(1)=m_j\}\slash\{{\rm homotopy}\}}
$$ 
{is isomorphic to the usual fundamental group and $\Pi_j = \gamma_{ij}^{-1} \Pi_i\gamma_{ij}$ for any arc $\gamma_{ij}\in \mathfrak U$ such that $\gamma_{ij}(0)=m_i$ and $\gamma_{ij}(1)=m_j$.}
 
Using the decoration at each cusp, we associate to each arc $\gamma_{ij}$ a matrix $M_{ij}\in SL_2(\mathbb R)$ as explained in the previous section, for example  $M_{ij}=X(k_j)LX(z_n)R\cdots L X(z_1)R X(k_i)$. In order to associate a matrix in $SL_2(\mathbb C)$, we complexify the coordinates
$Z_i\in {\mathbb C}$ in all formulas. We  define  the decorated character variety\footnote{The authors are grateful to A. Alexeev and  P. Severa for their helpful remarks on the formulation of this definition.} as:
\begin{equation}\label{fga}
{\rm Hom}\left(
\mathfrak U,SL_2(\mathbb C)
\right)\slash_{\prod_{j=1}^n B_j},
\end{equation}
where $B_j$ is the Borel unipotent subgroup in $SL_2(\mathbb C)$ - we mod out by one Borel unipotent subgroup for each cusp - this is very natural as the elements in the Borel unipotent subgroup naturally preserve parabolic the choice of horocycles. As a consequence, we define the characters:
$$
\begin{array}{ll}\tr_K:& SL_2(\mathbb C) \to \mathbb C\\
& M\mapsto \Tr(MK),\qquad \hbox{where}\quad K=\left(
\begin{array}{cc}
0  &  0   \\
-1 &   0  
\end{array}
\right).\\
\end{array}
$$
The complex dimension of the decorated character variety is $6g-6+3s+2n$. Infact, let us fix a bordered cusp as base point $c_0$, we have $2g$ matrices for the usual A- and B-cycles starting and terminating at $c_0$, $s-1$ matrices corresponding to going around all holes except the one to
which the cusp $c_0$ belongs, $n-1$ matrices corresponding to paths starting at $c_0$ and terminating at other
cusps. Each matrix depends on three independent complex coordinates, giving
$3(2g+s-1+n-1)$, by taking the quotient by ${\prod_{j=1}^n B_j}$ we obtain the final result.

Note that for arcs $\gamma_{jj}$ which are born and die in the same cusp, the usual character is also defined. Geometrically speaking $\tr_K(\gamma_{jj})$ is a function of the length  of the arc  which is born and dies in cusp $j$ (to be precise it is the exponential of the singed half length of the portion of the arc outside the fixed horocycle around the $j$-th cusp), while $\tr\gamma_{jj}$  is a function of the length (the double of the hyperbolic cosine of the half length) of the closed geodesic homotopic to it.

To equip the decorated character variety with a Poisson bracket, we extend Poisson brackets (\ref{WP-PB}) to the complexified shear coordinates.  For $Z_i,Z_j\in {\mathbb C}$ we postulate $\{ Z_i,Z_j\}=\{\overline Z_i,\overline Z_j\}:=\{Z_i,Z_j\}_{{\mathbb R}}$ and $\{Z_i,\overline Z_j\}\equiv 0$ or, explicitly, $\{\Re Z_i,\Re Z_j\}=-\{\Im Z_i, \Im Z_j\}=\frac 12 \{Z_i,Z_j\}_{{\mathbb R}}$, $\{\Re Z_i,\Im Z_j\}\equiv 0$ where we let  $\{Z_i,Z_j\}_{{\mathbb R}}$ denote the (constant) Poisson bracket (\ref{WP-PB}). All formulas for Poisson brackets between characters then remain valid irrespectively on whether we consider real or complexified generalized shear coordinates $Z_i$.

The representation space of the fundamental  groupoid of arcs (\ref{fga}) (before modding out by the product of Borel unipotent subgroups) carries a quasi-Hamiltonian structure. Originally, in \cite{AMM} and \cite{AKSM} it was shown that the moduli spaces of flat connections can be obtained by a  finite-dimensional reduction procedure  (rather than infinite dimensional reduction like in the classical Atiyah-Bott definition of symplectic structure) and are equipped with a quasi-Hamiltonian structure.
This construction was generalized in \cite{PB} for the case of the moduli space of flat connections with irregular singularities and  further by Li-Bland and Severa \cite{L-BS} who gave  almost all necessary 
definitions to enlarge the quasi-Hamiltonian technic to the moduli space of connections on a surfaces with several  marked points on its boundary components.

Our approach is very much in the spirit of the case of fundamental groupoid of ``quilted varieties" (see \cite{L-BS} ). 
We want to remind that in the case of four-holed sphere ($PVI$)  the quasi-Hamiltonian structure on the representation space
$${\rm Hom}\left(\pi_1(\mathbb P^1\setminus \lbrace 0,t,1,\infty \rbrace ,SL_2(\mathbb C)\right)$$
is given by the ``Korotkin-Samtleben" quadratic  brackets between matrix elements of the representation matrices which form a ``not closed" Poisson algebra \cite{KS}. The Jacobi identity is satisfied only modding the conjugation \cite{ChM}).

In our case the Poisson structure on the decorated character variety is an example of a ``Trace-Poisson" quadratic structure of \cite{MasTur} and \cite{ARR} defined on the representation space ${\rm Hom}\left(\mathfrak U,SL_2(\mathbb C)\right)$ \cite{CMR1}.

It is interesting to compare our approach to the one by Paul and Ramis who introduced the  ``wild fundamental groupoid" for $PV$ \cite{PR}. Of course in our case, the construction is not restricted to PV, but covers all possible Riemann surfaces. Moreover our parametrisation has the  advantage of making the confluence  evident in terms of
the Poisson structure.

In the next section we will provide a set of coordinates for the decorated character variety associated to each of the Riemann surfaces appearing in Fig. \ref{fi:confluences}. These coordinates are the $\lambda$--lengths (or indeed the characters) of $6g-6+3s+2n$ arcs which form a lamination in the given Riemann surface:

\begin{prop} 
For any Riemann surface $\Sigma_{g,s,n}$ of genus $g$ with $s$ holes and $n\geq 1$ cusps there exists a complete lamination consisting of  $6g-6+3s+2n$ arcs.
\end{prop}

This proposition was proved in \cite{ChM1} (see Prop. 4.1) . However, for convenience of the reader, in the next Section we will show in each case that the $\lambda$--lengths are indeed monomials in the exponentiated shear coordinates which are independent as proved \cite{ThSh} and \cite{Penn1}.

\section{Decorated character varieties and Painlev\'e monodromy manifolds}\label{se:ch-P-eqs}

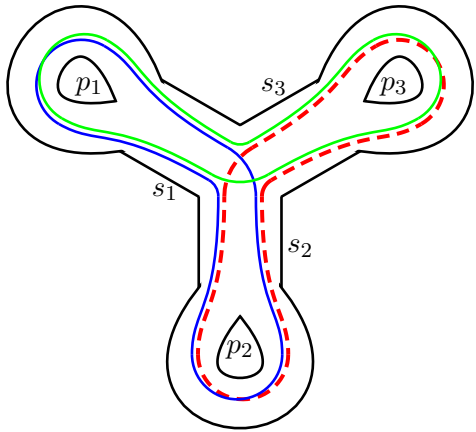
\begin{figure}
\psset{unit=0.5}
\begin{pspicture}(-5,-7)(7,5)
\newcommand{\LOOP}[1]{%
{\psset{unit=#1}
\pcline[linewidth=1pt,linecolor=black](0,0.5)(-1.5,0.5)
\pcline[linewidth=1pt,linecolor=black](0,-0.5)(-1.5,-0.5)
\psbezier[linewidth=20pt](-1,0)(-3,2)(-3,-2)(-1,0)
\psbezier[linewidth=18pt,linecolor=white](-1,0)(-3,2)(-3,-2)(-1,0)
\psframe[linewidth=1pt,linecolor=white,fillstyle=solid,fillcolor=white](0,0.47)(-1.3,-0.47)
}
}
\newcommand{\LASS}[2]{%
{\psset{unit=#1}
\psbezier[linecolor=#2, linewidth=1pt](0,.25)(-1.5,0.25)(-1.5,0.6)(-2.1,0.6)
\psbezier[linecolor=#2, linewidth=1pt](0,-.25)(-1.5,-0.25)(-1.5,-0.6)(-2.1,-0.6)
\psarc[linecolor=#2, linewidth=1pt](-2.1,0){0.6}{90}{270}
}
}
\newcommand{\LASSOP}[2]{%
{\psset{unit=#1}
\psbezier[linecolor=#2, linewidth=1pt](0,.25)(-1.5,0.25)(-1.5,0.6)(-2.1,0.6)
\psbezier[linecolor=#2, linewidth=1pt](0,-.25)(-1.5,-0.25)(-1.5,-0.6)(-2.1,-0.6)
\psarc[linecolor=#2, linewidth=1pt](-2.1,0){0.6}{90}{270}
\psarc[linecolor=#2, linewidth=1pt](0,-0.5){0.25}{30}{90}
\psarc[linecolor=#2, linewidth=1pt](0,-0.5){0.75}{30}{90}
}
}
\newcommand{\LASSD}[2]{%
{\psset{unit=#1}
\psbezier[linecolor=#2, linewidth=1.5pt,linestyle=dashed](0,.25)(-1.5,0.25)(-1.5,0.6)(-2.1,0.6)
\psbezier[linecolor=#2, linewidth=1.5pt,linestyle=dashed](0,-.25)(-1.5,-0.25)(-1.5,-0.6)(-2.1,-0.6)
\psarc[linecolor=#2, linewidth=1.5pt,linestyle=dashed](-2.1,0){0.6}{90}{270}
}
}
\newcommand{\LASSOPD}[2]{%
{\psset{unit=#1}
\psbezier[linecolor=#2, linewidth=1.5pt,linestyle=dashed](0,.25)(-1.5,0.25)(-1.5,0.6)(-2.1,0.6)
\psbezier[linecolor=#2, linewidth=1.5pt,linestyle=dashed](0,-.25)(-1.5,-0.25)(-1.5,-0.6)(-2.1,-0.6)
\psarc[linecolor=#2, linewidth=1.5pt,linestyle=dashed](-2.1,0){0.6}{90}{270}
\psarc[linecolor=#2, linewidth=1.5pt,linestyle=dashed](0,-0.5){0.25}{30}{90}
\psarc[linecolor=#2, linewidth=1.5pt,linestyle=dashed](0,-0.5){0.75}{30}{90}
}
}
\rput(0.9,2.3){$s_3$}
\rput(4.1,2.3){$p_3$}
\rput(-2,-0.5){$s_1$}
\rput(-4,2.3){$p_1$}
\rput(1.6,-2){$s_2$}
\rput(0,-4.7){$p_2$}
\rput{90}(0,0){\rput(-0.5,0){\LOOP{2.2}}}
\rput{210}(0,0){\rput(-0.5,0){\LOOP{2.2}}}
\rput{330}(0,0){\rput(-0.5,0){\LOOP{2.2}}}
\rput(0.08,-0.05){
\rput{90}(0,0){\rput(-0.45,0){\LASSOPD{2}{red}}}
\rput{210}(0,0){\rput(-0.45,0){\LASSD{2}{red}}}
}
\rput(-0.08,-0.05){
\rput{-30}(0,0){\rput(-0.45,0){\LASSOP{2}{blue}}}
\rput{90}(0,0){\rput(-0.45,0){\LASS{2}{blue}}}
}
\rput(0,0.1){
\rput{-150}(0,0){\rput(-0.45,0){\LASSOP{2}{green}}}
\rput{-30}(0,0){\rput(-0.45,0){\LASS{2}{green}}}
}
\end{pspicture}
\caption{The fat graph of the $4$ holed Riemann sphere. The dashed geodesic is $x_1$; the solid geodesics are $x_2$ and $x_3$.}
\label{fgPVI} 
\end{figure}

In the case of a Riemann sphere with $4$ holes and no cusps, the fat graph is given in Fig.~\ref{fgPVI} and the three geodesics  lengths $x_1,x_2, x_3$ of the three geodesics which go around two holes without self--intersections are enough to close the Poisson algebra.

By following the rules explained in Section \ref{se:background}, the following parameterisation of $x_1,x_2,x_3$ in shear coordinates on the fat-graph of a $4$--holed sphere was found in \cite{ChM}:
\begin{eqnarray}
\label{eq:shear-PVI}
x_1&=&-e^{s_2+ s_3+\frac{p_2}{2}+\frac{p_3}{2}}-e^{-  s_2-s_3-\frac{p_2}{2}-\frac{p_3}{2}}-e^{s_2- s_3+\frac{p_2}{2}-\frac{p_3}{2}}-G_2e^{-s_3-\frac{p_3}{2}}-G_3 e^{s_2+\frac{p_2}{2}},\nn\\
x_2&=&-e^{s_3+s_1+\frac{p_3}{2}+\frac{p_1}{2}}-e^{-s_3-s_1-\frac{p_3}{2}-\frac{p_1}{2}}-e^{s_3-s_1+\frac{p_3}{2}-\frac{p_1}{2}}-G_3 e^{-s_1-\frac{p_1}{2}}-G_1e^{s_3+\frac{p_3}{2}},\nn\\
x_3&=&-e^{s_1+ s_2+\frac{p_1}{2}+\frac{p_2}{2}}-e^{-s_1- s_2-\frac{p_1}{2}-\frac{p_2}{2}}-e^{s_1-s_2+\frac{p_1}{2}-\frac{p_2}{2}}-G_1e^{-s_2-\frac{p_2}{2}}-G_2 e^{s_1+\frac{p_1}{2}},\nn\\
\end{eqnarray}
where 
$$
G_i=e^{\frac{p_i}{2}}+e^{-\frac{p_i}{2}},\qquad i=1,2,3,
$$
and 
$$
G_\infty=e^{s_1+s_2+s_3+\frac{p_1}{2}+\frac{p_2}{2}+\frac{p_3}{2}}+e^{-s_1-s_2-s_3-\frac{p_1}{2}-\frac{p_3}{2}-\frac{p_3}{2}}.
$$
Note that by complexifying $s_1,s_2,s_3,p_1,p_2,p_3$, we obtain a parameterisation of the $PVI$ cubic, i.e. of the character variety of $SL_2(\mathbb C)$ character variety of a Riemann sphere with $4$ holes.

We are now going to produce a similar coordinate description of each of the other Painlev\'e cubics. We will provide a geometric description of the corresponding Riemann surface and its fat-graph and discuss the corresponding decorated character variety.

\subsection{Decorated character variety for $ PV$}

The confluence from the cubic associated to $PVI$ to the one associated to $PV$ is realised by 
$$
p_3\to p_3 -2\log[\epsilon],
$$
in the limit $\epsilon\to0$.  We obtain the following shear coordinate description for the $ PV$ cubic:
\begin{eqnarray}
\label{eq:shear-PV}
x_1&=&-e^{s_2+ s_3+\frac{p_2}{2}+\frac{p_3}{2}}-G_3 e^{s_2+\frac{p_2}{2}},\nn\\
x_2&=&-e^{s_3+s_1+\frac{p_3}{2}+\frac{p_1}{2}}-e^{s_3-s_1+\frac{p_3}{2}-\frac{p_1}{2}}-G_3 e^{-s_1-\frac{p_1}{2}}-G_1e^{s_3+\frac{p_3}{2}},\nn\\
x_3&=&-e^{s_1+ s_2+\frac{p_1}{2}+\frac{p_2}{2}}-e^{-s_1- s_2-\frac{p_1}{2}-\frac{p_2}{2}}-e^{s_1-s_2+\frac{p_1}{2}-\frac{p_2}{2}}-G_1e^{-s_2-\frac{p_2}{2}}-G_2 e^{s_1+\frac{p_1}{2}},\nn\\
\end{eqnarray}
where 
$$
G_i=e^{\frac{p_i}{2}}+e^{-\frac{p_i}{2}},\quad i=1,2,
\quad G_3=e^{\frac{p_3}{2}},\quad
G_\infty=e^{s_1+s_2+s_3+\frac{p_1}{2}+\frac{p_2}{2}+\frac{p_3}{2}}.
$$
These coordinates satisfy the following cubic relation:
\bea\label{eq:gen-p5}
&&
x_1x_2x_3 + x_1^2+x_2^2-(G_1 G_\infty+G_2 G_3) x_1 -(G_2 G_\infty+G_1 G_3) x_2  -\nn\\
&&-G_3 G_\infty x_3+ G_\infty^2+G_3^2+G_1 G_2G_3 G_\infty=0.
\eea
Note that the parameter $p_3$ is now redundant, we can eliminate it by rescaling.
To obtain the correct $PV$ cubic, we need to pick $p_3=-p_1-p_2-2 s_1-2 s_2-2 s_3$ so that $G_\infty=1$.

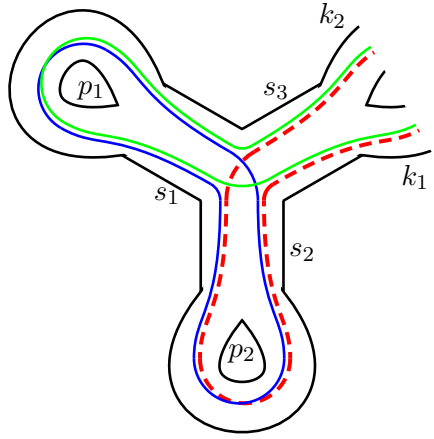
\begin{figure}
\psset{unit=0.5}
\begin{pspicture}(-5,-7)(7,5)
\newcommand{\LOOP}[1]{%
{\psset{unit=#1}
\pcline[linewidth=1pt,linecolor=black](0,0.5)(-1.5,0.5)
\pcline[linewidth=1pt,linecolor=black](0,-0.5)(-1.5,-0.5)
\psbezier[linewidth=20pt](-1,0)(-3,2)(-3,-2)(-1,0)
\psbezier[linewidth=18pt,linecolor=white](-1,0)(-3,2)(-3,-2)(-1,0)
\psframe[linewidth=1pt,linecolor=white,fillstyle=solid,fillcolor=white](0,0.47)(-1.3,-0.47)
}
}
\newcommand{\LASS}[2]{%
{\psset{unit=#1}
\psbezier[linecolor=#2, linewidth=1pt](0,.25)(-1.5,0.25)(-1.5,0.6)(-2.1,0.6)
\psbezier[linecolor=#2, linewidth=1pt](0,-.25)(-1.5,-0.25)(-1.5,-0.6)(-2.1,-0.6)
\psarc[linecolor=#2, linewidth=1pt](-2.1,0){0.6}{90}{270}
}
}
\newcommand{\LASSOP}[2]{%
{\psset{unit=#1}
\psbezier[linecolor=#2, linewidth=1pt](0,.25)(-1.5,0.25)(-1.5,0.6)(-2.1,0.6)
\psbezier[linecolor=#2, linewidth=1pt](0,-.25)(-1.5,-0.25)(-1.5,-0.6)(-2.1,-0.6)
\psarc[linecolor=#2, linewidth=1pt](-2.1,0){0.6}{90}{270}
\psarc[linecolor=#2, linewidth=1pt](0,-0.5){0.25}{30}{90}
\psarc[linecolor=#2, linewidth=1pt](0,-0.5){0.75}{30}{90}
}
}
\newcommand{\LASSD}[2]{%
{\psset{unit=#1}
\psbezier[linecolor=#2, linewidth=1.5pt,linestyle=dashed](0,.25)(-1.5,0.25)(-1.5,0.6)(-2.1,0.6)
\psbezier[linecolor=#2, linewidth=1.5pt,linestyle=dashed](0,-.25)(-1.5,-0.25)(-1.5,-0.6)(-2.1,-0.6)
\psarc[linecolor=#2, linewidth=1.5pt,linestyle=dashed](-2.1,0){0.6}{90}{270}
}
}
\newcommand{\LASSOPD}[2]{%
{\psset{unit=#1}
\psbezier[linecolor=#2, linewidth=1.5pt,linestyle=dashed](0,.25)(-1.5,0.25)(-1.5,0.6)(-2.1,0.6)
\psbezier[linecolor=#2, linewidth=1.5pt,linestyle=dashed](0,-.25)(-1.5,-0.25)(-1.5,-0.6)(-2.1,-0.6)
\psarc[linecolor=#2, linewidth=1.5pt,linestyle=dashed](-2.1,0){0.6}{90}{270}
\psarc[linecolor=#2, linewidth=1.5pt,linestyle=dashed](0,-0.5){0.25}{30}{90}
\psarc[linecolor=#2, linewidth=1.5pt,linestyle=dashed](0,-0.5){0.75}{30}{90}
}
}
\rput(0.9,2.3){$s_3$}
\rput(-2,-0.5){$s_1$}
\rput(-4,2.3){$p_1$}
\rput(1.6,-2){$s_2$}
\rput(0,-4.7){$p_2$}
\rput{90}(0,0){\rput(-0.5,0){\LOOP{2.2}}}
\rput{210}(0,0){\rput(-0.5,0){\LOOP{2.2}}}
\rput{330}(0,0){\rput(-0.5,0){\LOOP{2.2}}}
\rput(0.08,-0.05){
\rput{90}(0,0){\rput(-0.45,0){\LASSOPD{2}{red}}}
\rput{210}(0,0){\rput(-0.45,0){\LASSD{2}{red}}}
}
\rput(-0.08,-0.05){
\rput{-30}(0,0){\rput(-0.45,0){\LASSOP{2}{blue}}}
\rput{90}(0,0){\rput(-0.45,0){\LASS{2}{blue}}}
}
\rput(0,0.1){
\rput{-150}(0,0){\rput(-0.45,0){\LASSOP{2}{green}}}
\rput{-30}(0,0){\rput(-0.45,0){\LASS{2}{green}}}
}
\rput{30}(6,1){\psframe[linewidth=1pt,linecolor=white,fillstyle=solid,fillcolor=white](-1,-3)(2,4.5)}
\rput(2.4,4.3){$k_2$}
\rput(4.6,0){$k_1$}
\end{pspicture}
\caption{The fat graph corresponding to PV.   The geodesic $x_3$ remains closed, while $x_1$ (the dashed line) and $x_2$ become arcs.}
\label{fatgraphV}
\end{figure}

Geometrically speaking, sending the perimeter $p_3$ to infinity means that we are performing a chewing-gum move: two holes, one of perimeter $p_3$ and the other of perimeter $s_1+s_2+s_3+\frac{p_1}{2}+\frac{p_2}{2}+\frac{p_3}{2}$, become infinite, but the area between them remains finite, thus leading to a Riemann sphere with three holes and two cusps on one of them. In terms of the fat-graph, this is represented by Fig.~\ref{fatgraphV}. 

The geodesic $x_3$ corresponds to the closed loop obtained going around $p_1$ and $p_2$, while $x_1$ and $x_2$ are  arcs starting at one cusp, going around $p_1$ and $p_2$ respectively, and coming back to the other cusp.

\begin{figure}[h]
{\psset{unit=0.5}
\begin{pspicture}(-2,-8)(2,3)
\newcommand{\TAIL}[3]{%
{\psset{unit=#1}
\pcline[linewidth=1pt,linecolor=black](0,0.5)(0.5,0.5)
\pcline[linewidth=1pt,linecolor=black](0,-0.5)(0.5,-0.5)
\pcline[linewidth=1pt,linecolor=black](0.5,0.5)(1,1)
\pcline[linewidth=1pt,linecolor=black](0.5,-0.5)(1,-1)
\pcline[linewidth=1pt,linecolor=magenta](0,0.4)(0.5,0.4)
\pcline[linewidth=1pt,linecolor=green](0,0.3)(0.5,0.3)
\pcline[linewidth=1pt,linecolor=blue](0,0.1)(0.5,0.1)
\pcline[linewidth=1pt,linecolor=#2](0,0)(0.5,0)
\pcline[linewidth=1pt,linecolor=#3](0,-0.1)(0.5,-0.1)
\pcline[linewidth=1pt,linecolor=yellow](0,-0.2)(0.5,-0.2)
\pcline[linewidth=1pt,linecolor=green](0,-0.3)(0.5,-0.3)
\pcline[linewidth=1pt,linecolor=magenta](0,-0.4)(0.5,-0.4)
\psbezier[linewidth=1pt,linecolor=magenta](0.5,-0.4)(0.55,-0.4)(0.55,-0.4)(1.07,-0.93)
\psbezier[linewidth=1pt,linecolor=green](0.5,-0.3)(0.6,-0.3)(0.6,-0.3)(1.14,-0.86)
\psbezier[linewidth=1pt,linecolor=yellow](0.5,-0.2)(0.65,-0.2)(0.65,-0.2)(1.21,-0.79)
\psbezier[linewidth=1pt,linecolor=#3](0.5,-0.1)(0.7,-0.1)(0.7,-0.1)(1.28,-0.72)
\psbezier[linewidth=1pt,linecolor=#2](0.5,0)(0.75,0)(0.75,0)(1.35,-0.65)
\psbezier[linewidth=1pt,linecolor=blue](0.5,0.1)(0.8,0.1)(0.8,0.1)(1.42,-0.58)
\psbezier[linewidth=1pt,linecolor=green](0.5,0.3)(0.85,0.3)(0.85,0.3)(1.49,-0.51)
\psbezier[linewidth=1pt,linecolor=magenta](0.5,0.4)(0.55,0.4)(0.55,0.4)(1.07,0.93)
\psbezier[linewidth=1pt,linecolor=gray](1.56,-0.44)(1.12,0)(1.12,0)(1.56,0.44)
\pcline[linewidth=1pt,linecolor=black](1.65,0.35)(1.3,0)
\pcline[linewidth=1pt,linecolor=black](1.65,-0.35)(1.3,0)
\pcline[linewidth=0.5pt,linecolor=black]{->}(0.7,-1.5)(.77,-0.63)
\rput(.7,-1.62){$d$}
\pcline[linewidth=0.5pt,linecolor=black]{->}(0.9,-1.5)(.91,-0.63)
\rput(.9,-1.65){$c$}
\pcline[linewidth=0.5pt,linecolor=black]{->}(1.1,-1.5)(1.05,-0.63)
\rput(1.1,-1.65){$a$}
\pcline[linewidth=0.5pt,linecolor=black]{->}(1.3,-1.5)(1.33,-0.49)
\rput(1.3,-1.62){$b$}
\pcline[linewidth=0.5pt,linecolor=black]{->}(1.5,-1.5)(1.47,-0.35)
\rput(1.5,-1.65){$e$}
\rput(1.75,-0.2){$k_1$}
\rput(1.75,0.2){$k_2$}
\rput(0.25,0.65){$s_3$}
\rput(-1.3,0.65){$s_1$}
\rput(0.15,-0.75){$s_2$}
\rput(0.6,-1.95){$p_2$}
\rput(-2.35,-1.15){$p_1$}
}
}
\newcommand{\LOOPS}[1]{%
{\psset{unit=#1}
\pcline[linewidth=1pt,linecolor=black](0,0.5)(-1.5,0.5)
\pcline[linewidth=1pt,linecolor=black](0,-0.5)(0,-1)
\pcline[linewidth=1pt,linecolor=black](-1.1,-0.5)(-1.1,-1)
\pcline[linewidth=1pt,linecolor=black](-1.1,-0.5)(-1.5,-0.5)
\psarc[linecolor=black, linewidth=1pt](-0.55,-1.85){1}{-238}{58}
\pscircle[linecolor=black, linewidth=1pt](-0.55,-1.85){0.2}
\psarc[linecolor=black, linewidth=1pt](-2.35,0){1}{30}{330}
\pscircle[linecolor=black, linewidth=1pt](-2.35,0){0.2}
\psarc[linecolor=magenta, linewidth=1pt](0,-0.5){0.1}{90}{180}
\psarc[linecolor=green, linewidth=1pt](0,-0.5){0.2}{90}{180}
\psarc[linecolor=yellow, linewidth=1pt](0,-0.5){0.3}{90}{180}
\psarc[linecolor=yellow, linewidth=1pt](0,-0.5){0.4}{90}{180}
\psarc[linecolor=blue, linewidth=1pt](0,-0.5){0.5}{90}{180}
\psarc[linecolor=blue, linewidth=1pt](0,-0.5){0.6}{90}{180}
\psarc[linecolor=magenta, linewidth=1pt](-1.1,-0.5){0.1}{0}{90}
\psarc[linecolor=green, linewidth=1pt](-1.1,-0.5){0.2}{0}{90}
\psarc[linecolor=yellow, linewidth=1pt](-1.1,-0.5){0.3}{0}{90}
\psarc[linecolor=yellow, linewidth=1pt](-1.1,-0.5){0.4}{0}{90}
\pcline[linewidth=1pt,linecolor=magenta](0,0.4)(-1.5,0.4)
\pcline[linewidth=1pt,linecolor=green](0,0.3)(-1.5,0.3)
\pcline[linewidth=1pt,linecolor=magenta](-1.1,-0.4)(-1.5,-0.4)
\pcline[linewidth=1pt,linecolor=green](-1.1,-0.3)(-1.5,-0.3)
\pcline[linewidth=1pt,linecolor=magenta](-0.1,-0.5)(-0.1,-1)
\pcline[linewidth=1pt,linecolor=green](-0.2,-0.5)(-0.2,-1)
\pcline[linewidth=1pt,linecolor=magenta](-1,-0.5)(-1,-1)
\pcline[linewidth=1pt,linecolor=green](-0.9,-0.5)(-0.9,-1)
\pcline[linewidth=1pt,linecolor=blue](-0.5,-0.5)(-0.5,-1)
\pcline[linewidth=1pt,linecolor=blue](-0.6,-0.5)(-0.6,-1)
\pcline[linewidth=1pt,linecolor=yellow](-0.3,-0.5)(-0.3,-1)
\pcline[linewidth=1pt,linecolor=yellow](-0.4,-0.5)(-0.4,-1)
\pcline[linewidth=1pt,linecolor=yellow](-0.7,-0.5)(-0.7,-1)
\pcline[linewidth=1pt,linecolor=yellow](-0.8,-0.5)(-0.8,-1)
\pcline[linewidth=1pt,linecolor=yellow](-1.1,-0.2)(-1.5,-0.2)
\pcline[linewidth=1pt,linecolor=yellow](-1.1,-0.1)(-1.5,-0.1)
\psarc[linecolor=magenta, linewidth=1pt](-0.55,-1.85){0.85}{180}{360}
\psarc[linecolor=green, linewidth=1pt](-0.55,-1.85){0.75}{180}{360}
\psarc[linecolor=yellow, linewidth=1pt](-0.55,-1.85){0.6}{180}{360}
\psarc[linecolor=yellow, linewidth=1pt](-0.55,-1.85){0.5}{180}{360}
\psarc[linecolor=blue, linewidth=1pt](-0.55,-1.85){0.35}{180}{360}
\psarc[linecolor=magenta, linewidth=1pt](-2.35,0){0.8}{90}{270}
\psarc[linecolor=green, linewidth=1pt](-2.35,0){0.7}{90}{270}
\psarc[linecolor=yellow, linewidth=1pt](-2.35,0){0.4}{90}{270}
\psbezier[linewidth=1pt,linecolor=magenta](-0.1,-1)(-0.1,-1.2)(0.3,-1.5)(0.3,-1.85)
\psbezier[linewidth=1pt,linecolor=magenta](-1,-1)(-1,-1.2)(-1.4,-1.5)(-1.4,-1.85)
\psbezier[linewidth=1pt,linecolor=green](-0.2,-1)(-0.2,-1.2)(0.2,-1.5)(0.2,-1.85)
\psbezier[linewidth=1pt,linecolor=green](-0.9,-1)(-0.9,-1.2)(-1.3,-1.5)(-1.3,-1.85)
\psbezier[linewidth=1pt,linecolor=yellow](-0.3,-1)(-0.3,-1.2)(0.05,-1.5)(0.05,-1.85)
\psbezier[linewidth=1pt,linecolor=yellow](-0.4,-1)(-0.4,-1.2)(-0.05,-1.5)(-0.05,-1.85)
\psbezier[linewidth=1pt,linecolor=yellow](-0.8,-1)(-0.8,-1.2)(-1.15,-1.5)(-1.15,-1.85)
\psbezier[linewidth=1pt,linecolor=yellow](-0.7,-1)(-0.7,-1.2)(-1.05,-1.5)(-1.05,-1.85)
\psbezier[linewidth=1pt,linecolor=blue](-0.5,-1)(-0.5,-1.2)(-0.2,-1.5)(-0.2,-1.85)
\psbezier[linewidth=1pt,linecolor=blue](-0.6,-1)(-0.6,-1.2)(-0.9,-1.5)(-0.9,-1.85)
\psbezier[linewidth=1pt,linecolor=magenta](-2.35,0.8)(-2,0.8)(-1.7,0.4)(-1.5,0.4)
\psbezier[linewidth=1pt,linecolor=magenta](-2.35,-0.8)(-2,-0.8)(-1.7,-0.4)(-1.5,-0.4)
\psbezier[linewidth=1pt,linecolor=green](-2.35,0.7)(-2,0.7)(-1.7,0.3)(-1.5,0.3)
\psbezier[linewidth=1pt,linecolor=green](-2.35,-0.7)(-2,-0.7)(-1.7,-0.3)(-1.5,-0.3)
\psbezier[linewidth=1pt,linecolor=yellow](-2.35,0.4)(-2,0.4)(-1.7,-0.1)(-1.5,-0.1)
\psbezier[linewidth=1pt,linecolor=yellow](-2.35,-0.4)(-2,-0.4)(-1.7,-0.2)(-1.5,-0.2)
}
}
\rput(0,0){\TAIL{3}{blue}{yellow}}
\rput(0,0){\LOOPS{3}}
\end{pspicture}
}
\caption{\small The system of arcs for PV.}
\label{fi:PV}
\end{figure}
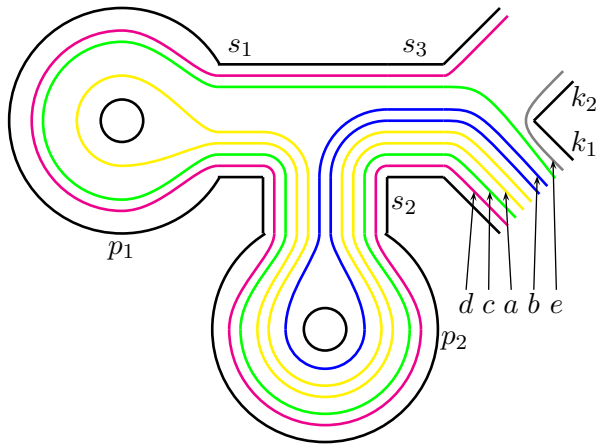

As explained in Section \ref{se:background}, according to \cite{ChM1}, the Poisson algebra related to the character variety of a Riemann sphere with three holes and two cusps on one boundary is $7$-dimensional. The fat-graph admits a complete cusped lamination as displayed in Fig.~\ref{fi:PV} 
so that a full set of coordinates on the character variety is given by the complexification of the five elements in the lamination and of the two parameters $G_1$ and $G_2$ which determine the perimeter of the two non-cusped holes.

\begin{remark}\label{rem:PV-SvdP}
Observe that in the  article  \cite{SvdP} the cubic corresponding to the $PV$ has exactly the same form as (\ref{eq:gen-p5}):
\bea\label{eq:gen-p5-1}
C&:=&d_1d_2d_3 + d_1^2+d_2^2-(s_1 + s_2 s_3) d_1 -(s_2 + s_1 s_3) d_2  -\nn\\
&&-s_3 d_3+ 1 +s_3^2+s_1 s_2s_3 =0,
\eea
The parameters $s_i$ in (\ref{eq:gen-p5-1})  coincide with $G_i$ in (\ref{eq:gen-p5}) for $i=1,2,3$ (recall that $G_{\infty} =1$), however the coordinates $ x_1, x_2, x_3$ are quite different from $ d_1, d_2, d_3$  in (\ref{eq:gen-p5-1}). Indeed the  parameters $d_i$ are the matrix elements of the triple of matrices $(M_1,M_2,M_{\infty})$ which parametrise the differential system data in \cite{SvdP} and satisfy the condition $M_1M_2M_{\infty}=\mathbb I$ . The matrix $M_1$ corresponds to our
loop $\gamma_b$, $M_2$ corresponds to $\gamma_a$ and $M_{\infty}$ is given by the topological monodromy $\exp(2\pi i \Lambda S_1 S_2)$ where $S_{1,2}$ are Stokes matrices and $\Lambda$ is diagonal ( the ``formal monodromy"). Our $x_1,x_2,x_3$ instead are  
$\lambda$-lengths of certain arcs, for example  $x_3 = \tr_k (\gamma_e\gamma_b)\neq eb = \tr_k e \tr_k b$. 
To compare the triples $d_i$ and $x_i$ for $i=1,2,3$ one should observe that both  $ x_1, x_2, x_3$ are $ d_1, d_2, d_3$  are Laurent polynomials of our $\lambda$-lengths $a,b,c,d,e$ so that indeed these is a bijection between these two sets of coordinates.
This phenomenon presents itself for all other cubics and we won't repeat this discussion in each case.
\end{remark}

\begin{remark}\label{rem:PV-fibration}
The family (\ref{eq:gen-p5-1}) appears in context of the Riemann-Hilbert fibraiton map and the corresponding affine coordinate rings admit the inclusion:
$$
\mathbb C[s_1,s_2,s_3,s_3^{-1}] \hookrightarrow \mathbb C[d_1,d_2,d_3,s_1,s_2,s_3,s_3^{-1}]/(C).
$$
This inclusion corresponds the projection of varieties:
$$
{\rm Spec}(\mathbb C[d_1,d_2,d_3,s_1,s_2,s_3,s_3^{-1}]/(C)) \mapsto {\rm Spec}(\mathbb C[s_1,s_2,s_3,s_3^{-1}]).
$$
This is a $5$--dimensional fibration whose fibers over the $3$--dimensional base $s_1,s_2,s_3 \in \mathbb C\times \mathbb C \times \mathbb C$ are singular cubic surfaces in $\mathbb C^3$, thus giving the $PV$--analogue of the fibrated family ${\mathcal S}(\bar x,\bar \omega)$ described in the subsection \ref{suse:natPF}.
\end{remark}

Notice that there are two shear coordinates associated to the two cusps, they are denoted by $k_1$ and $k_2$, their sum corresponds to what we call $p_3$ in (\ref{eq:shear-PV}). These shear coordinates do not commute with the other ones, they in fact satisfy the following relations:
$$
\{s_3,k_1\}=\{k_1,k_2\}=\{k_2,s_3\}=1.
$$
As a consequence, 
the elements $G_3$ and $G_\infty$ are not Casimirs in this Poisson algebra, despite being frozen variables in the cluster algebra setting (see Section \ref{se:P-cluster})

In terms of shear coordinates, the elements in the lamination correspond to two loops (whose hyperbolic cosin length is denoted by $G_1$ and $G_2$ respectively) and five arcs whose lengths are expressed as follows:
\bea\label{PV-lam}
&&
a= e^{k_1+s_1+2 s_2+s_3+\frac{p_1}{2}+p_2},\qquad
b= e^{k_1+s_2+s_3+\frac{p_2}{2}},\qquad e= e^{\frac{k_1}{2}+\frac{k_2}{2}},\nn\\
&&
c= e^{k_1+s_1+ s_2+s_3+\frac{p_1}{2}+\frac{p_2}{2}},\qquad
d= e^{\frac{k_1}{2}+\frac{k_2}{2}+s_1+ s_2+s_3+\frac{p_1}{2}+\frac{p_2}{2}}.
\eea
They satisfy the following Poisson relations, which can be deduced by formula (\ref{eq:comb-p}):
\bea\label{PV-lam-p}
&&
\{a,b\}= a b,\quad \{a,c\}=0,\quad\{a,d\}=-\frac{1}{2} a d,\quad \{a,e\}=\frac{1}{2} a e,\nn\\
&&
\{b,c\}= 0, \quad\{b,d\}=-\frac{1}{2} b d,\quad \{b,e\}=\frac{1}{2} b e,\\
&&
\{c,d\}= -\frac{1}{2} c d,\quad \{c,e\}=\frac{1}{2} c e,\quad\{d,e\}=0,\quad \{G_1,\cdot\}=\{G_2,\cdot\}=0,\nn
\eea
so that the elements $G_1$, $G_2$ and $G_3 G_\infty= d e$ are central. 

The generic family of symplectic leaves are determined by the common level set of the three Casimirs $G_1, G_2$ and $G_3 G_\infty= d e $ and are $4$-dimensional (rather than $2$-dimensional like in the $PVI$  case). 

On each symplectic leaf, the $PV$ monodromy manifold (\ref{eq:gen-p5}) is the subspace defined by those Laurent polynomials of $a,b,c,d$  that depend polynomially on the Casimir values $G_1,\,G_2,\, G_3 G_\infty=d e$ which commute with the frozen variables, i.e. with $G_3=e$ (and therefore with $d$ as well, since $d e$ is a Casimir). To see this, we can use relations (\ref{PV-lam}) to determine the exponentiated shear coordinates in terms of $a,b,c,d$, and then deduce the expressions of $x_1,x_2,x_3$ in terms of the lamination. We obtain the following expressions:
\bea\label{PV-lam-x1}
&&
x_1= - e \frac{a}{c}- d \frac{b}{c},\qquad
x_2= - e \frac{b}{c}- G_1 d \frac{b}{a} - d  \frac{b^2}{a c}-  d\frac{c}{a},\\
&&
x_3= -G_2  \frac{c}{b}- G_1  \frac{c}{a} -  \frac{b}{a }-  \frac{c^2}{a b}-\frac{a}{b},
\eea
which automatically satisfy (\ref{eq:gen-p5}).

Due to the Poisson relations (\ref{PV-lam-p}) the functions that commute with $e$ are exactly the functions of $\frac{a}{b},\frac{b}{c},\frac{c}{a},d$. Such functions may depend on the Casimir values $G_1,\,G_2$ and $G_3 G_\infty$ and $e$ itself, so that $d= G_\infty$ becomes a parameter now. With this in mind, it is straightforward to prove that $x_1,x_2,x_3$ are algebraically independent functions of $\frac{a}{b},\frac{b}{c},\frac{c}{a},d$ with polynomial coefficients in $G_1,\,G_2,\, G_3 G_\infty=d e$,
so that  $x_1,x_2,x_3$ form a basis in the space of functions which commute with $e$.

In similar fashion we can obtain also the character variety of $PVI$ as a Poisson sub--algebra of the $PV$ character variety defined by  those Laurent polynomials of $a,b,c,d$  that depend polynomially on the Casimir values $G_1,\,G_2$, and on $e+\frac{1}{e}$ (rather than polynomial in $e$ like for the $PV$ cubic) which commute with $e$. Setting:
\bea\label{PVI-lam-x1}
&&
\tilde x_1= - (e+\frac{1}{e}) \frac{a}{c}- d \frac{b}{c}-\frac{a^2}{b c d}-\frac{c}{b d}-G_2 \frac{a}{b d},\nn\\
&&
\tilde x_2= - (e+\frac{1}{e}) \frac{b}{c}- G_1 d \frac{b}{a} - d  \frac{b^2}{a c}-  d\frac{c}{a}-\frac{a}{c d},\nn\\
&&
\tilde x_3= -G_2  \frac{c}{b}- G_1  \frac{c}{a} -  \frac{b}{a }-  \frac{c^2}{a b}-\frac{a}{b},\nn
\eea
we obtain exactly formulae (\ref{eq:shear-PVI}) with the identification $x_i=\tilde x_i \big |_{k_1+k_3=p_3}$. This fact follows from the observation that $e^{\frac{p_3}{2}}+e^{-\frac{p_3}{2}}$ is the length function associated to the closed geodesic going around the third hole in $PVI$. This geodesic can be obtained as a concatenation of $e$ with $\frac{1}{e}$ and the identification ${k_1+k_3=p_3}$.

\subsection{Decorated character variety for $PV_{deg}$.}

The confluence from $PV$ to $PV_{deg}$  is realised by the substitution
$$
s_3\to s_3-\log[\epsilon],
$$
in formulae (\ref{eq:shear-PV}). In the limit $\epsilon\to0$ we obtain:
\begin{eqnarray}
\label{eq:shear-PVdeg}
x_1&=&-e^{s_2+ s_3+\frac{p_2}{2}+\frac{p_3}{2}},\nn\\
x_2&=&-e^{s_3+s_1+\frac{p_3}{2}+\frac{p_1}{2}}-e^{s_3-s_1+\frac{p_3}{2}-\frac{p_1}{2}}-G_1e^{s_3+\frac{p_3}{2}},\nn\\
x_3&=&-e^{s_1+ s_2+\frac{p_1}{2}+\frac{p_2}{2}}-e^{-s_1- s_2-\frac{p_1}{2}-\frac{p_2}{2}}-e^{s_1-s_2+\frac{p_1}{2}-\frac{p_2}{2}}-G_1e^{-s_2-\frac{p_2}{2}}-G_2 e^{s_1+\frac{p_1}{2}},\nn\\
\end{eqnarray}
where 
$$
G_i=e^{\frac{p_i}{2}}+e^{-\frac{p_i}{2}},\quad i=1,2,\quad
G_\infty=e^{s_1+s_2+s_3+\frac{p_1}{2}+\frac{p_2}{2}+\frac{p_3}{2}}.
$$
These coordinates satisfy the following cubic relation:
\begin{equation}\label{eq:gen-p5deg}
x_1 x_2 x_3 + x_1^2+x_2^2-G_1 G_\infty x_1 -G_2 G_\infty x_2 + G_\infty^2=0.
\end{equation}
Note that the parameter $p_3$ is now redundant, we can eliminate it by rescaling.
To obtain the correct $PV_{deg}$   cubic, we need to pick $p_3=-p_1-p_2-2 s_1-2 s_2-2 s_3$. 

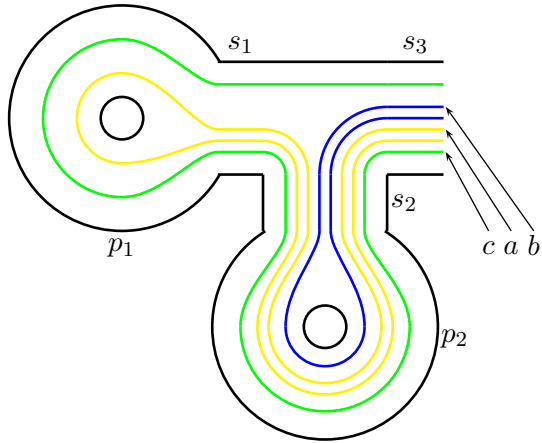
\begin{figure}[h]
{\psset{unit=0.5}
\begin{pspicture}(-2,-8)(2,3)
\newcommand{\TAIL}[3]{%
{\psset{unit=#1}
\pcline[linewidth=1pt,linecolor=black](0,0.5)(0.5,0.5)
\pcline[linewidth=1pt,linecolor=black](0,-0.5)(0.5,-0.5)
\pcline[linewidth=1pt,linecolor=green](0,0.3)(0.5,0.3)
\pcline[linewidth=1pt,linecolor=blue](0,0.1)(0.5,0.1)
\pcline[linewidth=1pt,linecolor=#2](0,0)(0.5,0)
\pcline[linewidth=1pt,linecolor=#3](0,-0.1)(0.5,-0.1)
\pcline[linewidth=1pt,linecolor=yellow](0,-0.2)(0.5,-0.2)
\pcline[linewidth=1pt,linecolor=green](0,-0.3)(0.5,-0.3)
%
\pcline[linewidth=0.5pt,linecolor=black]{->}(0.9,-1)(.52,-0.3)
\rput(.9,-1.15){$c$}
\pcline[linewidth=0.5pt,linecolor=black]{->}(1.1,-1)(.52,-0.1)
\rput(1.1,-1.15){$a$}
\pcline[linewidth=0.5pt,linecolor=black]{->}(1.3,-1)(.52,0.1)
\rput(1.3,-1.12){$b$}
\rput(0.25,0.65){$s_3$}
\rput(-1.3,0.65){$s_1$}
\rput(0.15,-0.75){$s_2$}
\rput(0.6,-1.95){$p_2$}
\rput(-2.35,-1.15){$p_1$}
}
}
\newcommand{\LOOPS}[1]{%
{\psset{unit=#1}
\pcline[linewidth=1pt,linecolor=black](0,0.5)(-1.5,0.5)
\pcline[linewidth=1pt,linecolor=black](0,-0.5)(0,-1)
\pcline[linewidth=1pt,linecolor=black](-1.1,-0.5)(-1.1,-1)
\pcline[linewidth=1pt,linecolor=black](-1.1,-0.5)(-1.5,-0.5)
\psarc[linecolor=black, linewidth=1pt](-0.55,-1.85){1}{-238}{58}
\pscircle[linecolor=black, linewidth=1pt](-0.55,-1.85){0.2}
\psarc[linecolor=black, linewidth=1pt](-2.35,0){1}{30}{330}
\pscircle[linecolor=black, linewidth=1pt](-2.35,0){0.2}
%
\psarc[linecolor=green, linewidth=1pt](0,-0.5){0.2}{90}{180}
\psarc[linecolor=yellow, linewidth=1pt](0,-0.5){0.3}{90}{180}
\psarc[linecolor=yellow, linewidth=1pt](0,-0.5){0.4}{90}{180}
\psarc[linecolor=blue, linewidth=1pt](0,-0.5){0.5}{90}{180}
\psarc[linecolor=blue, linewidth=1pt](0,-0.5){0.6}{90}{180}
%
\psarc[linecolor=green, linewidth=1pt](-1.1,-0.5){0.2}{0}{90}
\psarc[linecolor=yellow, linewidth=1pt](-1.1,-0.5){0.3}{0}{90}
\psarc[linecolor=yellow, linewidth=1pt](-1.1,-0.5){0.4}{0}{90}
%
\pcline[linewidth=1pt,linecolor=green](0,0.3)(-1.5,0.3)
\pcline[linewidth=1pt,linecolor=green](-1.1,-0.3)(-1.5,-0.3)
\pcline[linewidth=1pt,linecolor=green](-0.2,-0.5)(-0.2,-1)
\pcline[linewidth=1pt,linecolor=green](-0.9,-0.5)(-0.9,-1)
\pcline[linewidth=1pt,linecolor=blue](-0.5,-0.5)(-0.5,-1)
\pcline[linewidth=1pt,linecolor=blue](-0.6,-0.5)(-0.6,-1)
\pcline[linewidth=1pt,linecolor=yellow](-0.3,-0.5)(-0.3,-1)
\pcline[linewidth=1pt,linecolor=yellow](-0.4,-0.5)(-0.4,-1)
\pcline[linewidth=1pt,linecolor=yellow](-0.7,-0.5)(-0.7,-1)
\pcline[linewidth=1pt,linecolor=yellow](-0.8,-0.5)(-0.8,-1)
\pcline[linewidth=1pt,linecolor=yellow](-1.1,-0.2)(-1.5,-0.2)
\pcline[linewidth=1pt,linecolor=yellow](-1.1,-0.1)(-1.5,-0.1)
%
\psarc[linecolor=green, linewidth=1pt](-0.55,-1.85){0.75}{180}{360}
\psarc[linecolor=yellow, linewidth=1pt](-0.55,-1.85){0.6}{180}{360}
\psarc[linecolor=yellow, linewidth=1pt](-0.55,-1.85){0.5}{180}{360}
\psarc[linecolor=blue, linewidth=1pt](-0.55,-1.85){0.35}{180}{360}
%
\psarc[linecolor=green, linewidth=1pt](-2.35,0){0.7}{90}{270}
\psarc[linecolor=yellow, linewidth=1pt](-2.35,0){0.4}{90}{270}
%
\psbezier[linewidth=1pt,linecolor=green](-0.2,-1)(-0.2,-1.2)(0.2,-1.5)(0.2,-1.85)
\psbezier[linewidth=1pt,linecolor=green](-0.9,-1)(-0.9,-1.2)(-1.3,-1.5)(-1.3,-1.85)
\psbezier[linewidth=1pt,linecolor=yellow](-0.3,-1)(-0.3,-1.2)(0.05,-1.5)(0.05,-1.85)
\psbezier[linewidth=1pt,linecolor=yellow](-0.4,-1)(-0.4,-1.2)(-0.05,-1.5)(-0.05,-1.85)
\psbezier[linewidth=1pt,linecolor=yellow](-0.8,-1)(-0.8,-1.2)(-1.15,-1.5)(-1.15,-1.85)
\psbezier[linewidth=1pt,linecolor=yellow](-0.7,-1)(-0.7,-1.2)(-1.05,-1.5)(-1.05,-1.85)
\psbezier[linewidth=1pt,linecolor=blue](-0.5,-1)(-0.5,-1.2)(-0.2,-1.5)(-0.2,-1.85)
\psbezier[linewidth=1pt,linecolor=blue](-0.6,-1)(-0.6,-1.2)(-0.9,-1.5)(-0.9,-1.85)
%
\psbezier[linewidth=1pt,linecolor=green](-2.35,0.7)(-2,0.7)(-1.7,0.3)(-1.5,0.3)
\psbezier[linewidth=1pt,linecolor=green](-2.35,-0.7)(-2,-0.7)(-1.7,-0.3)(-1.5,-0.3)
\psbezier[linewidth=1pt,linecolor=yellow](-2.35,0.4)(-2,0.4)(-1.7,-0.1)(-1.5,-0.1)
\psbezier[linewidth=1pt,linecolor=yellow](-2.35,-0.4)(-2,-0.4)(-1.7,-0.2)(-1.5,-0.2)
}
}
\rput(0,0){\TAIL{3}{blue}{yellow}}
\rput(0,0){\LOOPS{3}}
\end{pspicture}
}
\caption{The fat graph corresponding to $PV_{deg}$ with the related system of arcs.}
\label{f:PVdeg-fg}
\end{figure}

Geometrically speaking, sending the shear coordinate $s_3$ to infinity means that we are performing a cusp-removing move. In terms of the fat-graph, this is represented by Fig.~\ref{f:PVdeg-fg}.

The character variety of a Riemann sphere with three holes and one cusp on one boundary is $5$-dimensional.
The fat-graph admits a complete cusped lamination so that a full set of coordinates on the character variety is given by the complexification of the geodesic length functions of the elements in the lamination. Now we have only one shear coordinate associated to the cusp,  denoted by $s_3$, which does not commute with the other shear coordinates. 

We omit the picture of the $PV_{deg}$ lamination as it is very similar to Fig.~\ref{fi:PV}, in which the edges labelled by  $k_1$ and $k_2$ are removed and  the geodesics $d$ and $e$ are lost.

In terms of shear coordinates, the elements in the lamination are two loops corresponding to the parameters $G_1$ and $G_2$ and three arcs for which the lengths are expressed as follows:
\be\label{PVdeg-lam}
a= e^{s_1+2 s_2+s_3+\frac{p_1}{2}+p_2},\qquad
b= e^{s_2+s_3+\frac{p_2}{2}},\qquad c= e^{s_1+ s_2+s_3+\frac{p_1}{2}+\frac{p_2}{2}},
\ee
They satisfy the following Poisson relations, which can be deduced by formula (\ref{eq:comb-p}):
\be\label{PVdeg-lam-p}
\{a,b\}= a b,\quad \{a,c\}=0,\quad \{b,c\}= 0,\quad \{G_1,\cdot\}=\{G_2,\cdot\}=0,
\ee
so that the element $c$ is a Casimir as well as the parameters $G_1, G_2$. Each symplectic leaf is two-dimensional and corresponds to the 
$PV_{deg}$ monodromy manifold (\ref{eq:gen-p5deg}). Indeed, we can use relations (\ref{PVdeg-lam}) to determine the exponentiated shear coordinates in terms of $a,b,c$, and then deduce the expressions of $x_1,x_2,x_3$ in terms of the lamination. We obtain the following expressions:
\bea\label{PV-lam-x2}
&&
x_1= -b,\qquad
x_2= - G_1 \frac{b c}{a} -   \frac{b^2}{a}-  \frac{c^2}{a},\\
&&
x_3= -G_2  \frac{c}{b}- G_1  \frac{c}{a} -  \frac{b}{a }-  \frac{c^2}{a b}-\frac{a}{b},
\eea
which automatically satisfy (\ref{eq:gen-p5deg}).

In terms of the lamination, the confluence from $PV$ to $PV_{deg}$  is given by the following prescription:
$$
a\to a,\quad b\to b,\quad c\to c,\quad d\to 0, \quad e\to 0.
$$
Note that the Poisson algebra (\ref{PVdeg-lam-p}) can be obtained as the Poisson sub-algebra of (\ref{PV-lam-p}) defined by the functions of $a,b,c$ only.

\subsection{Decorated character variety for $PIV$}

The confluence from the generic $PV$ cubic (\ref{eq:gen-p5}) to the $PIV$ one is realised by the substitution
$$
p_2\to p_2-2\log[\epsilon],
$$
in formulae (\ref{eq:shear-PV}). In the limit $\epsilon\to0$ we obtain:
\begin{eqnarray}
\label{eq:shear-PIV}
x_1&=&-e^{s_2+ s_3+\frac{p_2}{2}+\frac{p_3}{2}}-G_3 e^{s_2+\frac{p_2}{2}},\nn\\
x_2&=&-e^{s_3+s_1+\frac{p_3}{2}+\frac{p_1}{2}}-e^{s_3-s_1+\frac{p_3}{2}-\frac{p_1}{2}}-G_3 e^{-s_1-\frac{p_1}{2}}-G_1e^{s_3+\frac{p_3}{2}},\nn\\
x_3&=&-e^{s_1+ s_2+\frac{p_1}{2}+\frac{p_2}{2}}-G_2 e^{s_1+\frac{p_1}{2}},\nn\\
\end{eqnarray}
where 
$$
G_1=e^{\frac{p_1}{2}}+e^{-\frac{p_1}{2}},\quad 
G_2=e^{+\frac{p_2}{2}},\quad
G_3=e^{+\frac{p_3}{2}},\quad
G_\infty=e^{s_1+s_2+s_3+\frac{p_1}{2}+\frac{p_2}{2}+\frac{p_3}{2}}.
$$
These coordinates satisfy the following cubic relation:
\bea\label{eq:gen-p4}
&&
x_1x_2x_3 + x_1^2-(G_1 G_\infty+G_2 G_3) x_1 -G_2 G_\infty  x_2  -\nn\\
&&-G_3 G_\infty x_3+ G_\infty^2+G_1 G_2G_3 G_\infty=0.
\eea
Note that the parameters $p_3,p_2$ are now redundant, we can eliminate it by rescaling.
To obtain the correct $PIV$ cubic, we need to pick $p_2=p_3=-p_1-2 s_1-2 s_2-2 s_3$ so that $G_2=G_3=G_\infty$. 

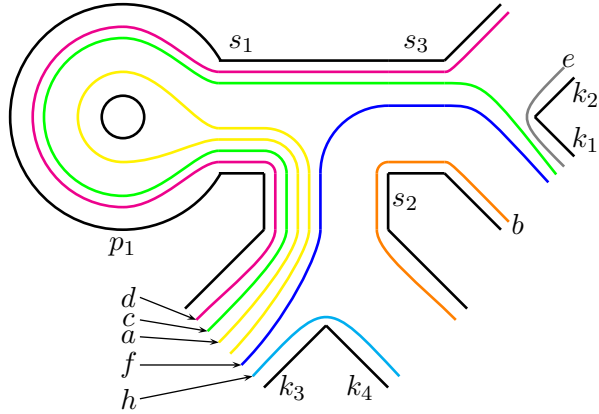
\begin{figure}[h]
{\psset{unit=0.5}
\begin{pspicture}(-2,-8)(2,3)
\newcommand{\TAIL}[3]{%
{\psset{unit=#1}
\pcline[linewidth=1pt,linecolor=black](0,0.5)(0.5,0.5)
\pcline[linewidth=1pt,linecolor=black](0,-0.5)(0.5,-0.5)
\pcline[linewidth=1pt,linecolor=black](0.5,0.5)(1,1)
\pcline[linewidth=1pt,linecolor=black](0.5,-0.5)(1,-1)
\pcline[linewidth=1pt,linecolor=magenta](0,0.4)(0.5,0.4)
\pcline[linewidth=1pt,linecolor=green](0,0.3)(0.5,0.3)
\pcline[linewidth=1pt,linecolor=blue](0,0.1)(0.5,0.1)
\pcline[linewidth=1pt,linecolor=white](0,0)(0.5,0)
\pcline[linewidth=1pt,linecolor=#3](0,-0.1)(0.5,-0.1)
\pcline[linewidth=1pt,linecolor=white](0,-0.3)(0.5,-0.3)
\pcline[linewidth=1pt,linecolor=orange](0,-0.4)(0.5,-0.4)
\psbezier[linewidth=1pt,linecolor=orange](0.5,-0.4)(0.55,-0.4)(0.55,-0.4)(1.07,-0.93)
\psbezier[linewidth=1pt,linecolor=white](0.5,-0.2)(0.65,-0.2)(0.65,-0.2)(1.21,-0.79)
\psbezier[linewidth=1pt,linecolor=#3](0.5,-0.1)(0.7,-0.1)(0.7,-0.1)(1.28,-0.72)
\psbezier[linewidth=1pt,linecolor=white](0.5,0)(0.75,0)(0.75,0)(1.35,-0.65)
\psbezier[linewidth=1pt,linecolor=blue](0.5,0.1)(0.8,0.1)(0.8,0.1)(1.42,-0.58)
\psbezier[linewidth=1pt,linecolor=green](0.5,0.3)(0.85,0.3)(0.85,0.3)(1.49,-0.51)
\psbezier[linewidth=1pt,linecolor=magenta](0.5,0.4)(0.55,0.4)(0.55,0.4)(1.07,0.93)
\psbezier[linewidth=1pt,linecolor=gray](1.56,-0.44)(1.12,0)(1.12,0)(1.56,0.44)
\pcline[linewidth=1pt,linecolor=black](1.65,0.35)(1.3,0)
\pcline[linewidth=1pt,linecolor=black](1.65,-0.35)(1.3,0)
\pcline[linewidth=0.5pt,linecolor=black]{->}(-2.2,-1.6)(-1.7,-1.8)
\rput(-2.3,-1.6){$d$}
\pcline[linewidth=0.5pt,linecolor=black]{->}(-2.2,-1.8)(-1.6,-1.9)
\rput(-2.3,-1.8){$c$}
\pcline[linewidth=0.5pt,linecolor=black]{->}(-2.2,-1.95)(-1.5,-2)
\rput(-2.3,-1.95){$a$}
\pcline[linewidth=0.5pt,linecolor=black]{->}(-2.2,-2.2)(-1.3,-2.2)
\rput(-2.3,-2.2){$f$}
\pcline[linewidth=0.5pt,linecolor=black]{->}(-2.2,-2.5)(-1.2,-2.3)
\rput(-2.3,-2.5){$h$}
\rput(1.15,-0.95){$b$}
\rput(1.6,.5){$e$}
\rput(1.75,-0.2){$k_1$}
\rput(1.75,0.2){$k_2$}
\rput(-.85,-2.4){$k_3$}
\rput(-.25,-2.4){$k_4$}
\rput(0.25,0.65){$s_3$}
\rput(-1.3,0.65){$s_1$}
\rput(0.15,-0.75){$s_2$}
\rput(-2.35,-1.15){$p_1$}
}
}
\newcommand{\LOOPS}[1]{%
{\psset{unit=#1}
\pcline[linewidth=1pt,linecolor=black](0,0.5)(-1.5,0.5)
\pcline[linewidth=1pt,linecolor=black](0,-0.5)(0,-1)
\pcline[linewidth=1pt,linecolor=black](-1.1,-0.5)(-1.1,-1)
\pcline[linewidth=1pt,linecolor=black](-1.1,-0.5)(-1.5,-0.5)
\pcline[linewidth=1pt,linecolor=black](-1.1,-1)(-1.8,-1.7)
\pcline[linewidth=1pt,linecolor=black](0,-1)(.7,-1.7)
\pcline[linewidth=1pt,linecolor=black](-.55,-1.85)(-1.1,-2.4)
\pcline[linewidth=1pt,linecolor=black](-.55,-1.85)(0,-2.4)
\psarc[linecolor=black, linewidth=1pt](-2.35,0){1}{30}{330}
\pscircle[linecolor=black, linewidth=1pt](-2.35,0){0.2}
\psarc[linecolor=orange, linewidth=1pt](0,-0.5){0.1}{90}{180}
\psarc[linecolor=blue, linewidth=1pt](0,-0.5){0.6}{90}{180}
\psarc[linecolor=magenta, linewidth=1pt](-1.1,-0.5){0.1}{0}{90}
\psarc[linecolor=green, linewidth=1pt](-1.1,-0.5){0.2}{0}{90}
\psarc[linecolor=yellow, linewidth=1pt](-1.1,-0.5){0.3}{0}{90}
\psarc[linecolor=yellow, linewidth=1pt](-1.1,-0.5){0.4}{0}{90}
\pcline[linewidth=1pt,linecolor=magenta](0,0.4)(-1.5,0.4)
\pcline[linewidth=1pt,linecolor=green](0,0.3)(-1.5,0.3)
\pcline[linewidth=1pt,linecolor=magenta](-1.1,-0.4)(-1.5,-0.4)
\pcline[linewidth=1pt,linecolor=green](-1.1,-0.3)(-1.5,-0.3)
\pcline[linewidth=1pt,linecolor=orange](-0.1,-0.5)(-0.1,-1)
\pcline[linewidth=1pt,linecolor=magenta](-1,-0.5)(-1,-1)
\pcline[linewidth=1pt,linecolor=green](-0.9,-0.5)(-0.9,-1)
\pcline[linewidth=1pt,linecolor=blue](-0.6,-0.5)(-0.6,-1)
\pcline[linewidth=1pt,linecolor=yellow](-0.7,-0.5)(-0.7,-1)
\pcline[linewidth=1pt,linecolor=yellow](-0.8,-0.5)(-0.8,-1)
\pcline[linewidth=1pt,linecolor=yellow](-1.1,-0.2)(-1.5,-0.2)
\pcline[linewidth=1pt,linecolor=yellow](-1.1,-0.1)(-1.5,-0.1)
%
%
\psarc[linecolor=magenta, linewidth=1pt](-2.35,0){0.8}{90}{270}
\psarc[linecolor=green, linewidth=1pt](-2.35,0){0.7}{90}{270}
\psarc[linecolor=yellow, linewidth=1pt](-2.35,0){0.4}{90}{270}
\psbezier[linewidth=1pt,linecolor=orange](-0.1,-1)(-0.1,-1.2)(0.3,-1.5)(0.6,-1.8)
\psbezier[linewidth=1pt,linecolor=magenta](-1,-1)(-1,-1.2)(-1.4,-1.5)(-1.7,-1.8)
\psbezier[linewidth=1pt,linecolor=green](-0.9,-1)(-0.9,-1.2)(-1.3,-1.6)(-1.6,-1.9)
\psbezier[linewidth=1pt,linecolor=yellow](-0.8,-1)(-0.8,-1.2)(-1.2,-1.7)(-1.5,-2)
\psbezier[linewidth=1pt,linecolor=yellow](-0.7,-1)(-0.7,-1.2)(-1.1,-1.8)(-1.4,-2.1)
\psbezier[linewidth=1pt,linecolor=blue](-0.6,-1)(-0.6,-1.2)(-0.9,-1.8)(-1.3,-2.2)
\psbezier[linewidth=1pt,linecolor=cyan](0.1,-2.3)(-0.55,-1.6)(-0.55,-1.6)(-1.2,-2.3)
\psbezier[linewidth=1pt,linecolor=magenta](-2.35,0.8)(-2,0.8)(-1.7,0.4)(-1.5,0.4)
\psbezier[linewidth=1pt,linecolor=magenta](-2.35,-0.8)(-2,-0.8)(-1.7,-0.4)(-1.5,-0.4)
\psbezier[linewidth=1pt,linecolor=green](-2.35,0.7)(-2,0.7)(-1.7,0.3)(-1.5,0.3)
\psbezier[linewidth=1pt,linecolor=green](-2.35,-0.7)(-2,-0.7)(-1.7,-0.3)(-1.5,-0.3)
\psbezier[linewidth=1pt,linecolor=yellow](-2.35,0.4)(-2,0.4)(-1.7,-0.1)(-1.5,-0.1)
\psbezier[linewidth=1pt,linecolor=yellow](-2.35,-0.4)(-2,-0.4)(-1.7,-0.2)(-1.5,-0.2)
}
}
\rput(0,0){\TAIL{3}{blue}{white}}
\rput(0,0){\LOOPS{3}}
\end{pspicture}
}
\caption{\small The system of arcs for $PIV$.}
\label{fi:PIV}
\end{figure}

Similarly to the previous case, this means that we send the perimeter $p_2$ to infinity, which is a chewing-gum move leading to a Riemann sphere with two holes, one of which has $4$ cusps on it. The corresponding fat-graph is given in Fig.~\ref{fi:PIV}, where we see $4$ new shear coordinates, one for each cusp, so that in formulae (\ref{eq:shear-PIV})  $p_2 =k_3+k_4$ and $p_3=k_1+k_2$.

The character variety is now $8$ dimensional and the complete cusped lamination is given in Fig.~\ref{fi:PIV}.

In terms of shear coordinates, the elements in the lamination are expressed as follows:
\bea\label{PIV-lam}
&&
\qquad a= e^{s_1+s_2+k_3+\frac{p_1}{2}},\quad
b= e^{\frac{s_2}{2}+\frac{s_3}{2}+\frac{k_1}{2}+\frac{k_4}{2}},\quad c= e^{s_1+ \frac{s_2}{2}+\frac{s_3}{2}+\frac{p_1}{2}+\frac{k_1}{2}+\frac{k_3}{2}},\\
&&
d= e^{s_1+ \frac{s_2}{2}+\frac{s_3}{2}+\frac{p_1}{2}+\frac{k_2}{2}+\frac{k_3}{2}},\quad
e = e^{\frac{k_1}{2}+\frac{k_2}{2}},\quad
f=  e^{ \frac{s_2}{2}+\frac{s_3}{2}+\frac{k_1}{2}+\frac{k_3}{2}},\quad
h = e^{\frac{k_3}{2}+\frac{k_4}{2}}.\nn
\eea
The Poisson brackets can be easily extracted from (\ref{eq:comb-p}):
\bea\label{PIV-lam-p}
&&
\{a,b\}=0,\quad\{a,c\}=-\frac{1}{2} a c,\quad \{a,d\}=-\frac{1}{2} a d,\quad\{a,e\}=0,\quad \{a,f\}=\frac{1}{2} a f,\nn\\
 &&
 \{a,h\}=\frac{1}{2} a h,\quad\{b,c\}=\frac{1}{4}b c, \quad \{b,d\}=0,\quad\{b,e\}=\frac{1}{4} b e,\quad  \{b,f\}=\frac{1}{4} b f,\nn\\
 &&
 \{b,h\}=-\frac{1}{4} b h,\quad
 \{c,d\}=-\frac{1}{4} c d,\quad \{c,e\}=\frac{1}{4} c e,\quad  \{c,f\}=0,\\
 &&
 \{c,h\}=\frac{1}{4} c h,\quad \{d,e\}=-\frac{1}{4} d e,\quad  \{d,f\}=\frac{1}{4} d f,\quad
 \{d,h\}=\frac{1}{4} d h,\nn\\
 &&
 \{e,f\}=-\frac{1}{4} e f,\quad
 \{e,h\}=0,\quad \{f,h\}=\frac{1}{4} f h.\nn
\eea
The element $b d e h$ is a Casimir as well as the perimeter $G_1$. Each symplectic leaf is six-dimensional and the 
$PIV$ monodromy manifold (\ref{eq:gen-p4}) is the subspace of those functions of $a,b,\dots,h$ which commute with $e$ and $h$. 
The proof of this statement is quite similar to the previously considered analogous assertion for the $PV$ case and we omit it.

In terms of Poisson manifolds, the confluence from $PV$ to $PIV$ is {\it reversed.}\/ Indeed the Poisson algebra (\ref{PV-lam-p}) is a Poisson sub-algebra of (\ref{PIV-lam-p}) specified by the functions of 
$$
a_V:=a_{IV} b_{IV}^2,\quad b_V:=b_{IV} f_{IV},\quad c_V:=b_{IV} c_{IV},\quad d_V:=b_{IV} d_{IV},
$$
where we have denoted with an index $V$ the lamination coordinates in  (\ref{PV-lam-p})  and by $IV$ the ones in  (\ref{PIV-lam-p}). Note that $h_{IV}$ is automatically a Casimir for the Poisson sub-algebra of (\ref{PV-lam-p}) and can be identified with $G_2$.

\subsection{Decorated character variety for $PIII^{D_6}$}
The confluence from $PV$ to $PIII^{D_6}$ is obtained by the following substitution:
\begin{equation}\label{eq:sub-epsV}
s_1\to s_1+2\log[\epsilon],\quad p_2\to p_2 -2\log[\epsilon],\quad
 p_1\to p_1-2\log[\epsilon].\nn
\end{equation}
In the limit as $\epsilon\to 0$ we obtain:
\begin{eqnarray}
\label{eq:shear-PIII}
x_1&=&-e^{s_2+ s_3+\frac{p_2}{2}+\frac{p_3}{2}}-G_3 e^{s_2+\frac{p_2}{2}},\nn\\
x_2&=&-e^{s_3-s_1+\frac{p_3}{2}-\frac{p_1}{2}}-G_3 e^{-s_1-\frac{p_1}{2}}-G_1e^{s_3+\frac{p_3}{2}},\nn\\
x_3&=&-e^{s_1+ s_2+\frac{p_1}{2}+\frac{p_2}{2}}-e^{-s_1- s_2-\frac{p_1}{2}-\frac{p_2}{2}}-G_1e^{-s_2-\frac{p_2}{2}}-G_2 e^{s_1+\frac{p_1}{2}},\nn\\
\end{eqnarray}
where 
$$
\widetilde G_i=e^{\frac{p_i}{2}},\quad i=1,2,3
\quad 
\widetilde G_\infty=e^{s_1+s_2+s_3+\frac{p_1}{2}+\frac{p_2}{2}+\frac{p_3}{2}}.
$$
These coordinates satisfy the following cubic relation:
\begin{equation}\label{eq:gen-p3}
x_1x_2x_3 + x_1^2+x_2^2-(\widetilde G_1 \widetilde G_\infty+\widetilde G_2  \widetilde G_3) x_1 -(\widetilde G_2 \widetilde G_\infty+\widetilde G_1 \widetilde G_3) x_2
+\widetilde G_1\widetilde G_2\widetilde G_3\widetilde G_\infty=0.
\end{equation}
We can pick $p_2=p_3=0$ in order to obtain the correct $PIII^{D_6}$ cubic. Note that there is a slight discrepancy between the $\widetilde G_i$s in the cubic (\ref{eq:gen-p3}) and the $G_i$s  dictated by our formulae (\ref{eq:G-formula}). This is easily solved by a simple transformation 
$$
G_\infty=\sqrt{\widetilde G_1 \widetilde G_\infty},\quad G_1= \sqrt{\widetilde G_1 \widetilde G_\infty}+\frac{1}{ \sqrt{\widetilde G_1 \widetilde G_\infty}},\quad
G_2=\sqrt{\frac{\widetilde G_\infty}{\widetilde G_1}}+\sqrt{\frac{\widetilde G_1}{\widetilde G_\infty}}.
$$

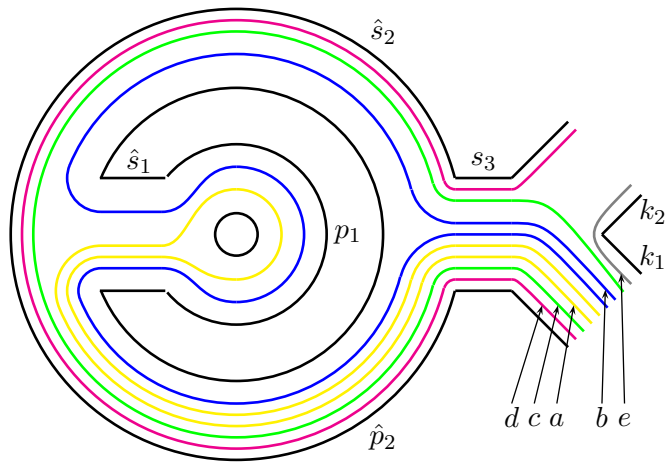
\begin{figure}[h]
{\psset{unit=0.5}
\begin{pspicture}(-6,-6)(2,6)
\newcommand{\TAIL}[3]{%
{\psset{unit=#1}
\pcline[linewidth=1pt,linecolor=black](0,0.5)(0.5,0.5)
\pcline[linewidth=1pt,linecolor=black](0,-0.5)(0.5,-0.5)
\pcline[linewidth=1pt,linecolor=black](0.5,0.5)(1,1)
\pcline[linewidth=1pt,linecolor=black](0.5,-0.5)(1,-1)
\pcline[linewidth=1pt,linecolor=magenta](0,0.4)(0.5,0.4)
\pcline[linewidth=1pt,linecolor=green](0,0.3)(0.5,0.3)
\pcline[linewidth=1pt,linecolor=blue](0,0.1)(0.5,0.1)
\pcline[linewidth=1pt,linecolor=#2](0,0)(0.5,0)
\pcline[linewidth=1pt,linecolor=#3](0,-0.1)(0.5,-0.1)
\pcline[linewidth=1pt,linecolor=yellow](0,-0.2)(0.5,-0.2)
\pcline[linewidth=1pt,linecolor=green](0,-0.3)(0.5,-0.3)
\pcline[linewidth=1pt,linecolor=magenta](0,-0.4)(0.5,-0.4)
\psbezier[linewidth=1pt,linecolor=magenta](0.5,-0.4)(0.55,-0.4)(0.55,-0.4)(1.07,-0.93)
\psbezier[linewidth=1pt,linecolor=green](0.5,-0.3)(0.6,-0.3)(0.6,-0.3)(1.14,-0.86)
\psbezier[linewidth=1pt,linecolor=yellow](0.5,-0.2)(0.65,-0.2)(0.65,-0.2)(1.21,-0.79)
\psbezier[linewidth=1pt,linecolor=#3](0.5,-0.1)(0.7,-0.1)(0.7,-0.1)(1.28,-0.72)
\psbezier[linewidth=1pt,linecolor=#2](0.5,0)(0.75,0)(0.75,0)(1.35,-0.65)
\psbezier[linewidth=1pt,linecolor=blue](0.5,0.1)(0.8,0.1)(0.8,0.1)(1.42,-0.58)
\psbezier[linewidth=1pt,linecolor=green](0.5,0.3)(0.85,0.3)(0.85,0.3)(1.49,-0.51)
\psbezier[linewidth=1pt,linecolor=magenta](0.5,0.4)(0.55,0.4)(0.55,0.4)(1.07,0.93)
\psbezier[linewidth=1pt,linecolor=gray](1.56,-0.44)(1.12,0)(1.12,0)(1.56,0.44)
\pcline[linewidth=1pt,linecolor=black](1.65,0.35)(1.3,0)
\pcline[linewidth=1pt,linecolor=black](1.65,-0.35)(1.3,0)
\pcline[linewidth=0.5pt,linecolor=black]{->}(0.5,-1.5)(.77,-0.63)
\rput(.5,-1.62){$d$}
\pcline[linewidth=0.5pt,linecolor=black]{->}(0.7,-1.5)(.91,-0.63)
\rput(.7,-1.65){$c$}
\pcline[linewidth=0.5pt,linecolor=black]{->}(0.9,-1.5)(1.05,-0.63)
\rput(.9,-1.65){$a$}
\pcline[linewidth=0.5pt,linecolor=black]{->}(1.3,-1.5)(1.33,-0.49)
\rput(1.3,-1.62){$b$}
\pcline[linewidth=0.5pt,linecolor=black]{->}(1.5,-1.5)(1.47,-0.35)
\rput(1.5,-1.65){$e$}
\rput(1.75,-0.2){$k_1$}
\rput(1.75,0.2){$k_2$}
\rput(0.25,0.65){$s_3$}
\rput(-0.64,1.8){$\hat s_2$}
\rput(-0.64,-1.8){$\hat p_2$}
\rput(-2.8,0.65){$\hat s_1$}
\rput(-0.95,0){$p_1$}
}
}
\newcommand{\BIGLOOP}[1]{%
{\psset{unit=#1}
\pcline[linewidth=1pt,linecolor=black](-3.15,0.5)(-2.57,0.5)
\pcline[linewidth=1pt,linecolor=black](-3.15,-0.5)(-2.57,-0.5)
\pscircle[linecolor=black, linewidth=1pt](-1.94,0){0.2}
\psarc[linecolor=black, linewidth=1pt](-1.94,0){2}{14.5}{345.5}
\psarc[linecolor=black, linewidth=1pt](-1.94,0){1.3}{-157.5}{157.5}
\psarc[linecolor=black, linewidth=1pt](-1.94,0){0.8}{-140}{140}
\psarc[linecolor=magenta, linewidth=1pt](-1.94,0){1.9}{14.5}{345.5}
\psarc[linecolor=green, linewidth=1pt](-1.94,0){1.8}{14.5}{345.5}
\psarc[linecolor=yellow, linewidth=1pt](-1.94,0){1.7}{202.5}{345.5}
\psarc[linecolor=yellow, linewidth=1pt](-1.94,0){1.6}{202.5}{345.5}
\psarc[linecolor=blue, linewidth=1pt](-1.94,0){1.5}{202.5}{345.5}
\psarc[linecolor=blue, linewidth=1pt](-1.94,0){1.6}{14.5}{157.5}
\psarc[linecolor=magenta, linewidth=1pt](0,-0.5){0.1}{90}{165}
\psarc[linecolor=green, linewidth=1pt](0,-0.5){0.2}{90}{165}
\psarc[linecolor=yellow, linewidth=1pt](0,-0.5){0.3}{90}{165.5}
\psarc[linecolor=yellow, linewidth=1pt](0,-0.5){0.4}{90}{165.5}
\psarc[linecolor=blue, linewidth=1pt](0,-0.5){0.5}{90}{165.5}
\psarc[linecolor=blue, linewidth=1pt](0,0.5){0.4}{195}{270}
\psarc[linecolor=magenta, linewidth=1pt](0,0.5){0.1}{195}{270}
\psarc[linecolor=green, linewidth=1pt](0,0.5){0.2}{195}{270}
\psarc[linecolor=yellow, linewidth=1pt](-3.14,-0.5){0.3}{90}{202.5}
\psarc[linecolor=yellow, linewidth=1pt](-3.14,-0.5){0.4}{90}{202.5}
\psarc[linecolor=blue, linewidth=1pt](-3.14,-0.5){0.2}{90}{202.5}
\psarc[linecolor=blue, linewidth=1pt](-3.14,0.5){0.3}{157.5}{270}
\pcline[linewidth=1pt,linecolor=blue](-3.14,0.2)(-2.6,0.2)
\pcline[linewidth=1pt,linecolor=blue](-3.14,-0.3)(-2.6,-0.3)
\pcline[linewidth=1pt,linecolor=yellow](-3.14,-0.2)(-2.6,-0.2)
\pcline[linewidth=1pt,linecolor=yellow](-3.14,-0.1)(-2.6,-0.1)
%
\psarc[linecolor=yellow, linewidth=1pt](-1.94,0){0.4}{-90}{90}
\psarc[linecolor=blue, linewidth=1pt](-1.94,0){0.6}{-90}{90}
\psbezier[linewidth=1pt,linecolor=yellow](-2.6,-0.2)(-2.3,-0.2)(-2.3,-0.4)(-1.94,-0.4)
\psbezier[linewidth=1pt,linecolor=yellow](-2.6,-0.1)(-2.35,-0.1)(-2.35,0.4)(-1.94,0.4)
\psbezier[linewidth=1pt,linecolor=blue](-2.6,0.2)(-2.3,0.2)(-2.3,0.6)(-1.94,0.6)
\psbezier[linewidth=1pt,linecolor=blue](-2.6,-0.3)(-2.3,-0.3)(-2.3,-0.6)(-1.94,-0.6)
}
}
\rput(0,0){\TAIL{3}{blue}{yellow}}
\rput(0,0){\BIGLOOP{3}}
\end{pspicture}
}
\caption{\small Transformation of the $PV$ system of arcs from Fig.~\ref{fi:PV} under a sequence of two flips.}
\label{fi:PV-transformed}
\end{figure}

To understand the geometry of this confluence, we first need to flip the $PV$ fat-graph to the equivalent graph given in Fig.~\ref{fi:PV-transformed}. 

The new shear coordinates are expressed in terms of the old ones as follows:
\bea\label{new:shear}
\hat s_1&=&-s_1-p_1- \log(1+e^{s_2}),\nn\\
\hat s_2 &=&-s_2+ \log\left((1+e^{p_1+s_1}+e^{p_1+s_1+s_2})(1+e^{s_1}+e^{s_1+s_2})\right),\nn\\
\hat s_3&=&s_3-\log(1+e^{-s_2}),\quad \hat p_1=p_1,\\
\hat p_2&=&p_2+s_2+ p_1+2 s_1+ 2 \log(1+e^{s_2})+ \nn\\
&+& \log\left((1+e^{p_1+s_1}+e^{p_1+s_1+s_2})(1+e^{s_1}+e^{s_1+s_2})\right).\nn
\eea

\begin{remark} Note that this flip is obtained by composing two mapping class group transformations described in Figures 3 and 4 of \cite{ChM1}. This means that the fat-graph of $PV$ is mapped to an intermediate fat-graph which does not satisfy the property that  the polygons containing holes with no cusps are  monogons. 
 This is not a problem as in fact we can map the lamination  in Fig.~\ref{fi:PV} to this new fat-graph and then to Fig.\ref{fi:PV-transformed}.
\end{remark}

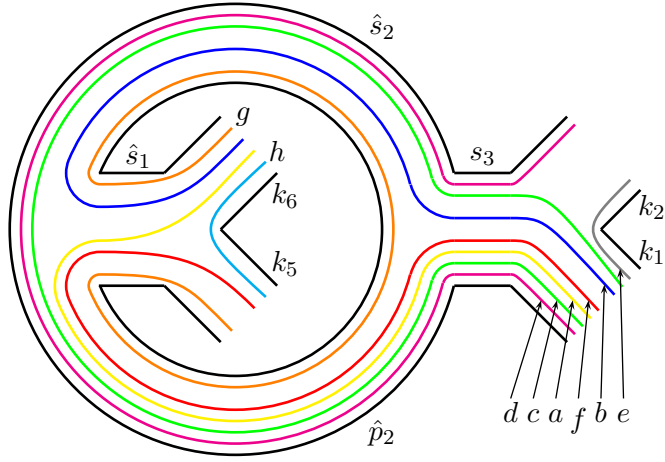
\begin{figure}[h]
{\psset{unit=0.5}
\begin{pspicture}(-6,-6)(2,6)
\newcommand{\TAIL}[3]{%
{\psset{unit=#1}
\pcline[linewidth=1pt,linecolor=black](0,0.5)(0.5,0.5)
\pcline[linewidth=1pt,linecolor=black](0,-0.5)(0.5,-0.5)
\pcline[linewidth=1pt,linecolor=black](0.5,0.5)(1,1)
\pcline[linewidth=1pt,linecolor=black](0.5,-0.5)(1,-1)
\pcline[linewidth=1pt,linecolor=magenta](0,0.4)(0.5,0.4)
\pcline[linewidth=1pt,linecolor=green](0,0.3)(0.5,0.3)
\pcline[linewidth=1pt,linecolor=blue](0,0.1)(0.5,0.1)
\pcline[linewidth=1pt,linecolor=#2](0,0)(0.5,0)
\pcline[linewidth=1pt,linecolor=#3](0,-0.1)(0.5,-0.1)
\pcline[linewidth=1pt,linecolor=yellow](0,-0.2)(0.5,-0.2)
\pcline[linewidth=1pt,linecolor=green](0,-0.3)(0.5,-0.3)
\pcline[linewidth=1pt,linecolor=magenta](0,-0.4)(0.5,-0.4)
\psbezier[linewidth=1pt,linecolor=magenta](0.5,-0.4)(0.55,-0.4)(0.55,-0.4)(1.07,-0.93)
\psbezier[linewidth=1pt,linecolor=green](0.5,-0.3)(0.6,-0.3)(0.6,-0.3)(1.14,-0.86)
\psbezier[linewidth=1pt,linecolor=yellow](0.5,-0.2)(0.65,-0.2)(0.65,-0.2)(1.21,-0.79)
\psbezier[linewidth=1pt,linecolor=#3](0.5,-0.1)(0.7,-0.1)(0.7,-0.1)(1.28,-0.72)
\psbezier[linewidth=1pt,linecolor=#2](0.5,0)(0.75,0)(0.75,0)(1.35,-0.65)
\psbezier[linewidth=1pt,linecolor=blue](0.5,0.1)(0.8,0.1)(0.8,0.1)(1.42,-0.58)
\psbezier[linewidth=1pt,linecolor=green](0.5,0.3)(0.85,0.3)(0.85,0.3)(1.49,-0.51)
\psbezier[linewidth=1pt,linecolor=magenta](0.5,0.4)(0.55,0.4)(0.55,0.4)(1.07,0.93)
\psbezier[linewidth=1pt,linecolor=gray](1.56,-0.44)(1.12,0)(1.12,0)(1.56,0.44)
\pcline[linewidth=1pt,linecolor=black](1.65,0.35)(1.3,0)
\pcline[linewidth=1pt,linecolor=black](1.65,-0.35)(1.3,0)
\pcline[linewidth=0.5pt,linecolor=black]{->}(0.5,-1.5)(.77,-0.63)
\rput(.5,-1.62){$d$}
\pcline[linewidth=0.5pt,linecolor=black]{->}(0.7,-1.5)(.91,-0.63)
\rput(.7,-1.65){$c$}
\pcline[linewidth=0.5pt,linecolor=black]{->}(0.9,-1.5)(1.05,-0.63)
\rput(.9,-1.65){$a$}
\pcline[linewidth=0.5pt,linecolor=black]{->}(1.1,-1.5)(1.19,-0.63)
\rput(1.1,-1.7){$f$}
\pcline[linewidth=0.5pt,linecolor=black]{->}(1.3,-1.5)(1.33,-0.49)
\rput(1.3,-1.62){$b$}
\pcline[linewidth=0.5pt,linecolor=black]{->}(1.5,-1.5)(1.47,-0.35)
\rput(1.5,-1.65){$e$}
\rput(1.75,-0.2){$k_1$}
\rput(1.75,0.2){$k_2$}
\rput(0.25,0.65){$s_3$}
\rput(-0.64,1.8){$\hat s_2$}
\rput(-0.64,-1.8){$\hat p_2$}
\rput(-2.8,0.65){$\hat s_1$}
\rput(-1.5,0.3){$k_6$}
\rput(-1.5,-0.3){$k_5$}
}
}
\newcommand{\BIGLOOP}[1]{%
{\psset{unit=#1}
\pcline[linewidth=1pt,linecolor=black](-3.15,0.5)(-2.57,0.5)
\pcline[linewidth=1pt,linecolor=black](-3.15,-0.5)(-2.57,-0.5)
\psarc[linecolor=black, linewidth=1pt](-1.94,0){2}{14.5}{345.5}
\psarc[linecolor=black, linewidth=1pt](-1.94,0){1.3}{-157.5}{157.5}
\pcline[linewidth=1pt,linecolor=black](-2.57,0.5)(-2.07,1)
\pcline[linewidth=1pt,linecolor=black](-2.57,-0.5)(-2.07,-1)
\pcline[linewidth=1pt,linecolor=black](-2.07,0)(-1.57,0.5)
\pcline[linewidth=1pt,linecolor=black](-2.07,0)(-1.57,-0.5)
%
\psarc[linecolor=magenta, linewidth=1pt](-1.94,0){1.9}{14.5}{345.5}
\psarc[linecolor=green, linewidth=1pt](-1.94,0){1.8}{14.5}{345.5}
\psarc[linecolor=yellow, linewidth=1pt](-1.94,0){1.7}{202.5}{345.5}
\psarc[linecolor=red, linewidth=1pt](-1.94,0){1.6}{202.5}{345.5}
\psarc[linecolor=white, linewidth=1pt](-1.94,0){1.5}{202.5}{345.5}
\psarc[linecolor=blue, linewidth=1pt](-1.94,0){1.6}{14.5}{157.5}
\psarc[linecolor=orange, linewidth=1pt](-1.94,0){1.4}{-157.5}{157.5}
\psarc[linecolor=magenta, linewidth=1pt](0,-0.5){0.1}{90}{165}
\psarc[linecolor=green, linewidth=1pt](0,-0.5){0.2}{90}{165}
\psarc[linecolor=yellow, linewidth=1pt](0,-0.5){0.3}{90}{165.5}
\psarc[linecolor=red, linewidth=1pt](0,-0.5){0.4}{90}{165.5}
\psarc[linecolor=white, linewidth=1pt](0,-0.5){0.5}{90}{165.5}
\psarc[linecolor=blue, linewidth=1pt](0,0.5){0.4}{195}{270}
\psarc[linecolor=magenta, linewidth=1pt](0,0.5){0.1}{195}{270}
\psarc[linecolor=green, linewidth=1pt](0,0.5){0.2}{195}{270}
\psarc[linecolor=red, linewidth=1pt](-3.14,-0.5){0.3}{90}{202.5}
\psarc[linecolor=yellow, linewidth=1pt](-3.14,-0.5){0.4}{90}{202.5}
\psarc[linecolor=white, linewidth=1pt](-3.14,-0.5){0.2}{90}{202.5}
\psarc[linecolor=orange, linewidth=1pt](-3.14,-0.5){0.1}{90}{202.5}
\psarc[linecolor=blue, linewidth=1pt](-3.14,0.5){0.3}{157.5}{270}
\psarc[linecolor=orange, linewidth=1pt](-3.14,0.5){0.1}{157.5}{270}
\psbezier[linewidth=1pt,linecolor=blue](-3.14,0.2)(-2.4,0.2)(-2.27,0.4)(-1.87,0.8)
\psbezier[linewidth=1pt,linecolor=white](-3.14,-0.3)(-2.4,-0.3)(-2.27,-0.4)(-1.87,-0.8)
\psbezier[linewidth=1pt,linecolor=yellow](-3.14,-0.1)(-2.4,-0.1)(-2.17,0.3)(-1.77,0.7)
\psbezier[linewidth=1pt,linecolor=red](-3.14,-0.2)(-2.4,-0.2)(-2.17,-0.3)(-1.77,-0.7)
\psbezier[linewidth=1pt,linecolor=cyan](-1.67,0.6)(-2.32,0)(-2.32,0)(-1.67,-0.6)
\psbezier[linewidth=1pt,linecolor=orange](-3.14,0.4)(-2.4,0.4)(-2.37,0.5)(-1.97,0.9)
\psbezier[linewidth=1pt,linecolor=orange](-3.14,-0.4)(-2.4,-0.4)(-2.37,-0.5)(-1.97,-0.9)
\rput(-1.87,1){$g$}
\rput(-1.57,.7){$h$}
}
}
\rput(0,0){\TAIL{3}{white}{red}}
\rput(0,0){\BIGLOOP{3}}
\end{pspicture}
}
\caption{\small The character variety for the $P{III}$ system.}
\label{fi:PIII}
\end{figure}

In the new shear coordinates the substitution (\ref{eq:sub-epsV}) becomes simply:
$$
\hat p_1\to \hat p_1-2\log[\epsilon],\nn
$$
which geometrically speaking corresponds to the fat-graph in Fig.~\ref{fi:PIII},  where we see $4$ new shear coordinates, one for each cusp, so that  $\hat p_2 =\hat k_3+\hat k_4$ and $\hat p_1=\hat k_5+\hat k_6$. This is the fat-graph of a Riemann sphere with two holes each of them with two cusps.

Note that the coordinates in Fig.~\ref{fi:PIII} are the true shear coordinates, namely they satisfy the Poisson brackets:
\bea
&&
\{\hat k_2,\hat s_3\}=\{\hat s_3,\hat k_1\}=\{\hat k_1,\hat k_2\}=\{\hat s_3,\hat s_2\}=\{\hat p_2,\hat s_3\}=1,\quad \{\hat s_2,\hat p_2\}=2,\nn\\
&&\{\hat s_1,\hat s_2\}=\{\hat p_2,\hat s_1\}=\{\hat s_1,\hat k_5\}=\{\hat k_6,\hat s_1\}=\{\hat k_5,\hat k_6\}=1,
\eea
If we use the limiting transformation of (\ref{new:shear}):
\bea\label{new:shearIII}
\hat s_1&=&-s_1-k_5-k_6- \log(1+e^{s_2}),\qquad \hat s_3=s_3-\log(1+e^{-s_2}),\nn\\
\hat s_2 &=&-s_2+ \log\left(1+e^{k_5+k_6+s_1}+e^{k_5+k_6+s_1+s_2})\right),\qquad \hat p_1=p_1,\nn\\
\hat k_1&=&k_1,\qquad \hat k_2=k_2,\qquad \hat k_5=k_5,\qquad \hat k_6=k_6,\nn\\
\hat p_2&=&p_2+s_2+ k_5+k_6+2 s_1+ 2 \log(1+e^{s_2})+ \\ 
&+&\log\left((1+e^{ k_5+k_6+s_1}+
e^{ k_5+k_6+s_1+s_2})(1+e^{s_1}+e^{s_1+s_2}\right).\nn
\eea
to go back to $s_1,s_2,s_3,p_2,k_1,k_2,k_5,k_6$, we see that $k_5$, $k_6$ have non standard Poisson brackets with $s_1,s_2,s_3$. This is due to the fact that this limiting transformation (\ref{new:shearIII}) destroys the geometry, as it essentially maps from a Riemann sphere with two holes each of them with two cusps to a Riemann sphere with two holes one of which has 4 cusps, and the other has no cusps (the $PIV$ case). This implies that the correct  coordinates to describe the character variety  of a Riemann sphere with two holes each of them with two cusps are the complexified   
$\hat s_1,\hat s_2,\hat s_3,\hat p_2,\hat k_1,\hat k_2,\hat k_5,\hat k_6$.

This character variety  is  $8$-dimensional.
The fat-graph admits a complete cusped lamination as displayed in Fig.~\ref{fi:PIII}
so that a full set of coordinates on the character variety is given by the eight complexified elements in the lamination.

In terms of the shear coordinates $\hat s_1,\hat s_2, \hat s_3,\hat p_2, \hat k_1,\hat k_2,\hat k_5,\hat k_6$ the elements in the $PIII^{D_6}$ lamination are expressed as follows:
\bea\label{PIII-hat-lam}
&&
a= e^{\frac{\hat k_1+\hat k_6-\hat s_1+\hat s_3+\hat p_2}{2}}+e^{\frac{\hat k_1+\hat k_6+\hat s_1+\hat s_3+\hat p_2}{2}},\qquad
e=e^{\frac{\hat k_1+\hat k_2}{2}},\qquad g=e^{\frac{\hat k_5+\hat k_6}{2}},\nn\\
&&
b= e^{\frac{\hat k_1+\hat k_6-\hat s_1-\hat s_2+\hat s_3}{2}}+e^{\frac{\hat k_1+\hat k_6+\hat s_1-\hat s_2+\hat s_3}{2}}+e^{\frac{\hat k_1+\hat k_6+\hat s_1+\hat s_2+\hat s_3}{2}},
\\
&&
c= e^{\frac{2\hat k_1+\hat s_2+2 \hat s_3+\hat p_2}{2}},\qquad
d= e^{\frac{\hat k_1+\hat k_2+\hat s_2+2 \hat s_3+\hat p_2}{2}},\nn\\
&&
f=e^{\frac{\hat k_1+\hat k_5+\hat s_1+\hat s_3+\hat p_2}{2}},\qquad 
h= e^{\frac{\hat k_5+\hat k_6+2\hat s_1+\hat s_2+\hat p_2}{2}}.\nn
\eea
The Poisson brackets admit two Casimirs, $d e$ and $h g$ so that the symplectic leaves are $6$-dimensional. Within each symplectic leaf, the $PIII^{D_6}$ monodromy manifold is given by the set of functions which commute with $d,e,g,h$. To see this, we proceed in the same way as in the case of PV. On the monodromy manifold we set $e=\tilde G_3$, $d=\tilde G_\infty$, $h=\tilde G_2$ and $g=\tilde G_1$.

\begin{figure}[h]
{\psset{unit=0.5}
\begin{pspicture}(-6,-6)(2,6)
\newcommand{\TAIL}[3]{%
{\psset{unit=#1}
\pcline[linewidth=1pt,linecolor=black](0,0.5)(0.5,0.5)
\pcline[linewidth=1pt,linecolor=black](0,-0.5)(0.5,-0.5)
\pcline[linewidth=1pt,linecolor=black](0.5,0.5)(1,1)
\pcline[linewidth=1pt,linecolor=black](0.5,-0.5)(1,-1)
\pcline[linewidth=1pt,linecolor=magenta](0,0.4)(0.5,0.4)
\pcline[linewidth=1pt,linecolor=green](0,0.3)(0.5,0.3)
\pcline[linewidth=1pt,linecolor=white](0,0.1)(0.5,0.1)
\pcline[linewidth=1pt,linecolor=#2](0,0)(0.5,0)
\pcline[linewidth=1pt,linecolor=#3](0,-0.1)(0.5,-0.1)
\pcline[linewidth=1pt,linecolor=yellow](0,-0.2)(0.5,-0.2)
\pcline[linewidth=1pt,linecolor=green](0,-0.3)(0.5,-0.3)
\pcline[linewidth=1pt,linecolor=magenta](0,-0.4)(0.5,-0.4)
\psbezier[linewidth=1pt,linecolor=magenta](0.5,-0.4)(0.55,-0.4)(0.55,-0.4)(1.07,-0.93)
\psbezier[linewidth=1pt,linecolor=green](0.5,-0.3)(0.6,-0.3)(0.6,-0.3)(1.14,-0.86)
\psbezier[linewidth=1pt,linecolor=yellow](0.5,-0.2)(0.65,-0.2)(0.65,-0.2)(1.21,-0.79)
\psbezier[linewidth=1pt,linecolor=#3](0.5,-0.1)(0.7,-0.1)(0.7,-0.1)(1.28,-0.72)
\psbezier[linewidth=1pt,linecolor=#2](0.5,0)(0.75,0)(0.75,0)(1.35,-0.65)
\psbezier[linewidth=1pt,linecolor=white](0.5,0.1)(0.8,0.1)(0.8,0.1)(1.42,-0.58)
\psbezier[linewidth=1pt,linecolor=green](0.5,0.3)(0.85,0.3)(0.85,0.3)(1.49,-0.51)
\psbezier[linewidth=1pt,linecolor=magenta](0.5,0.4)(0.55,0.4)(0.55,0.4)(1.07,0.93)
\psbezier[linewidth=1pt,linecolor=gray](1.56,-0.44)(1.12,0)(1.12,0)(1.56,0.44)
\pcline[linewidth=1pt,linecolor=black](1.65,0.35)(1.3,0)
\pcline[linewidth=1pt,linecolor=black](1.65,-0.35)(1.3,0)
\pcline[linewidth=0.5pt,linecolor=black]{->}(0.5,-1.5)(.77,-0.63)
\rput(.5,-1.62){$d$}
\pcline[linewidth=0.5pt,linecolor=black]{->}(0.7,-1.5)(.91,-0.63)
\rput(.7,-1.65){$c$}
\pcline[linewidth=0.5pt,linecolor=black]{->}(0.9,-1.5)(1.05,-0.63)
\rput(.9,-1.65){$a$}
\pcline[linewidth=0.5pt,linecolor=black]{->}(1.1,-1.5)(1.19,-0.63)
\rput(1.1,-1.7){$f$}
\pcline[linewidth=0.5pt,linecolor=black]{->}(1.3,-1.5)(1.33,-0.63)
\rput(1.3,-1.62){$b$}
\pcline[linewidth=0.5pt,linecolor=black]{->}(1.5,-1.5)(1.47,-0.35)
\rput(1.5,-1.65){$e$}
\rput(1.95,-0.2){$\tilde k_1{=}k_1$}
\rput(1.95,0.2){$\tilde k_2{=}k_2$}
\rput(0.25,0.65){$\tilde s_3$}
\rput(-0.34,1.6){$\tilde s_1$}
\rput(-0.34,-1.6){$\tilde s_2$}
\rput(-3,0){$\tilde p_2$}
\rput(-1.3,-0.9){$\tilde k_6$}
\rput(-1.3,0.9){$\tilde k_5$}
}
}
\newcommand{\BIGLOOP}[1]{%
{\psset{unit=#1}
\psarc[linecolor=black, linewidth=1pt](-1.94,0){2}{14.5}{345.5}
\psarc[linecolor=black, linewidth=1pt](-1.94,0){1.3}{-67.5}{67.5}
\psarc[linecolor=black, linewidth=1pt](-1.94,0){1.3}{112.5}{247.5}
\pcline[linewidth=1pt,linecolor=black](-2.45,1.19)(-2.45,0.7)
\pcline[linewidth=1pt,linecolor=black](-1.45,1.19)(-1.45,0.7)
\pcline[linewidth=1pt,linecolor=black](-2.45,-1.19)(-2.45,-0.7)
\pcline[linewidth=1pt,linecolor=black](-1.45,-1.19)(-1.45,-0.7)
%
\psarc[linecolor=magenta, linewidth=1pt](-1.94,0){1.9}{14.5}{345.5}
\psarc[linecolor=green, linewidth=1pt](-1.94,0){1.8}{14.5}{345.5}
\psarc[linecolor=yellow, linewidth=1pt](-1.94,0){1.7}{90}{345.5}
\psarc[linecolor=red, linewidth=1pt](-1.94,0){1.6}{90}{345.5}
\psarc[linecolor=blue, linewidth=1pt](-1.94,0){1.5}{270}{345.5}
%
\psarc[linecolor=magenta, linewidth=1pt](0,-0.5){0.1}{90}{165}
\psarc[linecolor=green, linewidth=1pt](0,-0.5){0.2}{90}{165}
\psarc[linecolor=yellow, linewidth=1pt](0,-0.5){0.3}{90}{165.5}
\psarc[linecolor=red, linewidth=1pt](0,-0.5){0.4}{90}{165.5}
\psarc[linecolor=blue, linewidth=1pt](0,-0.5){0.5}{90}{165.5}
\psarc[linecolor=magenta, linewidth=1pt](0,0.5){0.1}{195}{270}
\psarc[linecolor=green, linewidth=1pt](0,0.5){0.2}{195}{270}
\psarc[linecolor=red, linewidth=1pt](-1.94,1.5){0.1}{0}{90}
\pcline[linewidth=1pt,linecolor=red](-1.84,1.5)(-1.84,0.7)
\pcline[linewidth=1pt,linecolor=blue](-2.04,-1.4)(-2.04,-0.7)
\psarc[linecolor=blue, linewidth=1pt](-1.94,-1.4){0.1}{180}{270}
\psarc[linecolor=yellow, linewidth=1pt](-1.94,0){1.45}{-80}{0}
\psbezier[linewidth=1pt,linecolor=yellow](-1.94,1.7)(-1.2,1.7)(-0.49,1.1)(-0.49,0)
\psarc[linecolor=yellow, linewidth=1pt](-1.71,-1.23){0.2}{180}{280}
\pcline[linewidth=1pt,linecolor=yellow](-1.91,-1.23)(-1.91,-0.7)
%
\rput{90}(-1.94,0){
\psarc[linecolor=orange, linewidth=1pt](0,0){1.4}{22.5}{157.5}
\psarc[linecolor=orange, linewidth=1pt](-1.2,0.5){0.1}{157.5}{270}
\psarc[linecolor=orange, linewidth=1pt](1.2,0.5){0.1}{-90}{22.5}
\pcline[linewidth=1pt,linecolor=orange](-1.2,0.4)(-0.7,0.4)
\pcline[linewidth=1pt,linecolor=orange](1.2,0.4)(0.7,0.4)
}
\rput{-90}(-1.94,0){
\psarc[linecolor=cyan, linewidth=1pt](0,0){1.4}{22.5}{157.5}
\psarc[linecolor=cyan, linewidth=1pt](-1.2,0.5){0.1}{157.5}{270}
\psarc[linecolor=cyan, linewidth=1pt](1.2,0.5){0.1}{-90}{22.5}
\pcline[linewidth=1pt,linecolor=cyan](-1.2,0.4)(-0.7,0.4)
\pcline[linewidth=1pt,linecolor=cyan](1.2,0.4)(0.7,0.4)
}
\rput(-2.37,.55){$g$}
\rput(-1.57,.55){$h$}
}
}
\rput(0,0){\TAIL{3}{blue}{red}}
\rput(0,0){\BIGLOOP{3}}
\end{pspicture}
}
\caption{\small The character variety for the $P{III}$ system depicted on the fat graph dual to this variety.}
\label{fi:PIII-flip}
\end{figure}
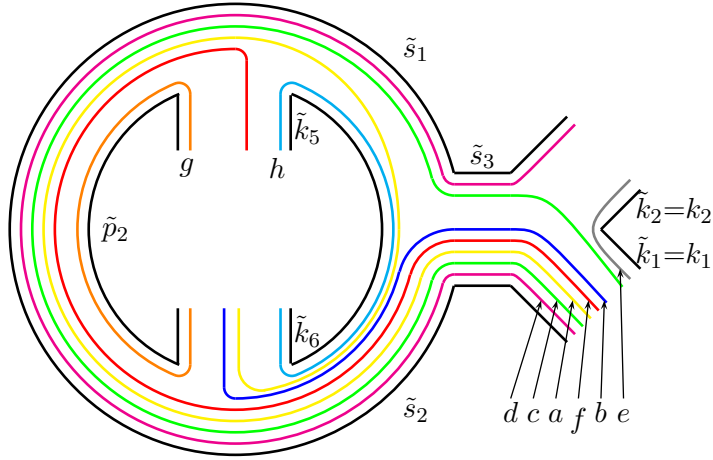

Note that, unlike the cases of $PV$ (Fig.~\ref{fi:PV}), $PIV$ (Fig.~\ref{fi:PIV}), and $PII$ (Fig.~\ref{fi:PIIFN} below), the expressions for the $\lambda$-lengths of arcs in 
(\ref{PIII-hat-lam}) are not monomials in the exponentiated shear coordinates. This is because, unlike the other named cases, the fat graph in Fig.~\ref{fi:PIII} is not dual to the maximum cusped lamination $\{a,b,c,d,e,f,g,h\}$. We obtain the dual graph out of the one in Fig.~\ref{fi:PIII} if we first flip the edge $\hat s_1$ subsequently flipping the edge $\hat s_2$. We depict the resulting fat graph and lamination in Fig.~\ref{fi:PIII-flip}. 
In the transformed shear coordinates (indicated by tildes), the elements of the $PIII^{D_6}$ lamination read
\bea\label{PIII-tilde-lam}
&&
a= e^{\frac{\tilde k_1+\tilde k_6+\tilde s_1+\tilde s_3+2\tilde s_2 +\tilde p_2}{2}},\qquad
e=e^{\frac{\tilde k_1+\tilde k_2}{2}},\qquad g=e^{\frac{\tilde k_5+\tilde k_6+\tilde p_2}{2}},\nn\\
&&
b= e^{\frac{\tilde k_1+\tilde k_6 +\tilde s_2+\tilde s_3}{2}},\qquad
c= e^{\frac{2\tilde k_1+\tilde s_1+\tilde s_2+2 \tilde s_3+\tilde p_2}{2}},\qquad
d= e^{\frac{\tilde k_1+\tilde k_2+\tilde s_1+\tilde s_2+2 \tilde s_3+\tilde p_2}{2}},
\\
&&
f=e^{\frac{\tilde k_1+\tilde k_5+\tilde s_2+\tilde s_3+\tilde p_2}{2}},\qquad 
h= e^{\frac{\tilde k_5+\tilde k_6+\tilde s_1+\tilde s_2}{2}}.\nn
\eea
In this form, the homogeneity of the Poisson 
relations on the set $\{a,b,c,d,e,f,g,h\}$ becomes obvious:
\bea\label{PIII-lam-p}
&&
\{a,b\}=\frac{1}{2}a b,\,\{a,c\}=0,\, \{a,d\}=-\frac{1}{4} a d,\,\{a,e\}=\frac{1}{4} a e,\, \{a,f\}=\frac{1}{4} a f,\nn\\
 &&
 \{a,g\}=-\frac{1}{4} a g,\ \{a,h\}=\frac{1}{4} a h,\,\{b,c\}=0, \, \{b,d\}= -\frac{1}{4} b d\,
 \{b,e\}=\frac{1}{4} b e,\,  \{b,f\}=-\frac{1}{4} b f,\nn\\
 &&
  \{b,g\}=-\frac{1}{4} b g,\,\{b,h\}=\frac{1}{4} b h,\,
 \{c,d\}=-\frac{1}{2} c d,\, \{c,e\}=\frac{1}{2} c e,\,  \{c,f\}=0,\,
 \{c,g\}=0,\\
 &&
\{c,h\}=0,\, \{d,e\}=0,\,  \{d,f\}=\frac{1}{4} d f,\, \{d,g\}=0,\,
 \{d,h\}=0,\nn\\
 &&
 \{e,f\}=-\frac{1}{4} e f,\,\{e,g\}=0,\
 \{e,h\}=0,\, \{f,g\}=\frac{1}{4}f g,\,
 \{f,h\}=-\frac{1}{4} f h,\,
  \{h,g\}=0 .\nn
\eea

It is interesting to construct the lamination and pattern in Fig.~\ref{fi:PIII} as a limit of the corresponding system of $PV$.
When we flip the fat-graph for $PV$, the lamination is flipped too as in Fig.~\ref{fi:PV-transformed}.

When we open the inside hole to obtain the fat-graph for $PIII^{D_6}$, the $PV$ arcs $a$ and $b$ break giving rise to the new arcs $a$, $b$ and $f$ and $G_1$ breaks up giving rise to $g$
so that the confluence from $PV$ to $PIII^{D_6}$ is again {\it reversed.}\/ Indeed the Poisson algebra (\ref{PV-lam-p}) is a Poisson sub-algebra of (\ref{PIII-lam-p}) 
specified by the functions of 
$$
a_V:=a_{III} f_{III},\quad b_V:=b_{III} f_{III},\quad c_V:=c_{III},\quad d_V:=d_{III},\quad 
$$
where we have denoted with an index $V$ the lamination coordinates in  (\ref{PV-lam-p})  and by $III$ the ones in  (\ref{PIII-lam-p}). Note that $h_{III}$ and $g_{III}$ are automatically a Casimir for the Poisson sub-algebra of (\ref{PV-lam-p}) and can be identified with $G_2$ and $G_1$ respectively.

\subsection{Decorated character variety for $PIII^{D_7}$}
The confluence from the generic $PIII^{D_6}$ cubic (\ref{eq:gen-p3}) to the $PIII^{D_7}$ one is realised by the substitution
$$
s_3\to s_3-\log[\epsilon],
$$
in formulae (\ref{eq:shear-PIII}). in the limit $\epsilon\to0$ we obtain:
\begin{eqnarray}
\label{eq:shear-PIIID7}
x_1&=&-e^{s_2+ s_3+\frac{p_2}{2}+\frac{p_3}{2}},\nn\\
x_2&=&-e^{s_3-s_1+\frac{p_3}{2}-\frac{p_1}{2}}-G_1e^{s_3+\frac{p_3}{2}},\nn\\
x_3&=&-e^{s_1+ s_2+\frac{p_1}{2}+\frac{p_2}{2}}-e^{-s_1- s_2-\frac{p_1}{2}-\frac{p_2}{2}}-G_1e^{-s_2-\frac{p_2}{2}}-G_2 e^{s_1+\frac{p_1}{2}},\nn\\\end{eqnarray}
where 
$$
G_i=e^{\frac{p_i}{2}},\quad i=1,2,\quad G_3=0,\quad
G_\infty=e^{s_1+s_2+s_3+\frac{p_1}{2}+\frac{p_2}{2}+\frac{p_3}{2}}.
$$
These same expressions can equivalently obtained by the substitution:
$$
s_1\to s_1+2 \log(\epsilon), \quad p_1\to p_1-2 \log(\epsilon), \quad \quad p_2\to p_2-2 \log(\epsilon),
$$
in formulae (\ref{eq:shear-PVdeg}) and by taking the limit as $\epsilon\to 0$.

These coordinates satisfy the following cubic relation:
\begin{equation}\label{eq:gen-p37}
x_1x_2x_3 + x_1^2+x_2^2-G_1 G_\infty x_1 -G_2 G_\infty x_2=0.
\end{equation}
We can pick $p_2=p_3=0$ and $s_3=-s_1-s_2-\frac{p_1}{2}$ in order to obtain the correct $PIII^{D_7}$ cubic.

In terms of fat graph, we need to work with the coordinates (\ref{new:shear}), for which the confluence gives 
$$
\hat s_3\to\hat s_3-\log[\epsilon],
$$
which corresponds to the fat-graph in Fig.~\ref{f.PIIID7fg}. This corresponds to a Riemann sphere with two holes, one with one cusp and one with two cusps.

\begin{figure}[h]
{\psset{unit=0.5}
\begin{pspicture}(-6,-6)(2,6)
\newcommand{\TAIL}[3]{%
{\psset{unit=#1}
\pcline[linewidth=1pt,linecolor=black](0,0.5)(0.5,0.5)
\pcline[linewidth=1pt,linecolor=black](0,-0.5)(0.5,-0.5)
\pcline[linewidth=1pt,linecolor=green](0,0.3)(0.5,0.3)
\pcline[linewidth=1pt,linecolor=blue](0,0.1)(0.5,0.1)
\pcline[linewidth=1pt,linecolor=#2](0,0)(0.5,0)
\pcline[linewidth=1pt,linecolor=#3](0,-0.1)(0.5,-0.1)
\pcline[linewidth=1pt,linecolor=yellow](0,-0.2)(0.5,-0.2)
\pcline[linewidth=1pt,linecolor=green](0,-0.3)(0.5,-0.3)
%
\pcline[linewidth=0.5pt,linecolor=black]{->}(0.7,-1)(.52,-0.3)
\rput(.7,-1.15){$c$}
\pcline[linewidth=0.5pt,linecolor=black]{->}(0.9,-1)(.52,-0.2)
\rput(.9,-1.15){$a$}
\pcline[linewidth=0.5pt,linecolor=black]{->}(1.1,-1)(.52,-0.1)
\rput(1.1,-1.2){$f$}
\pcline[linewidth=0.5pt,linecolor=black]{->}(1.3,-1)(.52,0.1)
\rput(1.3,-1.12){$b$}
\rput(0.25,0.65){$s_3$}
\rput(-0.64,1.8){$\hat s_2$}
\rput(-0.64,-1.8){$\hat p_2$}
\rput(-2.8,0.65){$\hat s_1$}
\rput(-1.5,0.3){$k_6$}
\rput(-1.5,-0.3){$k_5$}
}
}
\newcommand{\BIGLOOP}[1]{%
{\psset{unit=#1}
\pcline[linewidth=1pt,linecolor=black](-3.15,0.5)(-2.57,0.5)
\pcline[linewidth=1pt,linecolor=black](-3.15,-0.5)(-2.57,-0.5)
\psarc[linecolor=black, linewidth=1pt](-1.94,0){2}{14.5}{345.5}
\psarc[linecolor=black, linewidth=1pt](-1.94,0){1.3}{-157.5}{157.5}
\pcline[linewidth=1pt,linecolor=black](-2.57,0.5)(-2.07,1)
\pcline[linewidth=1pt,linecolor=black](-2.57,-0.5)(-2.07,-1)
\pcline[linewidth=1pt,linecolor=black](-2.07,0)(-1.57,0.5)
\pcline[linewidth=1pt,linecolor=black](-2.07,0)(-1.57,-0.5)
%
\psarc[linecolor=green, linewidth=1pt](-1.94,0){1.8}{14.5}{345.5}
\psarc[linecolor=yellow, linewidth=1pt](-1.94,0){1.7}{202.5}{345.5}
\psarc[linecolor=red, linewidth=1pt](-1.94,0){1.6}{202.5}{345.5}
\psarc[linecolor=white, linewidth=1pt](-1.94,0){1.5}{202.5}{345.5}
\psarc[linecolor=blue, linewidth=1pt](-1.94,0){1.6}{14.5}{157.5}
\psarc[linecolor=orange, linewidth=1pt](-1.94,0){1.4}{-157.5}{157.5}
%
\psarc[linecolor=green, linewidth=1pt](0,-0.5){0.2}{90}{165}
\psarc[linecolor=yellow, linewidth=1pt](0,-0.5){0.3}{90}{165.5}
\psarc[linecolor=red, linewidth=1pt](0,-0.5){0.4}{90}{165.5}
\psarc[linecolor=white, linewidth=1pt](0,-0.5){0.5}{90}{165.5}
\psarc[linecolor=blue, linewidth=1pt](0,0.5){0.4}{195}{270}
\psarc[linecolor=green, linewidth=1pt](0,0.5){0.2}{195}{270}
\psarc[linecolor=red, linewidth=1pt](-3.14,-0.5){0.3}{90}{202.5}
\psarc[linecolor=yellow, linewidth=1pt](-3.14,-0.5){0.4}{90}{202.5}
\psarc[linecolor=white, linewidth=1pt](-3.14,-0.5){0.2}{90}{202.5}
\psarc[linecolor=orange, linewidth=1pt](-3.14,-0.5){0.1}{90}{202.5}
\psarc[linecolor=blue, linewidth=1pt](-3.14,0.5){0.3}{157.5}{270}
\psarc[linecolor=orange, linewidth=1pt](-3.14,0.5){0.1}{157.5}{270}
\psbezier[linewidth=1pt,linecolor=blue](-3.14,0.2)(-2.4,0.2)(-2.27,0.4)(-1.87,0.8)
\psbezier[linewidth=1pt,linecolor=white](-3.14,-0.3)(-2.4,-0.3)(-2.27,-0.4)(-1.87,-0.8)
\psbezier[linewidth=1pt,linecolor=yellow](-3.14,-0.1)(-2.4,-0.1)(-2.17,0.3)(-1.77,0.7)
\psbezier[linewidth=1pt,linecolor=red](-3.14,-0.2)(-2.4,-0.2)(-2.17,-0.3)(-1.77,-0.7)
\psbezier[linewidth=1pt,linecolor=cyan](-1.67,0.6)(-2.32,0)(-2.32,0)(-1.67,-0.6)
\psbezier[linewidth=1pt,linecolor=orange](-3.14,0.4)(-2.4,0.4)(-2.37,0.5)(-1.97,0.9)
\psbezier[linewidth=1pt,linecolor=orange](-3.14,-0.4)(-2.4,-0.4)(-2.37,-0.5)(-1.97,-0.9)
\rput(-1.87,1){$g$}
\rput(-1.57,.7){$h$}
}
}
\rput(0,0){\TAIL{3}{white}{red}}
\rput(0,0){\BIGLOOP{3}}
\end{pspicture}
}
\caption{The  character variety of $ PIII^{D_7}$.}
\label{f.PIIID7fg}
\end{figure}

The character variety is $6$ dimensional; the picture of the $PIII^{D_7}$ lamination is very similar to Fig.~\ref{fi:PIII}, in which the edges labelled by  $k_1$ and $k_2$ are removed and  the arcs $d$ and $e$ are lost. Again the Poisson brackets can be calculated using (\ref{eq:comb-p}) and there are two Casimirs so that the symplectic leaves are $4$ dimensional and that the Poisson algebra associated to $PIII^{D_7}$ is the Poisson sub-algebra of  (\ref{PIII-lam-p}) given by the functions of $a,b,c,f,g,h$ only. The $PIII^{D_7}$ monodromy manifold is given by those functions of $a,b,c,d,f,h$ that commute with $h$.

\subsection{Decorated character variety for $PIII^{D_8}$}

The confluence from the generic $PIII^{D_7}$ cubic (\ref{eq:gen-p37}) to the $ PIII^{D_8}$ one is realised by the substitution
$$
s_1\to s_1+\log[\epsilon],\quad p_2\to p_2-2\log[\epsilon]
$$
in formulae (\ref{eq:shear-PIIID7}). In the limit $\epsilon\to0$ we obtain:
\begin{eqnarray}
\label{eq:shear-PIIID8}
x_1&=&-e^{s_2+ s_3+\frac{p_2}{2}+\frac{p_3}{2}},\nn\\
x_2&=&-e^{s_3-s_1+\frac{p_3}{2}-\frac{p_1}{2}},\nn\\
x_3&=&-e^{s_1+ s_2+\frac{p_1}{2}+\frac{p_2}{2}}-e^{-s_1- s_2-\frac{p_1}{2}-\frac{p_2}{2}}-G_2 e^{s_1+\frac{p_1}{2}},\nn\\\end{eqnarray}
where 
$$
G_1=0,\quad G_2=e^{\frac{p_2}{2}},\quad i=1,2,\quad G_3=0,\quad
G_\infty=e^{s_1+s_2+s_3+\frac{p_1}{2}+\frac{p_2}{2}+\frac{p_3}{2}}.
$$
These coordinates satisfy the following cubic relation:
\begin{equation}\label{eq:gen-p37-1}
x_1x_2x_3 + x_1^2+x_2^2 -G_2 G_\infty x_2=0.
\end{equation}
We can pick $p_2=p_3=0$ and $s_3=-s_1-s_2-\frac{p_1}{2}$ in order to obtain the correct $ PIII^{D_8}$ cubic.

In terms of fat graph, we need to work with the coordinates (\ref{new:shear}), for which the confluence gives 
$$
\hat s_1\to\hat s_1-\log[\epsilon],
$$
which corresponds to the fat-graph in Fig.~\ref{f.PIIID8fg}. This corresponds to a Riemann sphere with two holes, each with one cusp on it.

\begin{figure}[h]
{\psset{unit=0.5}
\begin{pspicture}(-6,-6)(2,6)
\newcommand{\TAIL}[3]{%
{\psset{unit=#1}
\pcline[linewidth=1pt,linecolor=black](0,0.5)(0.5,0.5)
\pcline[linewidth=1pt,linecolor=black](0,-0.5)(0.5,-0.5)
\pcline[linewidth=1pt,linecolor=green](0,0.3)(0.5,0.3)
\pcline[linewidth=1pt,linecolor=blue](0,0.1)(0.5,0.1)
\pcline[linewidth=1pt,linecolor=#2](0,0)(0.5,0)
\pcline[linewidth=1pt,linecolor=#3](0,-0.1)(0.5,-0.1)
\pcline[linewidth=1pt,linecolor=green](0,-0.3)(0.5,-0.3)
%
\pcline[linewidth=0.5pt,linecolor=black]{->}(0.7,-1)(.52,-0.3)
\rput(.7,-1.15){$a$}
\pcline[linewidth=0.5pt,linecolor=black]{->}(1.1,-1)(.52,-0.1)
\rput(1.1,-1.2){$f$}
\pcline[linewidth=0.5pt,linecolor=black]{->}(1.3,-1)(.52,0.1)
\rput(1.3,-1.12){$b$}
\rput(0.25,0.65){$s_3$}
\rput(-0.64,1.8){$\hat s_2$}
\rput(-0.64,-1.8){$\hat p_2$}
\rput(-2.8,0.65){$\hat s_1$}
}
}
\newcommand{\BIGLOOP}[1]{%
{\psset{unit=#1}
\pcline[linewidth=1pt,linecolor=black](-3.15,0.5)(-2.57,0.5)
\pcline[linewidth=1pt,linecolor=black](-3.15,-0.5)(-2.57,-0.5)
\psarc[linecolor=black, linewidth=1pt](-1.94,0){2}{14.5}{345.5}
\psarc[linecolor=black, linewidth=1pt](-1.94,0){1.3}{-157.5}{157.5}
\psarc[linecolor=green, linewidth=1pt](-1.94,0){1.8}{14.5}{345.5}
\psarc[linecolor=red, linewidth=1pt](-1.94,0){1.6}{202.5}{345.5}
\psarc[linecolor=white, linewidth=1pt](-1.94,0){1.5}{202.5}{345.5}
\psarc[linecolor=blue, linewidth=1pt](-1.94,0){1.6}{14.5}{157.5}
\psarc[linecolor=orange, linewidth=1pt](-1.94,0){1.4}{-157.5}{157.5}
\psarc[linecolor=green, linewidth=1pt](0,-0.5){0.2}{90}{165}
\psarc[linecolor=red, linewidth=1pt](0,-0.5){0.4}{90}{165.5}
\psarc[linecolor=white, linewidth=1pt](0,-0.5){0.5}{90}{165.5}
\psarc[linecolor=blue, linewidth=1pt](0,0.5){0.4}{195}{270}
\psarc[linecolor=green, linewidth=1pt](0,0.5){0.2}{195}{270}
\psarc[linecolor=red, linewidth=1pt](-3.14,-0.5){0.3}{90}{202.5}
\psarc[linecolor=white, linewidth=1pt](-3.14,-0.5){0.2}{90}{202.5}
\psarc[linecolor=orange, linewidth=1pt](-3.14,-0.5){0.1}{90}{202.5}
\psarc[linecolor=blue, linewidth=1pt](-3.14,0.5){0.3}{157.5}{270}
\psarc[linecolor=orange, linewidth=1pt](-3.14,0.5){0.1}{157.5}{270}
\psbezier[linewidth=1pt,linecolor=blue](-3.14,0.2)(-2.4,0.2)(-2.27,0.4)(-1.87,0.8)
\psbezier[linewidth=1pt,linecolor=white](-3.14,-0.3)(-2.4,-0.3)(-2.27,-0.4)(-1.87,-0.8)
\psbezier[linewidth=1pt,linecolor=red](-3.14,-0.2)(-2.4,-0.2)(-2.17,-0.3)(-1.77,-0.7)
\psbezier[linewidth=1pt,linecolor=orange](-3.14,0.4)(-2.4,0.4)(-2.37,0.5)(-1.97,0.9)
\psbezier[linewidth=1pt,linecolor=orange](-3.14,-0.4)(-2.4,-0.4)(-2.37,-0.5)(-1.97,-0.9)
\rput(0,0){\psframe[linecolor=white, fillstyle=solid, fillcolor=white](-2.57,-1)(-1.57,1)}
\rput(-2.5,0.5){$g$}
}
}
\rput(0,0){\TAIL{3}{white}{red}}
\rput(0,0){\BIGLOOP{3}}
\end{pspicture}
}
\caption{The character variety of $ PIII^{D_8}$.}\label{f.PIIID8fg}
\end{figure}

The character variety is $4$ dimensional and the Poisson brackets can be calculated using (\ref{eq:comb-p}).
and there are two Casimirs so that the symplectic leaves are $2$ dimensional and  coincide with the $PIII^{D_8}$ monodromy manifold.
The Poisson algebra associated to $PIII^{D_8}$ is the Poisson sub-algebra of  the one associated to  $PIII^{D_7}$ given by the functions of $a,b,c,h$ only.

\subsection{Decorated character variety for $ PII^{JM} $}

The confluence from the generic $PIV$ cubic (\ref{eq:gen-p4}) to the $ PII^{JM} $ cubic is realised by the substitution
$$
 p_1\to p_1 -2\log[\epsilon],
$$
in formulae (\ref{eq:shear-PIV}). In the limit $\epsilon\to0$ we obtain:
\begin{eqnarray}
\label{eq:shear-PII}
x_1&=&-e^{s_2+ s_3+\frac{p_2}{2}+\frac{p_3}{2}}-G_3 e^{s_2+\frac{p_2}{2}},\nn\\
x_2&=&-e^{s_3+s_1+\frac{p_3}{2}+\frac{p_1}{2}}-G_1e^{s_3+\frac{p_3}{2}},\nn\\
x_3&=&-e^{s_1+ s_2+\frac{p_1}{2}+\frac{p_2}{2}}-G_2 e^{s_1+\frac{p_1}{2}},\nn\\
\end{eqnarray}
where 
$$
G_i=e^{\frac{p_i}{2}},\, i=1,2,3,\quad
G_\infty=e^{s_1+s_2+s_3+\frac{p_1}{2}+\frac{p_2}{2}+\frac{p_3}{2}}.
$$
These coordinates satisfy the following cubic relation:
\bea\label{eq:gen-p2}
&&
x_1x_2x_3 -G_1 G_\infty x_1 -G_2 G_\infty  x_2  -\nn\\
&&-G_3 G_\infty x_3+ G_\infty^2+G_1 G_2G_3 G_\infty=0.
\eea
Note that the parameters $p_3,p_2,p_1$ are now redundant, we can eliminate it by rescaling.
To obtain the correct $PII^{JM}$ cubic, we need to pick $p_2=p_3=0$ and $p_1=-2 s_1-2 s_2-2 s_3$. 

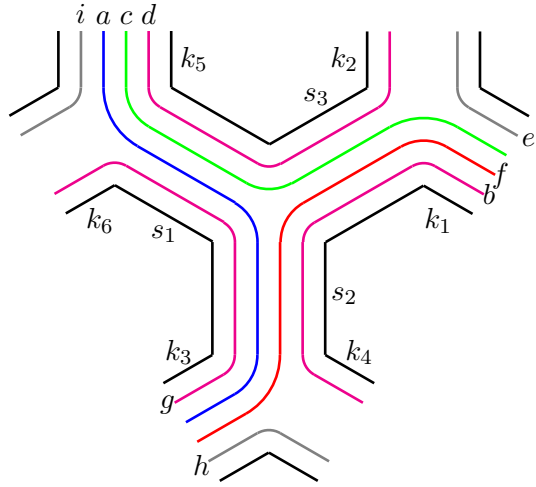
\begin{figure}[h]
{\psset{unit=0.5}
\begin{pspicture}(-6,-7)(2,6)
\newcommand{\TAIL}[3]{%
{\psset{unit=#1}
\pcline[linewidth=1pt,linecolor=black](-0.5,0)(-0.5,-1)
\pcline[linewidth=1pt,linecolor=black](0.5,0)(0.5,-1)
\pcline[linewidth=1pt,linecolor=magenta](0.3,0)(0.3,-1)
\pcline[linewidth=1pt,linecolor=#2](-0.1,0)(-0.1,-1)
\pcline[linewidth=1pt,linecolor=#3](0.1,0)(0.1,-1)
\pcline[linewidth=1pt,linecolor=magenta](-0.3,0)(-0.3,-1)
\psarc[linecolor=magenta, linewidth=1pt](-0.5,0){0.2}{0}{60}
\psarc[linecolor=#2, linewidth=1pt](-0.5,0){0.4}{0}{60}
\rput{120}(0,-1.29){\rput(0,0.29){
\pcline[linewidth=1pt,linecolor=black](-0.5,0)(-0.5,0.5)
\pcline[linewidth=1pt,linecolor=black](0.5,0)(0.5,0.5)
\pcline[linewidth=1pt,linecolor=magenta](0.3,0)(0.3,0.5)
\pcline[linewidth=1pt,linecolor=#2](0.1,0)(0.1,0.5)
\pcline[linewidth=1pt,linecolor=#3](-0.1,0)(-0.1,0.5)
\pcline[linewidth=1pt,linecolor=gray](-0.3,0)(-0.3,0.5)
\psarc[linecolor=magenta, linewidth=1pt](0.5,0){0.2}{180}{240}
\psarc[linecolor=#2, linewidth=1pt](0.5,0){0.4}{180}{240}
\psarc[linecolor=#3, linewidth=1pt](0.5,0){0.6}{180}{240}
}
}
\rput{240}(0,-1.29){\rput(0,0.29){
\pcline[linewidth=1pt,linecolor=black](-0.5,0)(-0.5,0.5)
\pcline[linewidth=1pt,linecolor=black](0.5,0)(0.5,0.5)
\pcline[linewidth=1pt,linecolor=gray](0.3,0)(0.3,0.5)
\pcline[linewidth=1pt,linecolor=magenta](-0.3,0)(-0.3,0.5)
\psarc[linecolor=magenta, linewidth=1pt](-0.5,0){0.2}{-60}{0}
\psarc[linecolor=gray, linewidth=1pt](0.5,0){0.2}{180}{240}
}
}
}
}
\rput(-5,5.2){$i$}
\rput(-4.4,5.1){$a$}
\rput(-3.8,5.1){$c$}
\rput(-3.2,5.2){$d$}
\rput(6.9,1.85){$e$}
\rput(6.15,.9){$f$}
\rput(5.85,.4){$b$}
\rput(-2.7,-5.2){$g$}
\rput(-1.8,-6.8){$h$}
\rput(4.5,-0.3){$k_1$}
\rput(2,4){$k_2$}
\rput(-2.4,-3.8){$k_3$}
\rput(2.4,-3.8){$k_4$}
\rput(1.25,3){$s_3$}
\rput(2,-2.3){$s_2$}
\rput(-2.8,-0.65){$s_1$}
\rput(-2,4){$k_5$}
\rput(-4.5,-0.3){$k_6$}
\rput(0,0){\rput(0.12,-0.87){\TAIL{3}{blue}{red}}}
\rput{120}(0,0){\rput(0.12,-0.87){\TAIL{3}{red}{green}}}
\rput{240}(0,0){\rput(0.12,-0.87){\TAIL{3}{green}{blue}}}
\end{pspicture}
}
\caption{The character variety of  $ PII^{JM} $.}
\label{f.PIIJMfg}
\end{figure}

Again, this means that we send the perimeter $p_1$ to infinity, which is a chewing-gum move leading to a Riemann sphere with one hole with $6$ cusps on it. The corresponding fat-graph and corresponding lamination is given in Fig.~\ref{f.PIIJMfg} where we have $6$ cusp shear coordinates $k_1,\dots,k_6$.

The elements in the lamination read:
\bea\label{PIIJM-lam}
&&
\qquad a= e^{\frac{s_1+s_2+k_3+k_5}{2}},\quad
b= e^{\frac{s_2+s_3+k_1+k_4}{2}},\quad c= e^{\frac{s_1+s_3+k_1+k_5}{2}},\quad d= e^{\frac{s_1+s_3+k_2+k_5}{2}},\\
&&
e = e^{\frac{k_1+k_2}{2}},\quad
f=  e^{\frac{s_2+s_3+k_1+k_3}{2}},\quad
g=e^{\frac{s_1+s_2+k_3+k_6}{2}},\quad
h = e^{\frac{k_3+ k_4}{2}},\quad 
i=e^{\frac{k_5+k_6}{2}}.\nn
\eea
The character variety is now $9$-dimensional, the Poisson brackets can be easily extracted from (\ref{eq:comb-p}):
\bea\label{PIIJM-lam-p}
&&
\{a,b\}=0,\quad \{a,c\}=-\frac{1}{4} a c,\quad \{a,d\}=-\frac{1}{4} a d,\quad\{a,e\}=0,\quad \{a,f\}=\frac{1}{4} a f,\nn\\
 &&
 \{a,g\}=-\frac{1}{4} a g,\quad
 \{a,h\}=\frac{1}{4} a h,\quad \{a,i\}=\frac{1}{4} a i,\quad
\{b,c\}=\frac{1}{4}b c, \quad \{b,d\}=0,\nn\\
 &&
\{b,e\}=\frac{1}{4} b e,\quad \{b,f\}=\frac{1}{4} b f,\quad  \{b,g\}=0,\quad
 \{b,h\}=-\frac{1}{4} b h,\quad \{b,i\}=0,\nn\\
 &&
 \{c,d\}=-\frac{1}{4} c d,\quad \{c,e\}=\frac{1}{4} c e,\quad  \{c,f\}=-\frac{1}{4} c f,\quad  \{c,g\}=0,\quad  \{c,h\}=0,\\
&&
 \{c,i\}=\frac{1}{4} c i,\quad 
\{d,e\}=-\frac{1}{4} d e,\quad  \{d,f\}=0,\quad  \{d,g\}=0,\quad
 \{d,h\}=0,\nn\\
 &&
  \{d,i\}=\frac{1}{4} d i,\quad \{e,f\}=-\frac{1}{4} e f,\quad
 \{e,g\}=0,\quad \{e,h\}=0,\quad  \{e,i\}=0,\nn\\ 
 &&
  \{f,g\}=-\frac{1}{4} f g,\quad  \{f,h\}=\frac{1}{4}  f h \quad \{f,i\}=0,\quad
   \{g,h\}=\frac{1}{4}  g h,\quad  \{g,i\}=-\frac{1}{4}  g i \quad \{h,i\}=0.\nn
\eea
so that there is only one Casimir $b d e g h i$.
The confluence from the $PIV$ character variety to the $PII^{JM}$ one is again reversed, so that the Poisson algebra (\ref{PIV-lam-p}) is the  sub-algebra of (\ref{PIIJM-lam-p}) defined by the functions of:
\bea
&&
a_{IV}:= a_{II^{JM}} g_{II^{JM}},\quad b_{IV}:= b_{II^{JM}},\quad c_{IV}:= c_{II^{JM}} g_{II^{JM}},\quad d_{IV}:= d_{II^{JM}} g_{II^{JM}},\nn\\
&&
e_{IV}:= e_{II^{JM}},\quad f_{IV}:= f_{II^{JM}},\quad h_{IV}:= h_{II^{JM}},\nn
\eea
where we have indicated with an index $IV$ the lamination coordinates associated to (\ref{PIV-lam-p}) and by an index $II^{JM}$ the ones associated to (\ref{PIIJM-lam-p}).

Once again, let us stress that the $PII^{JM}$ character variety (\ref{PIIJM-lam-p}) is the largest Poisson algebra appearing in the theory of the defferential Painlev\'e equations. This algebra corresponds to the triangulations of a hexagon ($3+6 =9$ variables) which, for the Casimir value $1$,  is exactly the $A_3$ quiver in Sutherland's confluenced Dynkin diagrams \cite{S}.

The $PII^{JM}$ monodromy manifold is given by the set of functions that Poisson commute with $p_1=k_5+k_6$, $p_2=k_3+k_4$ and $p_3=k_1+k_2$. We omit all details in this case.

\subsection{Decorated character variety for $ PII^{FN}$}
The confluence from the generic $PIV$ cubic (\ref{eq:gen-p4}) to the $ PII^{FN}$ cubic is realised by the substitution
$$
s_3\to s_3-\log[\epsilon],
$$
in formulae (\ref{eq:shear-PIV}). In the limit $\epsilon\to0$ we obtain:
\begin{eqnarray}
\label{eq:shear-PIIFN}
x_1&=&-e^{s_2+ s_3+\frac{p_2}{2}+\frac{p_3}{2}},\nn\\
x_2&=&-e^{s_3+s_1+\frac{p_3}{2}+\frac{p_1}{2}}-e^{s_3-s_1+\frac{p_3}{2}-\frac{p_1}{2}}-G_1e^{s_3+\frac{p_3}{2}},\nn\\
x_3&=&-e^{s_1+ s_2+\frac{p_1}{2}+\frac{p_2}{2}}-G_2 e^{s_1+\frac{p_1}{2}},\nn\\
\nn
\end{eqnarray}
where 
$$
G_1=e^{\frac{p_1}{2}}+e^{-\frac{p_1}{2}},\quad G_2=e^{\frac{p_2}{2}},\quad G_3=0,\quad
G_\infty=e^{s_1+s_2+s_3+\frac{p_1}{2}+\frac{p_2}{2}+\frac{p_3}{2}}..
$$
They satisfy the following cubic relation
\begin{equation}\label{eq:gen-p2fn}
x_1x_2x_3 + x_1^2-G_1 G_\infty x_1 -G_2 G_\infty  x_2  + G_\infty^2=0.
\end{equation}

Observe that we can obtain exactly the same formulate by starting from the generic $PV_{deg}$  cubic (\ref{eq:gen-p5deg})  by the substitution
$$
p_2\to p_2-2\log[\epsilon],
$$
and by taking the limit as $\epsilon\to 0$.

To obtain the $ PII^{FN}$ cubic  we pick $p_2=p_3=0$ and $p_1=-2 s_1-2 s_2-2 s_3$.

\begin{figure}[h]
{\psset{unit=0.5}
\begin{pspicture}(-2,-8)(2,3)
\newcommand{\TAIL}[3]{%
{\psset{unit=#1}
\pcline[linewidth=1pt,linecolor=black](0,0.5)(0.5,0.5)
\pcline[linewidth=1pt,linecolor=black](0,-0.5)(0.5,-0.5)
\pcline[linewidth=1pt,linecolor=magenta](0,0.4)(0.5,0.4)
\pcline[linewidth=1pt,linecolor=white](0,0.3)(0.5,0.3)
\pcline[linewidth=1pt,linecolor=blue](0,0.1)(0.5,0.1)
\pcline[linewidth=1pt,linecolor=orange](0,-0.4)(0.5,-0.4)
\pcline[linewidth=0.5pt,linecolor=black]{->}(-2.2,-1.6)(-1.7,-1.8)
\rput(-2.3,-1.6){$d$}
\pcline[linewidth=0.5pt,linecolor=black]{->}(-2.2,-1.95)(-1.5,-2)
\rput(-2.3,-1.95){$a$}
\pcline[linewidth=0.5pt,linecolor=black]{->}(-2.2,-2.2)(-1.3,-2.2)
\rput(-2.3,-2.2){$f$}
\pcline[linewidth=0.5pt,linecolor=black]{->}(-2.2,-2.5)(-1.2,-2.3)
\rput(-2.3,-2.5){$h$}
\rput(.65,-1.85){$b$}
%
\rput(-.85,-2.4){$k_3$}
\rput(-.25,-2.4){$k_4$}
\rput(0.25,0.65){$s_3$}
\rput(-1.3,0.65){$s_1$}
\rput(0.15,-0.75){$s_2$}
\rput(-2.35,-1.15){$p_1$}
}
}
\newcommand{\LOOPS}[1]{%
{\psset{unit=#1}
\pcline[linewidth=1pt,linecolor=black](0,0.5)(-1.5,0.5)
\pcline[linewidth=1pt,linecolor=black](0,-0.5)(0,-1)
\pcline[linewidth=1pt,linecolor=black](-1.1,-0.5)(-1.1,-1)
\pcline[linewidth=1pt,linecolor=black](-1.1,-0.5)(-1.5,-0.5)
\pcline[linewidth=1pt,linecolor=black](-1.1,-1)(-1.8,-1.7)
\pcline[linewidth=1pt,linecolor=black](0,-1)(.7,-1.7)
\pcline[linewidth=1pt,linecolor=black](-.55,-1.85)(-1.1,-2.4)
\pcline[linewidth=1pt,linecolor=black](-.55,-1.85)(0,-2.4)
\psarc[linecolor=black, linewidth=1pt](-2.35,0){1}{30}{330}
\pscircle[linecolor=black, linewidth=1pt](-2.35,0){0.2}
\psarc[linecolor=orange, linewidth=1pt](0,-0.5){0.1}{90}{180}
\psarc[linecolor=blue, linewidth=1pt](0,-0.5){0.6}{90}{180}
\psarc[linecolor=magenta, linewidth=1pt](-1.1,-0.5){0.1}{0}{90}
\psarc[linecolor=yellow, linewidth=1pt](-1.1,-0.5){0.3}{0}{90}
\psarc[linecolor=yellow, linewidth=1pt](-1.1,-0.5){0.4}{0}{90}
\pcline[linewidth=1pt,linecolor=magenta](0,0.4)(-1.5,0.4)
\pcline[linewidth=1pt,linecolor=magenta](-1.1,-0.4)(-1.5,-0.4)
\pcline[linewidth=1pt,linecolor=orange](-0.1,-0.5)(-0.1,-1)
\pcline[linewidth=1pt,linecolor=magenta](-1,-0.5)(-1,-1)
\pcline[linewidth=1pt,linecolor=blue](-0.6,-0.5)(-0.6,-1)
\pcline[linewidth=1pt,linecolor=yellow](-0.7,-0.5)(-0.7,-1)
\pcline[linewidth=1pt,linecolor=yellow](-0.8,-0.5)(-0.8,-1)
\pcline[linewidth=1pt,linecolor=yellow](-1.1,-0.2)(-1.5,-0.2)
\pcline[linewidth=1pt,linecolor=yellow](-1.1,-0.1)(-1.5,-0.1)
%
%
\psarc[linecolor=magenta, linewidth=1pt](-2.35,0){0.8}{90}{270}
\psarc[linecolor=yellow, linewidth=1pt](-2.35,0){0.4}{90}{270}
\psbezier[linewidth=1pt,linecolor=orange](-0.1,-1)(-0.1,-1.2)(0.3,-1.5)(0.6,-1.8)
\psbezier[linewidth=1pt,linecolor=magenta](-1,-1)(-1,-1.2)(-1.4,-1.5)(-1.7,-1.8)
\psbezier[linewidth=1pt,linecolor=yellow](-0.8,-1)(-0.8,-1.2)(-1.2,-1.7)(-1.5,-2)
\psbezier[linewidth=1pt,linecolor=yellow](-0.7,-1)(-0.7,-1.2)(-1.1,-1.8)(-1.4,-2.1)
\psbezier[linewidth=1pt,linecolor=blue](-0.6,-1)(-0.6,-1.2)(-0.9,-1.8)(-1.3,-2.2)
\psbezier[linewidth=1pt,linecolor=cyan](0.1,-2.3)(-0.55,-1.6)(-0.55,-1.6)(-1.2,-2.3)
\psbezier[linewidth=1pt,linecolor=magenta](-2.35,0.8)(-2,0.8)(-1.7,0.4)(-1.5,0.4)
\psbezier[linewidth=1pt,linecolor=magenta](-2.35,-0.8)(-2,-0.8)(-1.7,-0.4)(-1.5,-0.4)
\psbezier[linewidth=1pt,linecolor=yellow](-2.35,0.4)(-2,0.4)(-1.7,-0.1)(-1.5,-0.1)
\psbezier[linewidth=1pt,linecolor=yellow](-2.35,-0.4)(-2,-0.4)(-1.7,-0.2)(-1.5,-0.2)
}
}
\rput(0,0){\TAIL{3}{blue}{white}}
\rput(0,0){\LOOPS{3}}
\end{pspicture}
}
\caption{\small The system of arcs for $PII^{\text{FN}}$.}
\label{fi:PIIFN}
\end{figure}
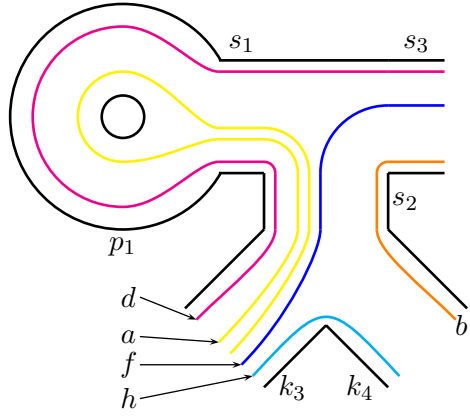

Geometrically speaking, sending the shear coordinate $s_3$ to infinity means that we are performing a cusp-removing move. This gives a Riemann sphere with two holes, one of them having three cusps on its boundary. In terms of the fat-graph, this is represented by Fig.  \ref{fi:PIIFN}, where we wee three cusps of coordinates $k_3,k_4$ and $s_3$, so that in formulae (\ref{eq:shear-PIIFN}) we must set $p_2=k_3+k_4$. .

The decorated character variety in this case is $6$ dimensional. The lamination is given by  the loop around the un-cusped-hole and the five arcs in Fig.~\ref{fi:PIIFN}.

The  lengths of the arcs are 
\bea\label{PIIFN-lam}
&&
a= e^{s_1+s_2+k_3+\frac{p_1}{2}},\quad
b= e^{\frac{s_2}{2}+\frac{s_3}{2}+\frac{k_4}{2}},\\
&&
d= e^{s_1+ \frac{s_2}{2}+\frac{s_3}{2}+\frac{p_1}{2}+\frac{k_3}{2}},\quad
f=  e^{ \frac{s_2}{2}+\frac{s_3}{2}+\frac{k_3}{2}},\quad
g = e^{\frac{k_3}{2}+\frac{k_4}{2}}.\nn
\eea
To show that our decorated character variety is not the same as the wild character variety (see  \cite{B2} for the $PII^{FN}$ case), we deal with this case in all details. The Poisson brackets among the complexified lamination arcs lengths are given by
\bea
&&
\{a,b\}=\{d,f\}=0,\quad \{a,d\}=-\frac{ad}{2},\quad \{a,f\}=\frac{af}{2},\quad \{a,h\}=\frac{af}{2},\nn\\
&&
\{b,d\}=\frac{bd}{4},
\quad \{b,f\}=\frac{bf}{4},\quad \{b,h\}=-\frac{bh}{4},\quad \{d,h\}=\frac{dh}{4},\quad \{f,h\}=\frac{fh}{4}.\nn
\eea
It is a straightforward computation to check that this is the Poisson sub-algebra of (\ref{PIV-lam-p}) given by the functions that do not depend on $c$ and $e$ and that
 there are two Casimirs, $G_1$ and $b d h$, so that the symplectic leaves are $4$-dimensional. The 
$PII^{FN}$ monodromy manifold (\ref{eq:gen-p2fn}) is the subspace of those functions of $a,b,d,f,h$ which commute with $h$. To check that $x_1,x_2$ and $x_3$ commute with $h$ it is enough to express them in terms of the lamination:
$$
x_1=-b f,\quad x_2= -G_1 \frac{d f}{a}-\frac{d^2}{a}-\frac{f^2}{a},\quad 
x_3= -\frac{d h}{f}-\frac{a b}{f}.
$$
Vice versa all functions commuting with $h$ must have the form $a^\alpha b^\beta d^\delta f^\phi$ where $\alpha,\beta,\gamma,\delta,\phi$ are some numbers satisfying $2\alpha-\beta+\delta+\phi=0$. Using this fact, it follows that on each symplectic leaf the set of functions which commute with $h$ is $2$-dimensional.

\subsection{Decorated character variety for $PI$}
The confluence from the generic $PII^{JM}$ cubic to the $PI$ one is realised by 
$$
s_3\to s_3-\log[\epsilon],
$$
in formulae (\ref{eq:shear-PII}). In  the limit $\epsilon\to0$ we obtain:
\begin{eqnarray}
\label{eq:shear-PI}
x_1&=&-e^{s_2+ s_3+\frac{p_2}{2}+\frac{p_3}{2}},\nn\\
x_2&=&-e^{s_3+s_1+\frac{p_3}{2}+\frac{p_1}{2}}-G_1e^{s_3+\frac{p_3}{2}},\nn\\
x_3&=&-e^{s_1+ s_2+\frac{p_1}{2}+\frac{p_2}{2}}-G_2 e^{s_1+\frac{p_1}{2}},\nn\\
\end{eqnarray}
where 
$$
G_i=e^{\frac{p_i}{2}},\, i=1,2,\quad G_3=0,\quad
G_\infty=e^{s_1+s_2+s_3+\frac{p_1}{2}+\frac{p_2}{2}+\frac{p_3}{2}}.
$$
These coordinates satisfy the following cubic relation:
\begin{equation}\label{eq:gen-p1}
x_1x_2x_3 -G_1 G_\infty x_1 -G_2 G_\infty  x_2 + G_\infty^2=0.
\end{equation}
Observe that we can obtain exactly the same formulate by starting from the generic $PII^{FN}$ cubic (\ref{eq:gen-p2fn}) in Flaschka-Newell form by the substitution
$$
p_1\to p_1-2\log[\epsilon],
$$
and by taking the limit as $\epsilon\to 0$.

Note that the parameters $p_3,p_2,p_1$ and $s_3$ are now redundant, we can eliminate them by rescaling. We pick $p_1=p_2=p_3=0$ and $s_3=-s_1-s_2$, we obtain the correct 
$PI$ by changing the sign of $x_1$ and $x_2$.

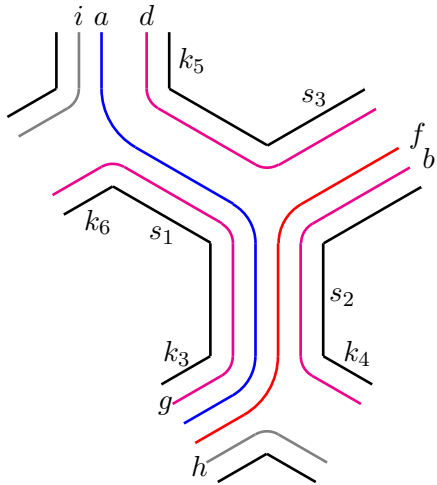
\begin{figure}[h]
{\psset{unit=0.5}
\begin{pspicture}(-6,-7)(2,6)
\newcommand{\TAIL}[3]{%
{\psset{unit=#1}
\pcline[linewidth=1pt,linecolor=black](-0.5,0)(-0.5,-1)
\pcline[linewidth=1pt,linecolor=black](0.5,0)(0.5,-1)
\pcline[linewidth=1pt,linecolor=magenta](0.3,0)(0.3,-1)
\pcline[linewidth=1pt,linecolor=#2](-0.1,0)(-0.1,-1)
\pcline[linewidth=1pt,linecolor=#3](0.1,0)(0.1,-1)
\pcline[linewidth=1pt,linecolor=magenta](-0.3,0)(-0.3,-1)
\psarc[linecolor=magenta, linewidth=1pt](-0.5,0){0.2}{0}{60}
\psarc[linecolor=#2, linewidth=1pt](-0.5,0){0.4}{0}{60}
\rput{120}(0,-1.29){\rput(0,0.29){
\pcline[linewidth=1pt,linecolor=black](-0.5,0)(-0.5,0.5)
\pcline[linewidth=1pt,linecolor=black](0.5,0)(0.5,0.5)
\pcline[linewidth=1pt,linecolor=magenta](0.3,0)(0.3,0.5)
\pcline[linewidth=1pt,linecolor=#2](0.1,0)(0.1,0.5)
\pcline[linewidth=1pt,linecolor=#3](-0.1,0)(-0.1,0.5)
\pcline[linewidth=1pt,linecolor=gray](-0.3,0)(-0.3,0.5)
\psarc[linecolor=magenta, linewidth=1pt](0.5,0){0.2}{180}{240}
\psarc[linecolor=#2, linewidth=1pt](0.5,0){0.4}{180}{240}
\psarc[linecolor=#3, linewidth=1pt](0.5,0){0.6}{180}{240}
}
}
\rput{240}(0,-1.29){\rput(0,0.29){
\pcline[linewidth=1pt,linecolor=black](-0.5,0)(-0.5,0.5)
\pcline[linewidth=1pt,linecolor=black](0.5,0)(0.5,0.5)
\pcline[linewidth=1pt,linecolor=gray](0.3,0)(0.3,0.5)
\pcline[linewidth=1pt,linecolor=magenta](-0.3,0)(-0.3,0.5)
\psarc[linecolor=magenta, linewidth=1pt](-0.5,0){0.2}{-60}{0}
\psarc[linecolor=gray, linewidth=1pt](0.5,0){0.2}{180}{240}
}
}
}
}
\rput(-5,5.2){$i$}
\rput(-4.4,5.1){$a$}
\rput(-3.2,5.2){$d$}
\rput(6.9,1.85){$e$}
\rput(-2.7,-5.2){$g$}
\rput(-1.8,-6.8){$h$}
\rput(-2.4,-3.8){$k_3$}
\rput(2.4,-3.8){$k_4$}
\rput(1.25,3){$s_3$}
\rput(2,-2.3){$s_2$}
\rput(-2.8,-0.65){$s_1$}
\rput(-2,4){$k_5$}
\rput(-4.5,-0.3){$k_6$}
\rput(0,0){\rput(0.12,-0.87){\TAIL{3}{blue}{red}}}
\rput{120}(0,0){\rput(0.12,-0.87){\TAIL{3}{red}{white}}}
\rput{240}(0,0){\rput(0.12,-0.87){\TAIL{3}{white}{blue}}}
\rput{30}(4,2){\psframe[linecolor=white, fillstyle=solid, fillcolor=white](-0.6,-3.2)(3,3.2)}
\rput(4,2){$f$}
\rput(4.3,1.4){$b$}
\end{pspicture}
}
\caption{The character variety of $PI$.}\label{f.PIfg}
\end{figure}

Geometrically speaking, sending the shear coordinate $s_3$ to infinity means that we are performing a cusp-removing move. In terms of the fat-graph, this is represented by Fig.  \ref{f.PIfg}.

The character variety is now $7$-dimensional, there is only one Casimir and the $PI$ monodromy manifold is given by the set of functions that Poisson commute with $p_1=k_5+k_6$, $p_2=k_3+k_4$. We omit all details as they are similar to the previous cases.

\subsection{Shear coordinates for the Weierstrass equation and Airy equation}

\subsubsection{Weierstrass equation}
We now remove one further cusp from the generic $PI$ cubic by replacing
$$
s_1\to s_1-\log[\epsilon],
$$
in formulae (\ref{eq:shear-PI}). In  the limit $\epsilon\to0$ we obtain:
\begin{eqnarray}
\label{eq:shear-PW}
x_1&=&-e^{s_2+ s_3+\frac{p_2}{2}+\frac{p_3}{2}},\nn\\
x_2&=&-e^{s_3+s_1+\frac{p_3}{2}+\frac{p_1}{2}},\nn\\
x_3&=&-e^{s_1+ s_2+\frac{p_1}{2}+\frac{p_2}{2}}-G_2 e^{s_1+\frac{p_1}{2}},\nn\\
\end{eqnarray}
where 
$$
G_2=e^{\frac{p_2}{2}},\quad G_1=G_3=0,\quad
G_\infty=e^{s_1+s_2+s_3+\frac{p_1}{2}+\frac{p_2}{2}+\frac{p_3}{2}}.
$$
These coordinates satisfy the following cubic relation:
\begin{equation}\label{eq:gen-w}
x_1x_2x_3  -G_2 G_\infty  x_2 + G_\infty^2=0.
\end{equation}

Note that the parameters $p_3,p_2,p_1$ and $s_3$ are now redundant, we can choose $p_1=p_2=p_3=0$ and $s_3=-s_1-s_2$ so that $G_2=G_\infty=1$. This gives us the following cubic
\begin{equation}\label{eq:-w}
x_1 x_2 x_3  -  x_2 +1=0.
\end{equation}
In order to relate this cubic to the Weiestrass elliptic curve, we need to projectivise it first:
\begin{equation}\label{eq:-wp}
x_1 x_2 x_3  -  x_2 x_0^2 +x_0^3=0.
\end{equation}
This cubic is now invariant under the following transformation
$$
x_0\to \alpha x_0,\quad x_1\to \beta x_1,\quad x_2\to \alpha x_2,\quad x_3\to \frac{\alpha^2}{\beta} x_3,
$$
so that we can rescale $x_2\to 1$ and $x_3\to x_1$, leading to the Weiestrass elliptic curve:
$$
x_1^2  - x_0^2 +x_0^3=0.
$$
The character variety in this case is a disk with $4$ cusps. A lamination can be produced along the same lines as before.

\subsubsection{Airy equation}
If we remove one further cusp we obtain the character variety of a  disk with $3$ cusps. This  corresponds to the famous Airy equation which the simplest linear ODE for which the Stokes phenomenon appears:
$$
\frac{{\rm d}^2 \psi}{{\rm d}z^2} + z\psi =0, \quad\hbox{or in system form:}\quad
   \left( \frac{{\rm d}}{{\rm d}z}+\begin{pmatrix} 
      0 & -1 \\
      z & 0\\
   \end{pmatrix}
\right)  \begin{pmatrix} 
        \psi\\
        \psi'\\
    \end{pmatrix} =0.
  $$

\section{Painlev\'e cluster algebras: braid-group and affine MCG actions}\label{se:P-cluster}

\subsection{Painlev\'e VI: analytic continuation and cluster mutations}\label{se:clusterPVI}

In \cite{DM,M1} it was proved that the procedure of analytic continuation of a local solution to the sixth Painlev\'e equation corresponds to the following action of the braid group on the monodromy manifold:
\be\label{eq:braid1}
\beta_1:\begin{array}{lcl}
x_1&\to&-x_1-x_2x_3+\omega_1,\\
x_2&\to&x_3,\\
x_3&\to&x_2,\\
\end{array}\ee
\be\label{eq:braid2}
\beta_2:\begin{array}{lcl}
x_1&\to&x_3,\\
x_2&\to&-x_2-x_1x_3+\omega_2,\\
x_3&\to&x_1,\\
\end{array}\ee
\be\label{eq:braid3}
\beta_3:\begin{array}{lcl}
x_1&\to&x_2,\\
x_2&\to&x_1,\\
x_3&\to&-x_3-x_1 x_2 +\omega_3.\\
\end{array}\ee
In \cite{ChM} it was shown that flips on the shear coordinates correspond to the action of the braid group on the cubic. The flips $f_1,f_2,f_3$ of the shear coordinates which give rise to the braid transformations $\beta_1\beta_2$ and $\beta_3$ respectively  have the following form
\be\label{eq:flip1}
f_1:\begin{array}{lcl}
s_1&\to&-p_1-s_1,\qquad p_2\to p_3,\qquad p_3\to p_2,\\
s_2&\to&s_3+\log\left[1+ e^{s_1}\right]+\log\left[1+e^{s_1+p_1}\right],\\
s_3&\to&s_2-\log\left[1+ e^{-s_1}\right]+\log\left[1+e^{-s_1-p_1}\right],\\
\end{array}\ee 
\be\label{eq:flip2}
f_2:\begin{array}{lcl}
s_1&\to&s_3-\log\left[1+ e^{-s_2}\right]-\log\left[1+e^{-s_2-p_2}\right],\\
s_2&\to&-p_2-s_2,\qquad p_1\to p_3,\qquad p_3\to p_1,\\
s_3&\to&s_1+\log\left[1+ e^{s_2}\right]+\log\left[1+e^{s_2+p_2}\right],\\
\end{array}\ee
\be\label{eq:flip3}
f_3:\begin{array}{lcl}
s_1&\to&s_2+\log\left[1+ e^{s_3}\right]+\log\left[1+e^{s_3+p_3}\right],\\
s_2&\to&s_1-\log\left[1+ e^{-s_3}\right]-\log\left[1+e^{-s_3-p_3}\right],\\
s_3&\to&-p_3-s_3\qquad p_1\to p_2,\qquad p_2\to p_1.\\
\end{array}\ee

\begin{remark}
Observe that in \cite{ChS} it was proved that shear coordinate flips (\ref{eq:flip1}), (\ref{eq:flip2}),  (\ref{eq:flip3}) are indeed {\it dual}\/ to the generalized cluster mutations (\ref{eq:tagcluster}) for the corresponding $\lambda$-lengths. 
\end{remark}

We are now going to show that when $G_\infty=2$ (geometrically this means that we have a puncture at infinity), the action of the braid group coincides with a {\it generalized cluster algebra structure} \cite{ChS}.

In order to see this let us compose each braid with a Okamoto symmetry in order to obtain the following
\be\label{eq:braid-generic}
\widetilde \beta_i:\begin{array}{lcl}
x_i&\to&-x_i-x_j x_k +\omega_i,\quad j,k\neq i,\\
x_j&\to&x_j,\quad\hbox{for}\quad j\neq i\\
\end{array}\ee
By using (\ref{eq:mon-mf}) this transformation acquires a cluster flavour:
\be\label{eq:braid-cluster}
\widetilde \beta_i: \, x_ix_i'=x_j^2+x_k^2+\omega_j x_j+\omega_k x_k +\omega_4\quad j,k\neq i.\\
\ee
Indeed let us introduce the shifted variables:
$$
y_i:=x_i-G_i,\quad i=1,2,3,
$$
they satisfy the {\it generalized cluster algebra relation:}
\be\label{eq:tagcluster}
\mu_i: \, y_iy_i'=y_j^2+y_k^2+G_i y_j y_k\quad j,k\neq i.\\
\ee
Note that generalized cluster algebras satisfy the Laurent phenomenon. In particular this result implies that procedure of analytic continuation of the solutions to the sixth Painlev\'e equation satisfies the Laurent phenomenon: if we start from a local solution corresponding to the point $(y_1^0,y_2^0,y_3^0)$ on the shifted Painlev\'e cubic
$$
y_1 y_2 y_3 + y_1^2+y_2^2 + y_3^2 + G_1 y_2 y_3+G_2 y_1 y_3+G_3 y_1 y_2=0
$$
any other branch of that solution will corresponds to points   $(y_1,y_2,y_3)$ on the same cubic such that each $y_i$ is a Laurent polynomial of the initial  $(y_1^0,y_2^0,y_3^0)$.

\vskip 2mm

\subsection{Generalized cluster algebra structure for $PV$ and $PV_{\text{deg}}$}\label{suse:tag1}

In this case, we have a Riemann surface $\Sigma_{0,3,2}$ with two bordered cusps on one hole. The only nontrivial Dehn twist is around the closed
geodesic $\gamma$ encircling these two holes (this geodesic is unique) { -- see the left side of Fig.18 here below.   }

We now consider the effect of this MCG transformation on the system of arcs in Fig.~\ref{fi:PV}. 
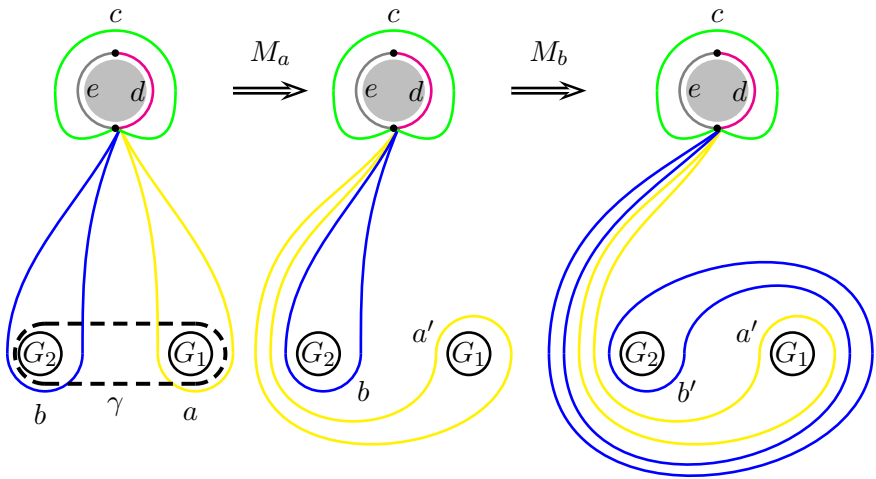
\begin{figure}[h]\label{fig:dehnPV}
$$
\begin{pspicture}(-1.7,-3.5)(1.7,2.4)
\newcommand{\PAT}[1]{%
{\psset{unit=#1}
\pscircle[linecolor=white, linewidth=1pt,fillstyle=solid,fillcolor=lightgray](0,1.5){0.45}
\psarc[linecolor=gray, linewidth=1pt](0,1.5){0.5}{90}{270}
\psarc[linecolor=magenta, linewidth=1pt](0,1.5){0.5}{-90}{90}
\psarc[linecolor=green, linewidth=1pt](0,1.5){0.8}{0}{180}
\psbezier[linewidth=1pt,linecolor=green](-0.8,1.5)(-0.8,0.9)(-0.6,0.7)(0,1)
\psbezier[linewidth=1pt,linecolor=green](0.8,1.5)(0.8,0.9)(0.6,0.7)(0,1)
\pscircle[linecolor=black, linewidth=1pt](-1,-2){0.3}
\pscircle[linecolor=black, linewidth=1pt](1,-2){0.3}
\pscircle*(0,1){0.05}
\pscircle*(0,2){0.05}
\rput(-0.3,1.5){$e$}
\rput(0.3,1.5){$d$}
\rput(0,2.5){$c$}
\rput(-1,-2.8){$b$}
\rput(1,-2.8){$a$}
\rput(0,-2.7){$\gamma$}
\rput(-1,-2){$ G_2$}
\rput(1,-2){$ G_1$}
}
}
\psbezier[linewidth=1pt,linecolor=yellow](0,1)(0.5,-0.2)(0.5,-1)(.5,-2)
\psbezier[linewidth=1pt,linecolor=yellow](0,1)(0.5,0)(1.5,-0.8)(1.5,-2)
\psarc[linecolor=yellow, linewidth=1pt](1,-2){0.5}{180}{360}
\psbezier[linewidth=1pt,linecolor=blue](0,1)(-0.5,-0.2)(-0.5,-1)(-.5,-2)
\psbezier[linewidth=1pt,linecolor=blue](0,1)(-0.5,0)(-1.5,-0.8)(-1.5,-2)
\psarc[linecolor=blue, linewidth=1pt](-1,-2){0.5}{180}{360}
\psarc[linecolor=black, linestyle=dashed, linewidth=1.5pt](-1,-2){0.4}{90}{270}
\psarc[linecolor=black, linestyle=dashed, linewidth=1.5pt](1,-2){0.4}{-90}{90}
\pcline[linewidth=1.5pt, linestyle=dashed, linecolor=black](-1,-1.6)(1,-1.6)
\pcline[linewidth=1.5pt, linestyle=dashed, linecolor=black](-1,-2.4)(1,-2.4)
\rput(0,0){\PAT{1}}
\rput(2,1.5){\pcline[linewidth=3pt, linecolor=black]{->}(-0.5,0)(0.5,0)}
\rput(2,1.5){\pcline[linewidth=1pt, linecolor=white]{->}(-0.5,0)(0.4,0)}
\rput(2,2){$M_a$}
\end{pspicture}
\begin{pspicture}(-2,-3.5)(2,2.4)
\newcommand{\PAT}[1]{%
{\psset{unit=#1}
\pscircle[linecolor=white, linewidth=1pt,fillstyle=solid,fillcolor=lightgray](0,1.5){0.45}
\psarc[linecolor=gray, linewidth=1pt](0,1.5){0.5}{90}{270}
\psarc[linecolor=magenta, linewidth=1pt](0,1.5){0.5}{-90}{90}
\psarc[linecolor=green, linewidth=1pt](0,1.5){0.8}{0}{180}
\psbezier[linewidth=1pt,linecolor=green](-0.8,1.5)(-0.8,0.9)(-0.6,0.7)(0,1)
\psbezier[linewidth=1pt,linecolor=green](0.8,1.5)(0.8,0.9)(0.6,0.7)(0,1)
\pscircle[linecolor=black, linewidth=1pt](-1,-2){0.3}
\pscircle[linecolor=black, linewidth=1pt](1,-2){0.3}
\pscircle*(0,1){0.05}
\pscircle*(0,2){0.05}
\rput(-0.3,1.5){$e$}
\rput(0.3,1.5){$d$}
\rput(0,2.5){$c$}
\rput(-0.4,-2.5){$b$}
\rput(0.4,-1.7){$a'$}
\rput(-1,-2){$ G_2$}
\rput(1,-2){$ G_1$}
}
}
\psbezier[linewidth=1pt,linecolor=yellow](0,1)(-0.7,-0.2)(-1.7,-0.4)(-1.7,-2)
\psbezier[linewidth=1pt,linecolor=yellow](0,1)(-0.7,0)(-1.9,-0.2)(-1.9,-2)
\psbezier[linewidth=1pt,linecolor=yellow](1.5,-2)(1.5,-3.4)(-1.9,-3.8)(-1.9,-2)
\psbezier[linewidth=1pt,linecolor=yellow](0.5,-2)(0.5,-3.2)(-1.7,-3.2)(-1.7,-2)
\psarc[linecolor=yellow, linewidth=1pt](1,-2){0.5}{0}{180}
\psbezier[linewidth=1pt,linecolor=blue](0,1)(-0.5,-0.2)(-0.5,-1)(-.5,-2)
\psbezier[linewidth=1pt,linecolor=blue](0,1)(-0.5,0)(-1.5,-0.8)(-1.5,-2)
\psarc[linecolor=blue, linewidth=1pt](-1,-2){0.5}{180}{360}
\rput(0,0){\PAT{1}}
\rput(2,1.5){\pcline[linewidth=3pt, linecolor=black]{->}(-0.5,0)(0.5,0)}
\rput(2,1.5){\pcline[linewidth=1pt, linecolor=white]{->}(-0.5,0)(0.4,0)}
\rput(2,2){$M_b$}
\end{pspicture}
\begin{pspicture}(-2.3,-3.5)(2.3,2.4)
\newcommand{\PAT}[1]{%
{\psset{unit=#1}
\pscircle[linecolor=white, linewidth=1pt,fillstyle=solid,fillcolor=lightgray](0,1.5){0.45}
\psarc[linecolor=gray, linewidth=1pt](0,1.5){0.5}{90}{270}
\psarc[linecolor=magenta, linewidth=1pt](0,1.5){0.5}{-90}{90}
\psarc[linecolor=green, linewidth=1pt](0,1.5){0.8}{0}{180}
\psbezier[linewidth=1pt,linecolor=green](-0.8,1.5)(-0.8,0.9)(-0.6,0.7)(0,1)
\psbezier[linewidth=1pt,linecolor=green](0.8,1.5)(0.8,0.9)(0.6,0.7)(0,1)
\pscircle[linecolor=black, linewidth=1pt](-1,-2){0.3}
\pscircle[linecolor=black, linewidth=1pt](1,-2){0.3}
\pscircle*(0,1){0.05}
\pscircle*(0,2){0.05}
\rput(-0.3,1.5){$e$}
\rput(0.3,1.5){$d$}
\rput(-0.4,-2.5){$b'$}
\rput(0.4,-1.7){$a'$}
\rput(0,2.5){$c$}
\rput(-1,-2){$ G_2$}
\rput(1,-2){$ G_1$}
}
}
\psbezier[linewidth=1pt,linecolor=yellow](0,1)(-0.7,-0.2)(-1.7,-0.4)(-1.7,-2)
\psbezier[linewidth=1pt,linecolor=yellow](0,1)(-0.7,0)(-1.9,-0.2)(-1.9,-2)
\psbezier[linewidth=1pt,linecolor=yellow](1.5,-2)(1.5,-3.4)(-1.9,-3.8)(-1.9,-2)
\psbezier[linewidth=1pt,linecolor=yellow](0.5,-2)(0.5,-3.2)(-1.7,-3.2)(-1.7,-2)
\psarc[linecolor=yellow, linewidth=1pt](1,-2){0.5}{0}{180}
\psbezier[linewidth=1pt,linecolor=blue](0,1)(-0.8,0.2)(-2.1,-0.4)(-2.1,-2)
\psbezier[linewidth=1pt,linecolor=blue](0,1)(-0.8,0.4)(-2.3,-0.2)(-2.3,-2)
\psbezier[linewidth=1pt,linecolor=blue](2,-2)(2,-3.8)(-2.3,-4.5)(-2.3,-2)
\psbezier[linewidth=1pt,linecolor=blue](1.7,-2)(1.7,-3.6)(-2.1,-4.3)(-2.1,-2)
\psbezier[linewidth=1pt,linecolor=blue](2,-2)(2,0)(-1.5,-0.8)(-1.5,-2)
\psbezier[linewidth=1pt,linecolor=blue](1.7,-2)(1.7,-0.6)(-0.5,-1)(-0.5,-2)
\psarc[linecolor=blue, linewidth=1pt](-1,-2){0.5}{180}{360}
\rput(0,0){\PAT{1}}
\end{pspicture}
$$ 
\caption{{The Dehn twist corresponding to $\gamma$ maps the arcs in the left-hand side of this picture to those on the right-hand side. It is obtained by composing two generalized mutations $M_a$ and $M_b$.}}
\end{figure}

The generalized mutations $M_a$ and $M_b$ are given by the formulas
$$
a'a=b^2+c^2+ G_1 bc;\qquad b'b=(a')^2+c^2+ G_2 a'c,
$$
or, explicitly,
\beq
\left[
\begin{array}{l}
  a   \\
    b 
\end{array}
\right] \to
\left[
\begin{array}{c}
  \dfrac{b^2+c^2+ G_1 bc}{a}   \\
    \dfrac{(b^2+c^2+ G_1bc)^2}{a^2b}+ G_2 c \dfrac{b^2+c^2+ G_1 bc}{ba}+\dfrac{c^2}{b}
\end{array}
\right].
\label{MCG1}
\eeq
The geodesic function of $\gamma$ is
\beq
G_\gamma =  G_2 \frac{c}{b}+ G_1 \frac{c}{a}+\frac{a}{b}+\frac{b}{a}+\frac{c^2}{ab} 
\eeq	
and this function is the so-called {\it Hamiltonian MCG invariant:}\/ { it is the only MCG-invariant that generates the corresponding Dehn twist}
(see \cite{Kashaev-Dehn}), has nontrivial Poisson brackets with $a$ and $b$, and is preserved by the MCG action (\ref{MCG1}).

In the case of $PV_{\text{deg}}$, all the above formulas remain valid provided we replace $c$ by the $\lambda$-length $d$ of the boundary arc.

\subsection{Generalized cluster algebra structure for $PIII^{D_6}$, $PIII^{D_7}$, and $PIII^{D_8}$}\label{suse:tag1-1}

In all cases of $PIII^{D_6}$, we have a Riemann surface $\Sigma_{0,2,n}$ with $n_1>0$ and $n_2>0$, $n_1+n_2=n$, bordered cusps on the respective holes. For any $n_1$ and $n_2$, the only nontrivial Dehn twist is around the closed
geodesic $\gamma$ separating the holes (this geodesic is unique). Its geodesic function $G_\gamma$ is the Hamiltonian MCG invariant. Besides 
this invariant, we have (non-Hamiltonian) invariants, which are $\lambda$-lengths of all arcs starting and terminating at the same boundary component.

We begin with the case of $PIII^{D_6}$ and consider the MCG action on the system of arcs in Fig.~\ref{fi:PIII}:
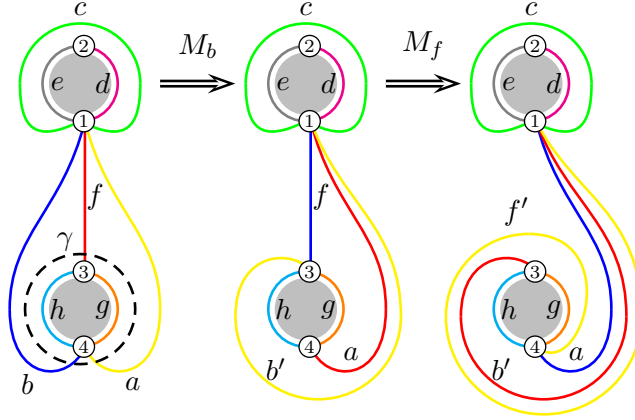
\begin{figure}[h]
$$
\begin{pspicture}(-1.5,-3)(1.5,2)
\newcommand{\PAT}[1]{%
{\psset{unit=#1}
\pscircle[linecolor=white, linewidth=1pt,fillstyle=solid,fillcolor=lightgray](0,1.5){0.45}
\psarc[linecolor=gray, linewidth=1pt](0,1.5){0.5}{90}{270}
\psarc[linecolor=magenta, linewidth=1pt](0,1.5){0.5}{-90}{90}
\psarc[linecolor=green, linewidth=1pt](0,1.5){0.8}{0}{180}
\psbezier[linewidth=1pt,linecolor=green](-0.8,1.5)(-0.8,0.9)(-0.6,0.7)(0,1)
\psbezier[linewidth=1pt,linecolor=green](0.8,1.5)(0.8,0.9)(0.6,0.7)(0,1)
\pscircle[linecolor=white, linewidth=1pt,fillstyle=solid,fillcolor=lightgray](0,-1.5){0.45}
\psarc[linecolor=cyan, linewidth=1pt](0,-1.5){0.5}{90}{270}
\psarc[linecolor=orange, linewidth=1pt](0,-1.5){0.5}{-90}{90}
\pscircle[linecolor=black, linewidth=1pt, linestyle=dashed](0,-1.5){0.75}
\pscircle[linecolor=black, linewidth=0.5pt,fillstyle=solid,fillcolor=white](0.05,1){0.15}
\rput(0.05,1){\tiny $1$}
\pscircle[linecolor=black, linewidth=0.5pt,fillstyle=solid,fillcolor=white](0.05,2){0.15}
\rput(0.05,2){\tiny $2$}
\pscircle[linecolor=black, linewidth=0.5pt,fillstyle=solid,fillcolor=white](0.05,-1){0.15}
\rput(0.05,-1){\tiny $3$}
\pscircle[linecolor=black, linewidth=0.5pt,fillstyle=solid,fillcolor=white](0.05,-2){0.15}
\rput(0.05,-2){\tiny $4$}
\rput(0,2.5){$c$}
\rput(-0.7,-2.5){$b$}
\rput(0.7,-2.5){$a$}
\rput(-0.2,-0.6){$\gamma$}
\rput(-0.3,-1.5){$h$}
\rput(0.3,-1.5){$g$}
\rput(-0.3,1.5){$e$}
\rput(0.3,1.5){$d$}
\rput(0.2,0){$f$}
}
}
\pcline[linewidth=1pt, linecolor=red](0,1)(0,-1)
\psbezier[linewidth=1pt,linecolor=yellow](0,1)(0.3,-0.2)(1,-0.5)(1,-1.5)
\psbezier[linewidth=1pt,linecolor=yellow](0,-2)(0.2,-2.5)(1,-2.5)(1,-1.5)
\psbezier[linewidth=1pt,linecolor=blue](0,1)(-0.3,-0.2)(-1,-0.5)(-1,-1.5)
\psbezier[linewidth=1pt,linecolor=blue](0,-2)(-0.2,-2.5)(-1,-2.5)(-1,-1.5)
\rput(0,0){\PAT{1}}
\rput(1.5,1.5){\pcline[linewidth=3pt, linecolor=black]{->}(-0.5,0)(0.5,0)}
\rput(1.5,1.5){\pcline[linewidth=1pt, linecolor=white]{->}(-0.5,0)(0.4,0)}
\rput(1.5,2){$M_b$}
\end{pspicture}
\begin{pspicture}(-1.5,-3)(1.5,2)
\newcommand{\PAT}[1]{%
{\psset{unit=#1}
\pscircle[linecolor=white, linewidth=1pt,fillstyle=solid,fillcolor=lightgray](0,1.5){0.45}
\psarc[linecolor=gray, linewidth=1pt](0,1.5){0.5}{90}{270}
\psarc[linecolor=magenta, linewidth=1pt](0,1.5){0.5}{-90}{90}
\psarc[linecolor=green, linewidth=1pt](0,1.5){0.8}{0}{180}
\psbezier[linewidth=1pt,linecolor=green](-0.8,1.5)(-0.8,0.9)(-0.6,0.7)(0,1)
\psbezier[linewidth=1pt,linecolor=green](0.8,1.5)(0.8,0.9)(0.6,0.7)(0,1)
\pscircle[linecolor=white, linewidth=1pt,fillstyle=solid,fillcolor=lightgray](0,-1.5){0.45}
\psarc[linecolor=cyan, linewidth=1pt](0,-1.5){0.5}{90}{270}
\psarc[linecolor=orange, linewidth=1pt](0,-1.5){0.5}{-90}{90}
\pscircle[linecolor=black, linewidth=0.5pt,fillstyle=solid,fillcolor=white](0.05,1){0.15}
\rput(0.05,1){\tiny $1$}
\pscircle[linecolor=black, linewidth=0.5pt,fillstyle=solid,fillcolor=white](0.05,2){0.15}
\rput(0.05,2){\tiny $2$}
\pscircle[linecolor=black, linewidth=0.5pt,fillstyle=solid,fillcolor=white](0.05,-1){0.15}
\rput(0.05,-1){\tiny $3$}
\pscircle[linecolor=black, linewidth=0.5pt,fillstyle=solid,fillcolor=white](0.05,-2){0.15}
\rput(0.05,-2){\tiny $4$}
\rput(0,2.5){$c$}
\rput(-0.4,-2.3){$b'$}
\rput(0.6,-2.1){$a$}
\rput(-0.3,-1.5){$h$}
\rput(0.3,-1.5){$g$}
\rput(-0.3,1.5){$e$}
\rput(0.3,1.5){$d$}
\rput(0.2,0){$f$}
}
}
\pcline[linewidth=1pt, linecolor=blue](0,1)(0,-1)
\psbezier[linewidth=1pt,linecolor=red](0,1)(0.3,-0.2)(1,-0.5)(1,-1.5)
\psbezier[linewidth=1pt,linecolor=red](0,-2)(0.2,-2.5)(1,-2.5)(1,-1.5)
\psbezier[linewidth=1pt,linecolor=yellow](0,1)(0.4,-0.1)(1.2,-0.4)(1.2,-1.6)
\psarc[linecolor=yellow, linewidth=1pt](0.1,-1.6){1.1}{180}{360}
\psbezier[linewidth=1pt,linecolor=yellow](0,-1)(-0.2,-0.7)(-1,-0.7)(-1,-1.6)
\rput(0,0){\PAT{1}}
\rput(1.5,1.5){\pcline[linewidth=3pt, linecolor=black]{->}(-0.5,0)(0.5,0)}
\rput(1.5,1.5){\pcline[linewidth=1pt, linecolor=white]{->}(-0.5,0)(0.4,0)}
\rput(1.5,2){$M_f$}
\end{pspicture}
\begin{pspicture}(-1.5,-3)(1.5,2)
\newcommand{\PAT}[1]{%
{\psset{unit=#1}
\pscircle[linecolor=white, linewidth=1pt,fillstyle=solid,fillcolor=lightgray](0,1.5){0.45}
\psarc[linecolor=gray, linewidth=1pt](0,1.5){0.5}{90}{270}
\psarc[linecolor=magenta, linewidth=1pt](0,1.5){0.5}{-90}{90}
\psarc[linecolor=green, linewidth=1pt](0,1.5){0.8}{0}{180}
\psbezier[linewidth=1pt,linecolor=green](-0.8,1.5)(-0.8,0.9)(-0.6,0.7)(0,1)
\psbezier[linewidth=1pt,linecolor=green](0.8,1.5)(0.8,0.9)(0.6,0.7)(0,1)
\pscircle[linecolor=white, linewidth=1pt,fillstyle=solid,fillcolor=lightgray](0,-1.5){0.45}
\psarc[linecolor=cyan, linewidth=1pt](0,-1.5){0.5}{90}{270}
\psarc[linecolor=orange, linewidth=1pt](0,-1.5){0.5}{-90}{90}
\pscircle[linecolor=black, linewidth=0.5pt,fillstyle=solid,fillcolor=white](0.05,1){0.15}
\rput(0.05,1){\tiny $1$}
\pscircle[linecolor=black, linewidth=0.5pt,fillstyle=solid,fillcolor=white](0.05,2){0.15}
\rput(0.05,2){\tiny $2$}
\pscircle[linecolor=black, linewidth=0.5pt,fillstyle=solid,fillcolor=white](0.05,-1){0.15}
\rput(0.05,-1){\tiny $3$}
\pscircle[linecolor=black, linewidth=0.5pt,fillstyle=solid,fillcolor=white](0.05,-2){0.15}
\rput(0.05,-2){\tiny $4$}
\rput(0,2.5){$c$}
\rput(-0.4,-2.3){$b'$}
\rput(0.6,-2.1){$a$}
\rput(-0.3,-1.5){$h$}
\rput(0.3,-1.5){$g$}
\rput(-0.3,1.5){$e$}
\rput(0.3,1.5){$d$}
\rput(-0.2,-0.2){$f'$}
}
}
\psbezier[linewidth=1pt,linecolor=blue](0,1)(0.3,-0.2)(1,-0.5)(1,-1.5)
\psbezier[linewidth=1pt,linecolor=blue](0,-2)(0.2,-2.5)(1,-2.5)(1,-1.5)
\psbezier[linewidth=1pt,linecolor=red](0,1)(0.4,-0.1)(1.2,-0.4)(1.2,-1.6)
\psarc[linecolor=red, linewidth=1pt](0.1,-1.6){1.1}{180}{360}
\psbezier[linewidth=1pt,linecolor=red](0,-1)(-0.2,-0.7)(-1,-0.7)(-1,-1.6)
\psbezier[linewidth=1pt,linecolor=yellow](0,1)(0.5,0)(1.4,-0.2)(1.4,-1.6)
\psarc[linecolor=yellow, linewidth=1pt](0.1,-1.6){1.3}{180}{360}
\psarc[linecolor=yellow, linewidth=1pt](-0.1,-1.6){1.1}{90}{180}
\psbezier[linewidth=1pt,linecolor=yellow](0,-2)(0.6,-2.5)(1.2,-0.5)(-0.1,-0.5)
\rput(0,0){\PAT{1}}
\end{pspicture}
$$
\caption{{The Dehn twist corresponding to $\gamma$ maps the arcs in the left-hand side of this picture to those on the right-hand side. It is obtained by composing two generalized mutations $M_b$ and $M_f$.}}
\end{figure}

These transformations are governed by the standard mutation rules,
$$
bb'=hc+fa,\qquad ff'=gc+ab';
$$
in order to describe them in a more regular way, let us introduce the notation: we let $\lambda_{\alpha,\beta}^{(i)}$ denote the $\lambda$-length of the arc that goes between bordered cusps $\alpha$ and $\beta$ (belonging to different boundary components) winding $i$ times around the lower hole. For example,
$$
f=\lambda_{1,3}^{(0)},\quad b=\lambda_{1,4}^{(0)}, \quad a=\lambda_{1,4}^{(1)}, \quad b'=\lambda_{1,3}^{(1)},\quad f'=\lambda_{1,4}^{(2)},\quad c=\lambda_{1,1}\ \hbox{etc.}
$$
Note that $\lambda_{\alpha,\beta}$ with the labels $\alpha$ and $\beta$ pertaining to the same boundary component are unique and invariant under the MCG action.

The net result of the Dehn twist on the triple $\{\lambda_{1,4}^{(i-1)},\lambda_{1,3}^{(i-1)},\lambda_{1,4}^{(i)}\}$ reads:
\beq
\left[
\begin{array}{l}
    \lambda_{1,4}^{(i-1)} \\
    \lambda_{1,3}^{(i-1)} \\
    \lambda_{1,4}^{(i)}
\end{array}
\right] \to
\left[
\begin{array}{c}
 \lambda_{1,4}^{(i)}\\
  \dfrac{  \lambda_{1,3}^{(i-1)}\lambda_{1,4}^{(i)}+h \lambda_{1,1}}{ \lambda_{1,4}^{(i-1)}}   \\
    \dfrac{( \lambda_{1,4}^{(i)})^2}{ \lambda_{1,4}^{(i-1)}}+ \dfrac{ h \lambda_{1,1} \lambda_{1,4}^{(i)}}{ \lambda_{1,4}^{(i-1)} \lambda_{1,3}^{(i-1)}}
   +\dfrac{g  \lambda_{1,1}}{ \lambda_{1,3}^{(i-1)}}
\end{array}
\right].
\label{MCG2}
\eeq
This action admits two invariants: $G_\gamma$ and $\lambda_{4,4}$ (the latter is obtained by the mutation of the element $f$, or $\lambda_{1,3}^{(i)}$):
\bea
G_\gamma&=& \frac{ \lambda_{1,4}^{(i)}}{ \lambda_{1,4}^{(i-1)}}+\frac{ \lambda_{1,4}^{(i-1)}}{ \lambda_{1,4}^{(i)}}
+\frac{h  \lambda_{1,1}}{ \lambda_{1,3}^{(i-1)} \lambda_{1,4}^{(i-1)}}+\frac{g  \lambda_{1,1}}{ \lambda_{1,3}^{(i-1)} \lambda_{1,4}^{(i)}}
\label{inv-1}\\
\lambda_{4,4}&=&\frac{g \lambda_{1,4}^{(i-1)}}{\lambda_{1,3}^{(i-1)}}+\frac{h \lambda_{1,4}^{(i)}}{\lambda_{1,3}^{(i-1)}}\label{inv-2}
\eea

The case of $PIII^{D_7}$ coincides with that of $PIII^{D_6}$, the geodesic $\lambda_{1,1}$ now becomes the boundary geodesic after erasing the bordered cusp $2$. 

{In the case of $PIII^{D_8}$ we erase bordered cusps $2$ and $3$; the only MCG transformation is:
\beq
\left[
\begin{array}{l}
    \lambda_{1,4}^{(i-1)} \\
    \lambda_{1,4}^{(i)}
\end{array}
\right] \to
\left[
\begin{array}{c}
 \lambda_{1,4}^{(i)}\\
  \dfrac{  (\lambda_{1,4}^{(i)})^2+\lambda_{1,1}\lambda_{4,4}}{ \lambda_{1,4}^{(i-1)}} 
 \end{array}
\right].
\label{MCG3}
\eeq
and the Hamiltonian MCG--invariant is
\beq
G_\gamma=\frac{ \lambda_{1,4}^{(i)}}{ \lambda_{1,4}^{(i-1)}}+\frac{ \lambda_{1,4}^{(i-1)}}{ \lambda_{1,4}^{(i)}}
+\frac{ \lambda_{1,1}\lambda_{4,4}}{ \lambda_{1,4}^{(i-1)} \lambda_{1,4}^{(i)}}.
\label{inv-3}
\eeq}

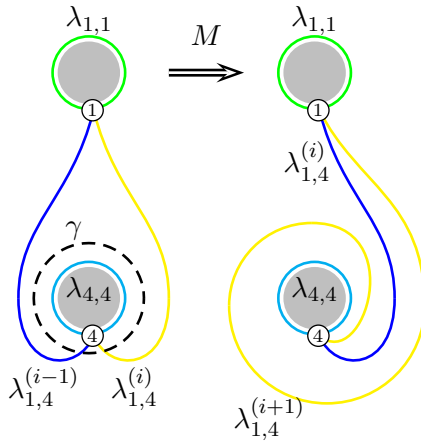
\begin{figure}[h]
$$
\begin{pspicture}(-1.5,-3.5)(1.5,2)
\newcommand{\PAT}[1]{%
{\psset{unit=#1}
\pscircle[linecolor=white, linewidth=1pt,fillstyle=solid,fillcolor=lightgray](0,1.5){0.45}
\pscircle[linecolor=green, linewidth=1pt](0,1.5){0.5}
\pscircle[linecolor=white, linewidth=1pt,fillstyle=solid,fillcolor=lightgray](0,-1.5){0.45}
\pscircle[linecolor=cyan, linewidth=1pt](0,-1.5){0.5}
\pscircle[linecolor=black, linewidth=1pt, linestyle=dashed](0,-1.5){0.75}
\pscircle[linecolor=black, linewidth=0.5pt,fillstyle=solid,fillcolor=white](0.05,1){0.15}
\rput(0.05,1){\tiny $1$}
\pscircle[linecolor=black, linewidth=0.5pt,fillstyle=solid,fillcolor=white](0.05,-2){0.15}
\rput(0.05,-2){\tiny $4$}
\rput(0,2.2){$\lambda_{1,1}$}
\rput(0,-1.4){$\lambda_{4,4}$}
\rput(-0.6,-2.7){$\lambda_{1,4}^{(i-1)}$}
\rput(0.6,-2.7){$\lambda_{1,4}^{(i)}$}
\rput(-0.2,-0.6){$\gamma$}
}
}
\psbezier[linewidth=1pt,linecolor=yellow](0,1)(0.3,-0.2)(1,-0.5)(1,-1.5)
\psbezier[linewidth=1pt,linecolor=yellow](0,-2)(0.2,-2.5)(1,-2.5)(1,-1.5)
\psbezier[linewidth=1pt,linecolor=blue](0,1)(-0.3,-0.2)(-1,-0.5)(-1,-1.5)
\psbezier[linewidth=1pt,linecolor=blue](0,-2)(-0.2,-2.5)(-1,-2.5)(-1,-1.5)
\rput(0,0){\PAT{1}}
\rput(1.5,1.5){\pcline[linewidth=3pt, linecolor=black]{->}(-0.5,0)(0.5,0)}
\rput(1.5,1.5){\pcline[linewidth=1pt, linecolor=white]{->}(-0.5,0)(0.4,0)}
\rput(1.5,2){$M$}
\end{pspicture}
\begin{pspicture}(-1.5,-3.5)(1.5,2)
\newcommand{\PAT}[1]{%
{\psset{unit=#1}
\pscircle[linecolor=white, linewidth=1pt,fillstyle=solid,fillcolor=lightgray](0,1.5){0.45}
\pscircle[linecolor=green, linewidth=1pt](0,1.5){0.5}
\pscircle[linecolor=white, linewidth=1pt,fillstyle=solid,fillcolor=lightgray](0,-1.5){0.45}
\pscircle[linecolor=cyan, linewidth=1pt](0,-1.5){0.5}
\pscircle[linecolor=black, linewidth=0.5pt,fillstyle=solid,fillcolor=white](0.05,1){0.15}
\rput(0.05,1){\tiny $1$}
\pscircle[linecolor=black, linewidth=0.5pt,fillstyle=solid,fillcolor=white](0.05,-2){0.15}
\rput(0.05,-2){\tiny $4$}
\rput(0,2.2){$\lambda_{1,1}$}
\rput(0,-1.4){$\lambda_{4,4}$}
\rput(-0.6,-3.1){$\lambda_{1,4}^{(i+1)}$}
\rput(-0.1,0.3){$\lambda_{1,4}^{(i)}$}
}
}
\psbezier[linewidth=1pt,linecolor=blue](0,1)(0.3,-0.2)(1,-0.5)(1,-1.5)
\psbezier[linewidth=1pt,linecolor= blue](0,-2)(0.2,-2.5)(1,-2.5)(1,-1.5)
\psbezier[linewidth=1pt,linecolor=yellow](0,1)(0.5,0)(1.4,-0.2)(1.4,-1.6)
\psarc[linecolor=yellow, linewidth=1pt](0.1,-1.6){1.3}{180}{360}
\psarc[linecolor=yellow, linewidth=1pt](-0.1,-1.6){1.1}{90}{180}
\psbezier[linewidth=1pt,linecolor=yellow](0,-2)(0.6,-2.5)(1.2,-0.5)(-0.1,-0.5)
\rput(0,0){\PAT{1}}
\end{pspicture}
$$ 
\caption{{The Dehn twist corresponding to $\gamma$ maps the arcs in the left-hand side of this picture to those on the right-hand side and it is given by one  generalized  mutation M.}}
\end{figure}

The cases of $PIV$, $PII$, and $PI$ correspond to finite cluster algebras admitting no nontrivial modular transformations.


\setcounter{section}{0}

\def\thetheorem{A.\arabic{theorem}}
\def\theprop{A.\arabic{prop}}
\def\thelemma{A.\arabic{lm}}
\def\thecor{A.\arabic{cor}}
\def\theexam{A.\arabic{exam}}
\def\theremark{A.\arabic{remark}}
\def\theequation{A.\arabic{equation}}

\appendix{Katz invariant and its geometric meaning}

In this appendix we consider the Painlev\'e monodromy manifolds as moduli spaces of flat connections on $\mathbb P\setminus S$, where $S$ is a set of isolated singularities ($S=\{0,1,t,\infty\}$ for $PVI$, $S=\{0,1,\infty\}$ for $PV,PIII^{D_6}$, $S=\{0,\infty\}$ for $PIV$, $S=\{\infty\}$ for $PII$ and $PI$). Points in this singular locus  $S$ may be regular (for example in the case of PVI) or irregular.  Irregular singular points are called ``non-ramified"  when the connection has non resonant residue at those points or ``ramified"  when the residue is resonant.

These moduli spaces have complex dimension 2 and are classified by the so-called {\it Katz invariants} associated to the singular locus $S$ of the flat connection.  
Informally speaking, let  $z$ be the coordinate in the punctured Riemann sphere and denote by $z_i$ the elements of $S$, then 
one can consider our flat connection as a local system whose local sections $q(z)$ have the following form: 
$$
\begin{array}{lr}
q(z) = a_1 (z-z_i)^{-k-1} +\ldots a_k (z-z_i)^{-1},&\hbox{non-ramified case,}\\
q(z) = (z-z_i)^{\frac{1}{2}}[a_1( z-z_i)^{-k-\frac{1}{2}} +\ldots a_{\ast} (z-z_i)^{-1}], &\hbox{ramified case}\\
\end{array}
$$
The number $k$ is called Katz invariant of the singular point $z_i$.

The simple poles have Katz invariant $k=0$ so that the $PVI$ linear differential system has the Katz invariants $(0,0,0,0).$  The confluence leads to two types of irregualr singularities: non-ramified with $k\in \mathbb Z$ and ramified ones with $k \in\frac{1}{2} + \mathbb Z.$ For example the $PV$ linear system has two simple poles $(0,1)$ and the non-resonant irregular point $\infty$ with $k=1$, so the Katz invariants are $(0,0,1)$, while the system for $PI$ has one ramified singular point $\infty$ with $k=\frac{5}{2}.$ See the precise definition of $k$ and all computational details can be found in  (\cite{SvdP}).

Our important observation is that the Katz invariants are given by the number of cusps on the corresponding hole divided by $2$.  So for example, $PVI$ corresponds to a Riemann sphere with $4$ holes and no cusps, so we have $0$ cusps on each hole, giving $(0,0,0,0)$ for the Katz invariants. For $PV$, we have two holes with no cusps and one hole with $2$ cusps on it, dividing by $2$ we obtain the Katz invariants $(0,0,1)$. The complete classification of the Katz invariants and the corresponding numbers of cusps is reported in the first three columns of  table \ref{tab:katz-1}.

\begin{table}[h]
\begin{center} 
\begin{tabular}{|c||c||c||c||c|c|} \hline 
 Painlev\'e eqs & no. of cusps & Katz invariants & no. Stokes rays & pole--orders for $\phi$ \\ \hline 
 $P{VI}$ &  $(0,0,0,0)$ &$(0,0,0,0)$&$(0,0,0,0)$ & $(2,2,2,2)$ \\ \hline 
 $P{V}$ & $(0,0,2)$& $(0,0,1)$& $(0,0,2)$& $(2,2,4)$\\ \hline 
 $PV_{deg}$& $(0,0,1)$ &$(0,0,1/2)$& $(0,0,1)$& $(2,2,3)$\\ \hline 
 $ P{IV}$     &  $(0,4)$& $(0,2)$ &  $(0,4)$   &  $(2,6)$\\ \hline
 $P{III}^{D_6}$ & $(0,2,2)$ & $(0,1,1)$& $(0,2,2)$& $(2,4,4)$\\ \hline 
 $P{III}^{ D_7}$ & $(0,1,2)$ &$(0,1/2,1)$& $(0,1,2)$& $(2,3,4)$\\ \hline 
 $P{III}^{ D_8}$ & $(0,1,1)$& $(0,1/2,1/2)$ & $(0,1,1)$& $(2,3,3)$\\ \hline  
 $P{II}^{FN}$ & $(0,3)$&  $(0,3/2)$& $(0,3)$& $(2,5)$\\ \hline  
$P{II}^{MJ}$ & $6$ & $3$& $6$&$8$ \\ \hline 
 $P{I}$  & $5$ & $5/2$& $5$&$7$\\ \hline 
\end{tabular}
\vspace{0.2cm}
\end{center}
\caption{For each Painlv\'e isomonodromic problem, this table displays the number of cusps on each hole for the corresponding Riemann surface, the Katz invariants associated to the corresponding flat connection, the number of Stokes rays in the linear system defined by the flat connection and the number of poles of the quadratic differential $\phi$ defined by the linear system.}
\label{tab:katz-1}
\end{table}

As a consequence of this observation, our cusps indeed correspond to Stokes rays at the irregular singular points. Indeed in 
 \cite{SvdP}  $2k$  Stokes rays are attached to each singular point with Katz invariant $k$. Moreover this relation between the number of cusps and the Katz invariant fits well with the computation of orders of poles for the corresponding quadratic differential (= Hitchin systems) in  the work by Gaiotto, Moore  and Neitzke \cite{GMN} : if the linear differential system for each isomonodromy problem has the form $\frac{dY}{dz} = A(z)Y$ then the corresponding quadratic differential 
is given by $\varphi (z) = {\rm det} A(z) dz^{\otimes 2}$. This means that our decorated character variety relates to the analytic properties of the quadratic differential $\varphi$: the number of the poles of $\varphi$ defines  the number of holes and the order of a $\varphi$ pole minus 2 defines the number of cusps on the corresponding hole. 
If the matrix $A(z)$ has a pole of order $n$ with diagonalisable leading residue matrix then the $\varphi$ has a pole of order $2n$. If the residue leading matrix is non--diagonalisable then the quadratic differential has a pole of order $2n-1.$  For example, for $PIII^{D_8}$ there are double poles in $0$ and in $\infty$ for $A(z)$, both have non--diagonalisable leading residue matrix. Hence, $\varphi$ has two third order poles in these points and the corresponding
monodromy surface is $\mathbb P^1$ with two holes and each of them has $3-2 =1$ bordered cusps. The matrix $A(z)$ for $PI$ has one pole of order 4, the quadratic differential has one pole with order $7=2\times 4-1$ 
and the surface has $7-2 =5$ cusps on the one hole (see table \ref{tab:katz-1}).
Note that these results agree with the work by T. Sutherland \cite{S} who used the auxiliary linear problem to produce a quadratic differential with high order poles on a punctured Riemann sphere. In his work, Sutherland associated a quiver to each of the above  
Painlev\'e cusped Riemann spheres and explicitly exhibit the canonical connected component of the space of numerical stability conditions of the Painlev\'e quivers.

\appendix{Singularity theory approach to the Painlev\'e cubics}\label{se:sing}

\def\thetheorem{B.\arabic{theorem}}
\def\theprop{B.\arabic{prop}}
\def\thelemma{B.\arabic{lm}}
\def\thecor{B.\arabic{cor}}
\def\theexam{B.\arabic{exam}}
\def\theremark{B.\arabic{remark}}
\def\theequation{B.\arabic{equation}}

As mentioned above,  for special values of $\omega_1^{(d)},\dots,\omega_4^{(d)}$ the fibre  may have a singularity. Such singularities were classified in \cite{IIS}  for $PVI$ and in \cite{SvdP} for all other Painlev\'e equations. These results can be summarised in the following table:

\begin{table}[h]
\begin{center} 
\begin{tabular}{|c||c|c|} \hline 
 Painlev\'e equations & Surface singularity type \\ \hline 
 $P_{VI}$ &  $D_4$ \\ \hline 
 $P_{V}$ & $A_3$ \\ \hline 
  deg $P_V$=$P_{III}(\widetilde D_6)$ & $A_1$ \\ \hline 
 $P_{III}$ & $A_1$\\ \hline  
 $P_{III}^{\widetilde D_7}$ & non-singular \\ \hline 
 $P_{III}^{\widetilde D_8}$ & non-singular \\ \hline  
$ P_{IV}$     &  $A_2$ \\ \hline
 $P_{II}^{FN}$ & $A_1$ \\ \hline  
$P_{II}^{MJ}$ & $A_1$ \\ \hline 
 $P_{I}$  & non-singular \\ \hline 
\end{tabular}
\vspace{0.2cm}
\end{center}
\caption{}
\label{tab:sing-1}
\end{table}

The meaning of the table is the following: for each Painlev\'e equation of type specified by the first column in the table,  there is at least one singular fibre with singularity of the type given in the second column of the table, and at least one singular fibre with singularity of type specified by any Dynkin sub-diagram of the Dynkin diagram given in the second column of the table. For example $PIV$ has a two singular fibres with singularity of type $A_2$ and at three singular fibres with singularity of type $A_1$.

The scope of this section is to show that the non singular fibres of each family of affine cubics are locally diffeomorphic to the versal unfolding \cite{Arnold} of the singularity of the type given in the second column of the table.

\subsection{$PVI$}
The cubic in this case is:
\be\label{eq:mon-mf-PVI}
x_1 x_2 x_3 + x_1^2+  x_2^2+  x_3^2 + \omega_1 x_1  + \omega_2  x_2 + \omega_3 x_3+
\omega_4 =0.
\ee
To show that this is diffeomorphic to the versal unfolding of  the $D_4$ we need to map this cubic to Arnol'd form. To this aim we first shift all variables by $x_i\to x_i +2$, $i=1,2,3$ to obtain
\be\label{eq:PVI-shifted}
x_1^2+x_2^2+x_3^2 + 2 x_1 x_2+ 2 x_2 x_3 + 2 x_1 x_3 + x_1 x_2 x_3 + \widetilde\omega_1 x_1  + \widetilde\omega_2  x_2 + \widetilde\omega_3 x_3+\widetilde\omega_4 =0,
\ee
where 
$$
 \widetilde\omega_i = \omega_i + 8, \quad\hbox{for } i=1,2,3,\qquad
  \widetilde\omega_4= \omega_4+2 (\omega_1+\omega_2+\omega_3)+20.
 $$
As a second step we use the following diffeomorphism around the origin:
$$
x\to x-\frac{1}{2} y, \quad y\to x+\frac{1}{2} x, \quad z\to z+\frac{y^2}{8}-2 x-\frac{x^2}{2} -\frac{\widetilde\omega_3}{2}
$$
so that the new cubic (up to a Morse singularity that we throw away and after a shift $x\to x-\frac{\omega_3}{4}$) becomes indeed the versal unfolding of a $D_4$ singularity in Arnol'd form:
$$
-2 x_1^3+ \frac{x_1 x_2^2}{2} +\widehat\omega_1 x_1+\widehat\omega_2 x_2+ \widehat\omega_3 x_1^2+\widehat\omega_4,
$$
where 
\bea
&&
\widehat\omega_1 = \omega_1+\omega_2 -8-4 \omega_3-\frac{\omega_3^2}{8}, \quad
\widehat\omega_2= \frac{\omega_2-\omega_1}{2},\nn\\
&&
\widehat\omega_3=8+\omega_3,\quad
\widehat\omega_4= \omega_4+2\omega_3-\frac{\omega_3(\omega_1+\omega_2-\omega_3)}{4}+4.\nn
 \eea
The above formulae show that the versal unfolding parameters $\widehat\omega_1,\dots,\widehat\omega_4$ are independent as long as $\omega_1,\dots,\omega_4$ are.

\subsection{$PV$}
The cubic in this case is:
\be\label{eq:mon-mf-PV}
x_1 x_2 x_3 + x_1^2+  x_2^2+ \omega_1 x_1  + \omega_2  x_2 + \omega_3 x_3+
\omega_4 =0,
\ee
where only three parameters are free:
$$
\omega_1=-G_2 G_3-G_1,\quad \omega_2=-G_1 G_3-G_2,\quad \omega_3=- G_3,\quad 
\omega_4=1+G_3^2+G_1 G_2 G_3.
$$
Again we want to show that this is diffeomorphic to the versal unfolding of $A_3$. To this aim we impose the following change of variables: 
\be\label{transf}
x_1\to u(x_2), \quad x_2\to x_1-x_3+\frac{G_3}{u(x_2)},\quad
x_3\to 2\frac{x_3}{u(x_2)}+\frac{G_2 + G_1 G_3}{u(x_2)}-\frac{2 G_3}{u(x_2)^2}, 
\ee
where $u(x_2)$ is a function to be determined. This maps the $PV$  cubic to:
$$
x_1^2-x_3^2 + 1+G_1 G_2 G_3+ G_3^2+ \frac{G_3^2}{u^2}-\frac{G_3(G_2+G_1 G_3)}{u}-(G_1+G_2 G_3) u + u^2.
$$
It is easy to prove that any solution $u(x_2)$ of the equation
$$
 \frac{G_3^2}{u^2}-\frac{G_3(G_2+G_1 G_3)}{u}-(G_1+G_2 G_3) u + u^2= x_2^4 + (G_2+G_1 G_3)x_2 ^2+ (G_1+G_2 G_3)x_2
 $$
will define a diffeomorphism by (\ref{transf}) mapping (\ref{eq:mon-mf-PV}) to the versal unfolding of $A_3$.

\subsection{$PIV$} 
The cubic in this case is:
\be\label{eq:mon-mf-PIV}
x_1 x_2 x_3 + x_1^2+ \omega_1 x_1  + \omega_2  x_2 + \omega_3 x_3+
\omega_4 =0,
\ee
where only two parameters are free:
$$
\omega_1=-G_1 G_\infty-G_\infty^2,\quad \omega_2=- G_\infty^2,\quad \omega_3=- G_\infty^2,\quad 
\omega_4=G_\infty^2+G_1 G_\infty^3.
$$
Again we want to show that this is diffeomorphic to the versal unfolding of $A_2$. To this aim we impose the following change of variables: 
\be\label{transf1}
x_1\to  x_1-x_3+\frac{G_\infty^2}{u},\quad 
x_2\to u,\quad x_3\to \frac{2 x_3}{u}+ \frac{G_\infty}{u}(G_1+G_\infty)-\frac{2 G_\infty^2}{u^2}
\ee
where $u$ is  function of $x_3$ satisfying the following 
$$
\frac{G_\infty^4}{u^2} -\frac{G_\infty^3(G_\infty+G_1)}{u}-G_\infty^2 u= x_2^3+ G_\infty x_2.
$$
It is easy to prove that this transformation is a local diffeomorphism mapping our cubic to 
$$
 x_1^2-x_3^2 +x_2^3+ G_\infty x_2+G_\infty+G_1 G_\infty^3,
$$
the versal unfolding of the $A_2$ singularity.

\subsection{$PIII^{D_6}$ and $PV_{deg}$}
The two cubics for $PIII^{D_6}$ and $PV_{deg}$ are equivalent. We choose to work with the $PV_{deg}$ one:
\be\label{eq:mon-mf-PV-1}
x_1 x_2 x_3 + x_1^2+  x_2^2+ \omega_1 x_1  + \omega_2  x_2+
1 =0,
\ee
where only two parameters are free:
$$
\omega_1=-G_1 ,\quad \omega_2=-G_2.
$$
The most singular fibre is given by $G_1=2$ and $G_2=2$ and has two singular points at $(1, 0, 2)$ and $(0, 1, 2)$ respectively. We can define two local diffeomorphisms, one around $(1, 0, 2)$, the other around $(0, 1, 2)$, which map our cubic to the versal unfolding of  a $A_1$ singularity. 

The first diffeomorphism is given by: 
$$
x_1\to x_1- \frac{\omega_1}{2},\quad 
x_2\to -x_2+ x_3 ,\quad x_3\to -\frac{2(2 x_3 +\omega_2  )}{2 x_1-\omega_1}
$$
The second diffeomorphism is:
$$
x_1\to -x_1+x_3,\quad 
x_2\to -x_2-\frac{\omega_2}{2} ,\quad x_3\to \frac{2( 2 x_3+\omega_1 )}{2 x_2+\omega_2}.
$$

\subsubsection{$PII$}
The two $PII$ cases are equivalent via a simple transformation (see Remark \ref{rem:PII}), so  we choose to work with the Jimbo-Miwa case:
\be\label{eq:mon-mf-PII}
x_1 x_2 x_3 - x_1  - x_2 -  x_3+
\omega_4 =0,
\ee
where:
$$
\omega_4=1+G_1.
$$
The following change of variables: 
$$
x_1\to  x_1-x_3+\frac{1}{u}, \quad x_2\to u, \quad 
x_3\to \frac{x_1+ x_3+1}{u},
$$
where $u$ is a function of $x_2$ satisfying 
$$
-\frac{1}{u}-u= x_2^2,
$$
is a local diffeomorphism mapping our cubic to the versal unfolding of the $A_1$ singularity:
$$
x_1^2-x_3^2+x_2^2+\omega_4.
$$



\begin{thebibliography}{99}

\footnotesize\itemsep=0pt


\bibitem{AKSM}
Alekseev A., Kosmann-Schwarzbach Y., Meinrenken E., Quasi-Poisson manifolds, {\it Canad. J. Math.,}\/ {\bf  54} (2002), no. 1:3--29.

\bibitem{AMM}
Alekseev A., Malkin A., Meinrenken E., Lie group valued moment maps., {\it J. Differential Geom.,}\/ {\bf 48} (1998), no. 3:445--495.

\bibitem{Arnold}
Arnol'd V.~I., Critical points of smooth functions and their normal forms, {\it Russian Math. Surveys,}\/ {\bf 30} (1975), no.5:3--65.

\bibitem{ARR}
Avan J., Ragoucy E., Rubtsov V., Quantization and Dynamisation of Trace-Poisson Brackets, {\it Commun. Math. Phys.,}\/ {\bf 341} (2016), no. 1:263--287. 


\bibitem{PB}
Boalch P., Geometry and braiding of Stokes data; fission and wild character varieties, {\it Ann. of Math. (2),} {\bf 179} (2014), no. 1:301--365. 

\bibitem{B2}
Boalch P., Wild Character Varieties, points on the Riemann sphere and Calabi's examples, {\it arXiv:1501.00930} (2015).


\bibitem{CantLor}
Cantat S., Loray, F., Dynamics on character varieties and Malgrange irreducibility of Painlev\'e VI equation. {\it Ann. Inst. Fourier (Grenoble)} \/{\bf 59} (2009), no. 7:2927--2978.


\bibitem{ChF1}
 Chekhov L., Fock V., {A quantum Techm\"uller space}, {\it Theor. and Math.
Phys.,}\/  {\bf120} (1999), 1245--1259.

\bibitem{ChF2}
Chekhov L., Fock V., {Quantum mapping class group,
pentagon relation, and geodesics},
{\it Proc. Steklov Math. Inst.,}\/
{\bf 226} (1999), 149--163.


\bibitem{ChM} 
Chekhov L., Mazzocco M.,
Shear coordinates on the versal unfolding of the $D_4$ singularity,
{\it J. Phys. A. Math. Gen.,}\/ {\bf 43}, (2010), 1--13.

\bibitem{ChM1} 
Chekhov L., Mazzocco M., Colliding holes in Riemann surfaces and quantum cluster algebras, {\it arXiv:1509.07044} (2015).

\bibitem{CMR1}
Chekhov L., Mazzocco M., Rubtsov V., quantized Painleve monodromy manifolds and Calabi-Yau algebras, {\it in progress}.

\bibitem{CMR}
Chekhov L., Mazzocco M., Rubtsov V., Quantum Stokes Phenomenon, {\it in progress}.

\bibitem{ChS}
Chekhov L., Shapiro M., Teichm\"uller spaces of Riemann surfaces with orbifold points of arbitrary order and cluster variables,
{\it Int. Math. Res. Not. IMRN,} {\bf 2014}, no. 10:2746--2772. 

\bibitem{DM}
Dubrovin B.A., Mazzocco M., Monodromy of certain Painlev\'e-VI transcendents and ref\/lection group,
{\it Invent. Math.} {\bf 141} (2000), 55--147.

\bibitem{FN}
Flaschka H. and Newell A.C.,
Monodromy and spectrum preserving
deformations I,
 {\it Commun. Math. Phys.}{\bf 76} (1980) 65--116.

\bibitem{Fock1}
Fock V.V., Combinatorial description of the moduli space of
projective structures, {http://arxiv.org/abs/hep-th/9312193}{hep-th/9312193}.

\bibitem{Fock2}
Fock V.V., Dual Teichm\"uller spaces, {http://arxiv.org/abs/dg-ga/9702018}{dg-ga/9702018}.

\bibitem{GMN}
Gaiotto D., Moore G.~W., Neitzke A.,
Wall-crossing, Hitchin systems, and the WKB approximation,
{\it Adv. Math.,}\/ {\bf 234} (2013), 239--403. 

\bibitem{Gold}
Goldman W.M., {\it Invariant functions on Lie groups and Hamiltonian
f\/lows of surface group representations}, {\it Invent. Math.}\/ {\bf85}
(1986), 263--302.

\bibitem{Crelle2015}
Gualtieri M., Li S., Pym B., The Stokes groupoids,
{\it J. Reine Angew. Math.,}\/ to appear in (2016)


\bibitem{GrossHK}
Gross M.,  Hacking P., Keel S., Mirror symmetry for log Calabi-Yau surfaces I,
{\it Publ. Math. Inst. Hautes \'Etudes Sci.,}\/{\bf 122} (2015), 65--168. 


\bibitem{Hit}
Hitchin N., 
Frobenius manifolds (with notes by David Calderbank),
{\it NATO Adv. Sci. Inst. Ser. C Math. Phys. Sci.,}\/  {\bf 488} (1997) 69--112.

\bibitem{IIS}
Inaba M., Iwasaki K., Saito M., 
Dynamics of the sixth Painlev\'e equation,
in {\it Th\'eories asymptotiques et \'equations de Painlev\'e,}\/ {\sl S\'emin. Congr.,}\/ {\bf 14} (2006) 103--167.

\bibitem{J}
Jimbo M., Monodromy Problem and the Boundary Condition for Some Painlev\'e Equations,
{\it Publ. RIMS, Kyoto Univ.,}\/ {\bf 18} (1982) 1137--1161.

\bibitem{MJ1}
Jimbo M. and Miwa T.,
Monodromy preserving deformations of linear ordinary differential
equations with rational coefficients \text{II},
{\it Physica 2D,}\/ 
{\textbf{2}} (1981),
no. 3, 407--448.

\bibitem{Kashaev-Dehn}
Kashaev R.~M., {\it On the spectrum of Dehn
twists in quantum Teichm\"uller theory},
in: {\sl Physics and Combinatorics}, (Nagoya 2000). River Edge, NJ, World Sci. Publ.,
2001, 63--81; math.QA/0008148.


\bibitem{KS}
Korotkin D. and Samtleben H.,
Quantization of coset space $\sigma$-models coupled to two-dimensional gravity,
{\it Comm. Math. Phys.}\/ {\bf 190} (1997), no. 2, 411--457.


\bibitem{L-BS}
Li-Bland D., Severa P.,
Symplectic and  Poisson  geometry of the moduli spaces
of flat connections over quilted surfaces.
{\it http://xxx.lanl.gov/abs/1304.0737v.2}

\bibitem{MasTur}
Massuyeau G., Turaev V., Quasi-Poisson structures on representation spaces of surfaces, {\it IMRN,}\/ {\bf 2014} no1:1--64 (2014).

\bibitem{M1}
Mazzocco M., 
Rational solutions of the Painleve' VI equation, 
Kowalevski Workshop on Mathematical Methods of Regular Dynamics (Leeds, 2000). 
{\it J. Phys. A} {\bf 34} (2001), no. 11, 2281--2294. 


\bibitem{MM}
Mazzocco M.,
Confluences of the Painlev\'e equations, Cherednik algebras and q-Askey scheme, {\it http://arxiv.org/abs/1307.6140}  to appear on {\it Nonlinearity} (2016).

\bibitem{Obl}
Oblomkov A., Double Affine Hecke Algebras of Rank 1 and Affine Cubic Surfaces,
{\it IMRN,}\/ {\bf 2004} no.18:877--912.

\bibitem{PR}
Paul E., Ramis, Jean-Pierre, Dynamics on Wild Character Varieties,
{\it SIGMA} {\bf}11 (2015), 068, 21 pages 


\bibitem{Penn1}
Penner R.C., {\it The decorated Teichm\"uller space of Riemann surfaces},
{\sl Comm. Math. Phys.} {\bf113} (1988), 299--339.


\bibitem{SvdP}
Saito M., van der Put M.,
Moduli spaces for linear differential equations and the Painlev\'e equations,
{\it arXiv:0902.1/02v5} (2009).

\bibitem{sakai}
Sakai H., Rational Surfaces Associated with Affine Root Systems and Geometry of the Painlev\'e Equations,
{\it Commun. Math. Phys.} {\bf 220}, (2001)  165--229. 

\bibitem{S}
Sutheralnd T., PhD thesis, {\it work in progress} (2013).


\bibitem{ThSh}
Thurston W.~P., {\it Minimal stretch maps between
hyperbolic surfaces}, preprint (1984),
math.GT/9801039.




\end{thebibliography}
\end{document}